\documentclass[reqno,11pt]{article}
\pdfoutput=1
\usepackage{ifpdf}
\usepackage[usenames,dvipsnames]{xcolor}
\usepackage{slashed}
\usepackage[T1]{fontenc}
\usepackage[ansinew]{inputenc}
\usepackage{amsmath}
\usepackage{amssymb}
\usepackage{booktabs}
\usepackage{siunitx}
\usepackage{graphicx}
\usepackage{color}
\definecolor{darkblue}{cmyk}{0.9,0.9,0,0}
\definecolor{darkgreen}{rgb}{0,0.55,0}
\usepackage[colorlinks=true,linkcolor=darkblue,citecolor=darkblue,urlcolor=darkblue]{hyperref}
\usepackage{simpler-wick}
\usepackage[export]{adjustbox}
\usepackage{mathtools}
\usepackage{minitoc}
\usepackage{mathtools}
\usepackage{subcaption}
\usepackage[normalem]{ulem}
\usepackage{bbm}
\usepackage{cite}

\usepackage{cancel}
\usepackage{physics}
\usepackage{tensind}
\usepackage{tensor}

\tensordelimiter{?}

\makeatletter
\DeclareRobustCommand\widecheck[1]{{\mathpalette\@widecheck{#1}}}
\def\@widecheck#1#2{
    \setbox\z@\hbox{\m@th$#1#2$}
    \setbox\tw@\hbox{\m@th$#1
       \widehat{
          \vrule\@width\z@\@height\ht\z@
          \vrule\@height\z@\@width\wd\z@}$}
    \dp\tw@-\ht\z@
    \@tempdima\ht\z@ \advance\@tempdima2\ht\tw@ \divide\@tempdima\thr@@
    \setbox\tw@\hbox{
       \raise\@tempdima\hbox{\scalebox{1.2}[-1]{\lower\@tempdima\box
\tw@}}}
    {\ooalign{\box\tw@ \cr \box\z@}}}
\makeatother

\newcommand{\comment}[1]{}

\newcommand{\beq}{\begin{equation}}
\newcommand{\eeq}{\end{equation}}
\newcommand{\beqq}{\begin{equation*}}
\newcommand{\eeqq}{\end{equation*}}
\newcommand\beqa{\begin{eqnarray}}
\newcommand\eeqa{\end{eqnarray}}
\newcommand\beqaa{\begin{eqnarray*}}
	\newcommand\eeqaa{\end{eqnarray*}}
\newcommand\bea{\begin{array}}
	\newcommand\eea{\end{array}}

\newcommand{\ii}{\operatorname{i}}

\def\XXint#1#2#3{{\setbox0=\hbox{$#1{#2#3}{\int}$ }
		\vcenter{\hbox{$#2#3$ }}\kern-.5\wd0}}

\def\XXint#1#2#3{{\setbox0=\hbox{$#1{#2#3}{\int}$}
		\vcenter{\hbox{$#2#3$}}\kern-.5\wd0}}

\newcommand{\nn}{\nonumber}

\newcommand{\neqa}{\nonumber\end{eqnarray}}
\newcommand{\la}[1]{\label{#1}}

\newcommand{\p}{\partial}

\def\tr{{\rm tr\,}}

\newcommand{\hs}{\frac{\sqrt{3}}{2}}
\renewcommand{\d}{\partial}

\newcommand{\<}{{\langle}}
\renewcommand{\>}{{\rangle}}

\newcommand{\cL}{{\cal L}}

\newcommand{\re}{\relax{\rm I\kern-.18em R}}

\renewcommand{\sp}{p\hspace{-.40em}/}

\def\su2{{SU(2)}}

\def\[{\left[}
\def\]{\right]}

\def\s{\sigma}

\def\({\left(}
\def\){\right)}
\def\[{\left[}
\def\]{\right]}

\def\<{\langle}
\def\>{\rangle}

\def\cO{{\cal O}}

\def\s*{\ *_{\!\!\!\!\!\!\!\!\!\,_{\,_\text{\scriptsize{sym}}}}}
\def\hs*{\ \hat{*}_{\!\!\!\!\!\!\!\!\!\,_{\,_\text{\scriptsize{sym}}}}}
\def\d{\partial}

\def\i2{\frac{i}{2}}

\def\spi{\relax{\rm \pi\kern-0.5em /}}
\def\sA{\relax{\rm A\kern-0.5em /}}
\def\sp{\relax{\rm p\kern-0.5em /}}
\def\sd{\relax{\rm \d\kern-0.5em /}}
\def\sk{\relax{\rm k\kern-0.5em /}}
\def\sn{\relax{\rm n\kern-0.5em /}}
\def\sl{\relax{\rm l\kern-0.5em /}}
\def\sP{\relax{\rm P\kern-0.7em /}}
\def\sBethe{\relax{\rm \Bethe\kern-0.5em /}}

\def\be#1\ee{\begin{equation}\begin{aligned}
#1
\end{aligned}
\end{equation}}

\usepackage{caption}
\numberwithin{equation}{section}

\usepackage{varioref}
\usepackage{makeidx}
\usepackage{cleveref}
\makeindex

\begin{document}
\begin{center}
$$$$
{\Large\textbf{\mathversion{bold}
A Large Twist Limit for Any Operator
}\par}

\vspace{1.0cm}

\textrm{Gwena\"el Ferrando$^a$, Amit Sever$^a$, 
Adar Sharon$^b$, and Elior Urisman$^a$}
\\ \vspace{1.2cm}
\footnotesize{

\textit{$^{a}$ School of Physics and Astronomy, Tel Aviv University, Ramat Aviv 69978, Israel\\
$^{b}$ Simons Center for Geometry and Physics, SUNY, Stony Brook, NY 11794, USA}}

\par\vspace{1.5cm}

\textbf{Abstract}\vspace{2mm}
\end{center}

\noindent

We argue that for any single-trace operator in ${\cal N}=4$ SYM theory there is a large twist double-scaling limit in which the Feynman graphs have an iterative structure. Such structure can be recast using a graph-building operator. Generically, this operator mixes between single trace operators with different scaling limits. The mixing captures both the finite coupling spectrum and corrections away from the large twist limit. We first consider a class of short operators with gluons and fermions for which such mixing problems do not arise, and derive their finite coupling spectra. We then focus on a class of long operators with gluons that do mix. We invert their graph-building operator and prove its integrability. The picture that emerges from this work opens the door to a systematic expansion of ${\cal N}=4$ SYM theory around the large twist limit.

\vspace*{\fill}

\setcounter{page}{1}
\renewcommand{\thefootnote}{\arabic{footnote}}
\setcounter{footnote}{0}

\newpage

 \def\nref#1{{(\ref{#1})}}

\tableofcontents

\newpage

\section{Introduction}

Due to the enormous attention the holographic principle has received in the last two decades, we have a fairly complete dictionary between the bulk and boundary observables, at least in the large $N$ limit. Yet knowing and testing the holographic map is not the same as understanding it. Even in the planar limit, we are still lacking a first-principle derivation of the duality. 
Such a derivation, however, was achieved for a simplifying limit of $\gamma$-deformed ${\cal N}=4$ SYM theory, called the fishnet theory \cite{Gurdogan:2015csr}. 
Taking such a limit, with simplified dynamics, comes at a price -- it only captures a subset of the observables and has more restrictive physical phenomena. 
Hence, the question remains how to extend this derivation to the full theory. In this paper, we take the first step in this direction. Explicitly, we show that the introduction of a so-called ``twist'' into the planar diagrams allows us to extend the results from the fishnet theory to any operator in ${\cal N}=4$ SYM theory.

A twist is defined using a symmetry generator and a continuous parameter, $e^{\ii\theta}$, \cite{Cavaglia:2020hdb}. It has the effect of giving different weights to different diagrams in a physically meaningful way, see appendix \ref{app:gammadeformed} and 
\cite{Gurdogan:2015csr, Cavaglia:2020hdb, Correa:2012nk} for details. Taking a large twist double-scaling limit, in particular, makes it possible to project to a certain subclass of diagrams. This limit is the limit in which we simultaneously take the twist parameter to be large and the 't Hooft coupling $g^2=g_\text{YM}^2N/(4\pi)^2$ to be small as
\beq\la{fnlimit}
e^{\ii\theta}\to\infty\,,\qquad\text{with}\quad\xi_n^2\equiv {g^2\,e^{\ii\frac{\theta}{n}} \over8\pi^2}\quad\text{fixed}\,,
\eeq
where 
$n>0$ depends on the operator. The remaining diagrams enter in a systematic expansion around this double-scaling limit.

The case of $n=1$ with a certain $R$-symmetry twist is the so-called ``fishnet'' limit that has been extensively studied in the literature \cite{Gurdogan:2015csr,Grabner:2017pgm,Gromov:2018hut,Kazakov:2018gcy,Basso:2019xay,Caetano:2016ydc,Basso:2017jwq,Mamroud:2017uyz,Basso:2018agi,Korchemsky:2018hnb,Basso:2018cvy,Karananas:2019fox,Ipsen:2018fmu,Cavaglia:2021mft,Basso:2021omx,Shahpo:2021xax,Olivucci:2021pss}. It was found to preserve the integrability of the planar theory, \cite{Gromov:2017cja,Chicherin:2017cns,Chicherin:2017frs,Kazakov:2018ugh,Loebbert:2019vcj}. Moreover, it turned out to have a holographic dual description in terms of a chain of point particles in $AdS_5$, \cite{Gromov:2019aku,Gromov:2019bsj,Gromov:2019jfh,Gromov:2021ahm}. Notably, both of these properties have been derived from first principles.

To extend these derivations to the mother ${\cal N}=4$ SYM theory, we need to gradually turn on the 't Hooft coupling $g^2$ while holding $\xi_1^2$ fixed. However, we then have to face the following problem. Most of the operators of the ${\cal N}=4$ SYM theory remain trivial in the fishnet limit. The perturbative (in $g$) expansion around the fishnet limit is analogous to the perturbative expansion around the free theory for them. Hence, we did not gain much by first taking the limit (\ref{fnlimit}). Here, we suggest an alternative constructive path towards ${\cal N}=4$ SYM theory. It involves a reorganisation of the perturbative expansion in an operator-dependent way. If all orders in this expansion are resumed, the full (twisted) ${\cal N}=4$ SYM result is reconstructed.

Concretely, we consider two-point functions of single-trace operators. We argue that, for any operator $\cO$, there is some twist generator and some $n_\cO>0$ such that in the corresponding double-scaling limit (\ref{fnlimit}), the diagrams that survive have an iterative structure. For most of the operators, taking their double-scaling limit involves a mixing problem with sub-leading loop corrections to operators with the same twist generator, but lower $n_\cO$. 

Diagrams with an iterative structure can be generated by a so-called {\it graph-building operator}. Such an operator is expected to exist even for the full ${\cal N}=4$ SYM mother theory. However, in that case, it is an infinite by infinite matrix, and therefore of very little practical use. Nevertheless, turning on a twist gives a grading on this infinite matrix, such that at any fixed order only a finite matrix remains. This property then allows us to analyse the matrix in steps. At each step we have to deal with a finite block of this matrix, which corresponds to a subset of operators. Gradually, more and more operators are added. Each time an operator $\cO$ is added, it receives all loop corrections in $\xi_{n_\cO}$ and has a corresponding holographic description. At the same time, we capture higher loop corrections in $g$ to operators $\cO'$ that were already included to all orders in their $\xi_{n_{\cO'}}$.

\begin{figure}[t]
\centering
\includegraphics[width=0.85\textwidth]{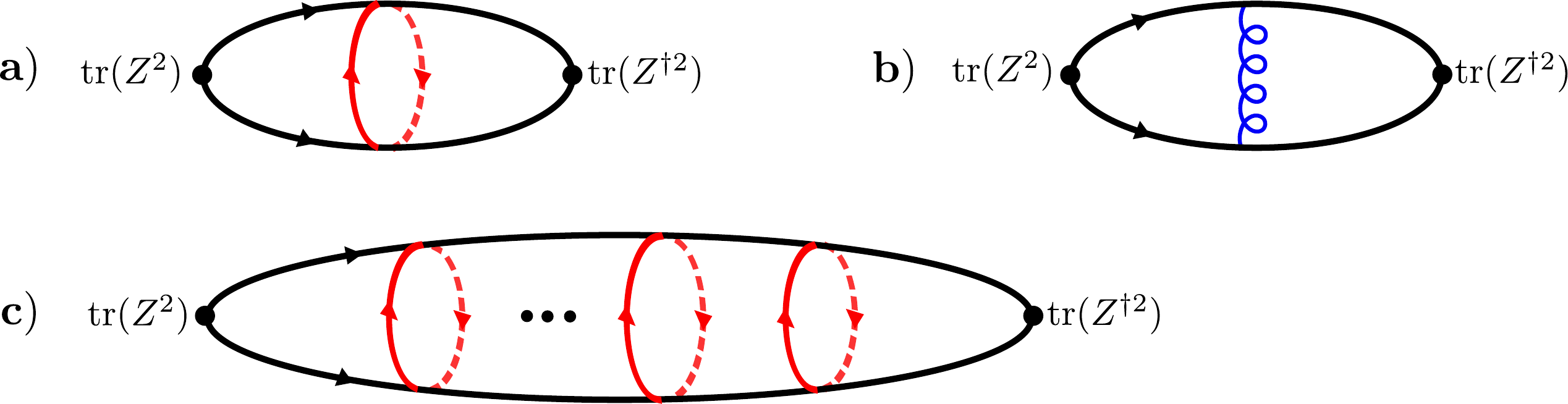}
\caption{\small In the fishnet limit, diagrams that contribute to the two-point function of the operator $\tr(Z^2)$ have wheel, or fishnet \cite{Gurdogan:2015csr,Grabner:2017pgm}, structure that can be resummed. The diagram in (a) has one wheel and is of order $\xi_1^4$. The diagram in (b), on the other hand, is of order $g^2$ and does not survive the fishnet limit. A typical diagram that does survive and has many wheels is plotted in (c). Here, the black lines denote $Z$-propagators and red lines denote $X$-propagators. The arrows indicate the flow of the $U(1)$ charge of these operators and are pointing towards $X^\dagger$, $Z^\dagger$.}\label{fig:introfigs}
\end{figure}

To demonstrate this structure, consider a toy example of an $SU(N)$ gauge theory consisting of two complex scalars, $Z$ and $X$, transforming in the adjoint representation:
\beq\la{simpleL}
\cL=- N\,\tr\Big[\frac{1}{4}F_{\mu\nu}F^{\mu\nu} + (D^{\mu}Z)^{\dagger}D_{\mu}Z+ (D^{\mu}X)^{\dagger}D_{\mu}X-2g^2
(X^\dagger Z^\dagger XZ+
Z^\dagger X^\dagger ZX)\Big]\,,
\eeq
where $D_{\mu}=\partial_{\mu}+\ii g[A_{\mu},\cdot]$ is the covariant derivative. In the planar limit, the rank $N\to\infty$ while the 't Hooft coupling $g^2$ is held fixed. In general, twisting a single-trace operator by a symmetry transformation is defined at the diagrammatic level, see \cite{Cavaglia:2020hdb} for details. Here, we choose to twist by an internal symmetry under which the $X$ and $Z$ have charge one. For this choice of symmetry, it can be shown that twisting is equivalent to deforming the interaction term in the Lagrangian (\ref{simpleL}) as
\beq
2g^2
(X^\dagger Z^\dagger XZ+
Z^\dagger X^\dagger ZX)\quad\rightarrow\quad 2g^2
(e^{\ii\theta}X^\dagger Z^\dagger XZ+e^{-\ii\theta}
Z^\dagger X^\dagger ZX)\,,
\eeq
where $\theta$ is the twist angle. 
Consider for example the diagrams in figure \ref{fig:introfigs} that contribute to the two-point function of the operator $\tr\!\!\(Z^2\)$. The diagram in figure \ref{fig:introfigs}.a scales as $g^4e^{2\ii\theta}$ whereas the diagram in \ref{fig:introfigs}.b scales as $g^2$. Hence, only the first survives the fishnet double-scaling limit, (\ref{fnlimit}) with $n=1$. At higher loop orders, the only diagrams that survive in the fishnet limit are the wheel-type diagrams in figure \ref{fig:introfigs}.c . They are all generated by iteration of the fishnet graph-building operator. This operator, that acts on two scalars, adds one more wheel to the diagram. The gauge field, on the other hand, decouples from any other correlator and we remain with the two complex scalars and the first quartic interaction term in (\ref{simpleL}) only. The corresponding non-trivial local single-trace operators are made from these two scalars, with no gauge fields.

\begin{figure}[t]
\centering
\includegraphics[width=0.6\textwidth]{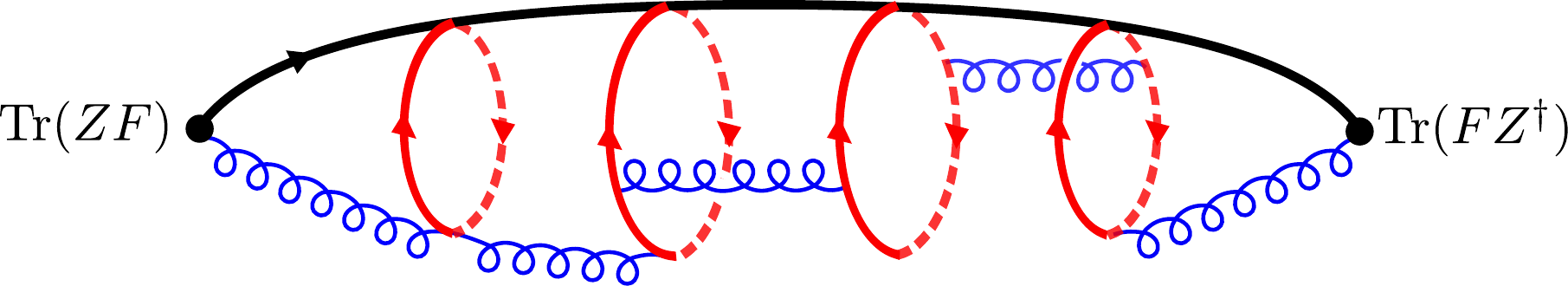}
\caption{\small Example of a high-loop diagram that contributes to the two-point function of the operator $\tr(F_{\mu\nu}Z)$ in the double-scaling limit (\ref{fnlimit}) with $n=2$.}\label{fig:intro_gluon_diagram}
\end{figure}

Suppose that instead of the operator $\tr\!\!\(Z^2\)$ we start with the operator
\beq\la{simpleop}
\cO_{\mu\nu}(x)=\tr\!\big(F_{\mu\nu}(x)Z(x)\big)\,.
\eeq
This is an example of an operator that decouples in the fishnet limit because it includes gluons. However, if we take the limit (\ref{fnlimit}) with $n=2$ instead of the fishnet limit, the gauge field no longer decouples and the operator (\ref{simpleop}) receives loop corrections in $\xi_2$. The corresponding diagrams that survive in this limit are of wheel-type with gluons, see figure \ref{fig:intro_gluon_diagram}. In section \ref{sec:FZ} we study the two-point function between operators of this type in detail and use the corresponding graph-building operator to resum them exactly. In particular, we find that in the limit (\ref{fnlimit}) with $n=2$, the exact conformal dimension of $\cO_{\mu\nu}$ is
\beq
\Delta_\cO=2+\sqrt{5-4\sqrt{1+\xi_2^4}}\,.
\eeq

It is worth noting that the operator $\cO_{\mu\nu}$ in (\ref{simpleop}) is in an antisymmetric representation of the Lorentz group. Consequently, it has zero overlap with the trace of two $Z$-scalars at arbitrary positions. As a result, the fishnet wheel diagrams that contribute to the two-point function of $\tr\!\!\(Z^2\)$ and are associated with the limit (\ref{fnlimit}) with $n=1$, decouple from the two-point function of $\cO_{\mu\nu}$. This is no longer the case for operators of the form $\tr\!\!\(F_{\mu\nu}Z^J\)$ with $J>1$. For them, we have to deal with a mixing problem between operators with different scaling limits. This mixing is studied in section \ref{sec:mixing} and involves the operators $\tr\!\!\(Z^J\)$ and $\tr\!\!\(X X^\dagger Z^J\)$.\footnote{These two operators have different tree-level dimension. Therefore, they do not mix with each other at the level of the dilatation operator. Instead, the mixing alluded to here happens at the level of the graph-building operator. It represents a mixing between these operators with any number of derivatives acting on the fields in the trace.}

In this paper, we focus on a relatively simple subset of single trace operators, with scalars, gluons, and fermions. The strategy we follow for studying their two-point functions involves two steps. We first twist them and then take the appropriate large twist double-scaling limit. We expect that the structure we find is general. That is, a single-trace local operator can be twisted by any internal or space-time global symmetry that leaves it invariant. Using the internal $SU(4)$ $R$-symmetry, an appropriate double-scaling limit can be applied to any operator. For simplicity, we will only consider a subset of twists for which the computation of the two-point function of the twisted operators in the ${\cal N}=4$ SYM theory is the same as the computation of the untwisted correlator in the $\gamma$-deformed theory. We refer the reader to \cite{Cavaglia:2020hdb} for a detailed definition of the colour-twist method and proof of its equivalence to a $\gamma$-deformation for the cases we study here.

In section \ref{section-shortoperators}, we present three examples of short operators whose scaling limit is different from the fishnet one. In section \ref{sec:mixing}, we tackle the issue of mixing alluded to above. In section \ref{section-longeroperators}, we study another infinite family of operators with fermions and a different scaling limit. Lastly, in section \ref{sec:holography}, we briefly discuss the holographic dual description of these operators. Various appendices contain additional details.

\section{Short Operators} \label{section-shortoperators}

For the examples studied in this paper, it is enough to consider correlators of operators in the $\gamma$-deformed ${\cal N}=4$ SYM theory, whose Lagrangian we recall in appendix \ref{app:gammadeformed}. In this section, we will focus on the operators
\beq\la{threeop}
\tr ( XX^\dagger Z )\,,\qquad\tr (X^\dagger X Z )\,,\qquad
\text{and}\qquad \tr(F_{\mu\nu} Z)\,,
\eeq
as well as their generalisations with derivatives. 
In the fishnet limit ((\ref{fnlimit}) with $n=1$) all of the loop corrections to their conformal dimensions vanish. Hence, these are examples of operators that do not survive the fishnet limit.

Instead of the fishnet limit, we consider the limit (\ref{fnlimit}) with $n=2$. More explicitly, in terms of the $\gamma$-deformation parameters that we review in appendix \ref{app:gammadeformed}, the limit we consider is
\beq\la{fnlimitgamma2}
\gamma_{1} = \gamma_{2} = 0\,, \qquad e^{-\ii\gamma_3}\to\infty\,, \qquad\text{with}\qquad  \xi_2^2\equiv{g^2\,e^{-\ii\frac{\gamma_3}{2}}\over8\pi^2}\quad\text{fixed}\,.
\eeq
In this limit, the correlators involving the operators in (\ref{threeop}) receive loop corrections of arbitrarily high order. As in the fishnet theory, these loop corrections have an iterative structure. In this section, we use the corresponding graph-building operator to compute the operators' conformal dimensions at finite $\xi_2$.

\subsection{The Operator $\tr(XX^\dagger Z)$}
\label{sec:ZXX}

Single-trace operators composed of a single $Z$, $X$, and an $X^\dagger$ field are the simplest examples of operators with a double-scaling limit different from that of the fishnet. 
For every spin, we find two twist three primary operators with a non-trivial anomalous dimension, while the other operators of this form remain trivial. In order to analyse all non-trivial operators at once, it is sufficient to study an operator that has some overlap with all of them. This is achieved by choosing 
one ordering of the fields in the trace and placing them at two different spacetime points. The result does not depend on these choices, as explained later. 
We choose to use the operators
\beq\la{pointsplit}
\tr \big(Z(x)(X X^\dagger)(y) \big)\qquad\text{and}\qquad\tr\big((Z^\dagger X)(x)X^\dagger(y)\big)\,.
\eeq
At finite coupling and twist, operators of this form are not gauge invariant. 
Instead, similar operators with Wilson lines between the fields in the trace can be thought of. However, in the limit \eqref{fnlimitgamma2}, the Wilson lines decouple, and one remains with the operators in (\ref{pointsplit}). 

To compute the non-trivial conformal dimensions of the local operators in this sector,  
we study the two-point function of the non-local ones in (\ref{pointsplit})
\begin{equation}\label{Gcorrelator}
    G(x_1,x_2|x_3,x_4)\equiv\langle \tr\big(Z(x_1)(X X^\dagger)(x_2)\big)\, \tr\big((Z^\dagger X) (x_3)X^\dagger(x_4)\big)\rangle = \sum_{L=0}^{\infty} \xi_2^{4L}G^{(L)}(x_1,x_2|x_3,x_4)\, .
\end{equation}
We will sometimes abuse the notation and refer to this correlator as a ``four-point function'', even though it is really a two-point function of two non-local single-trace operators. 

\begin{figure}[t]
\centering
\includegraphics[width=0.8\textwidth]{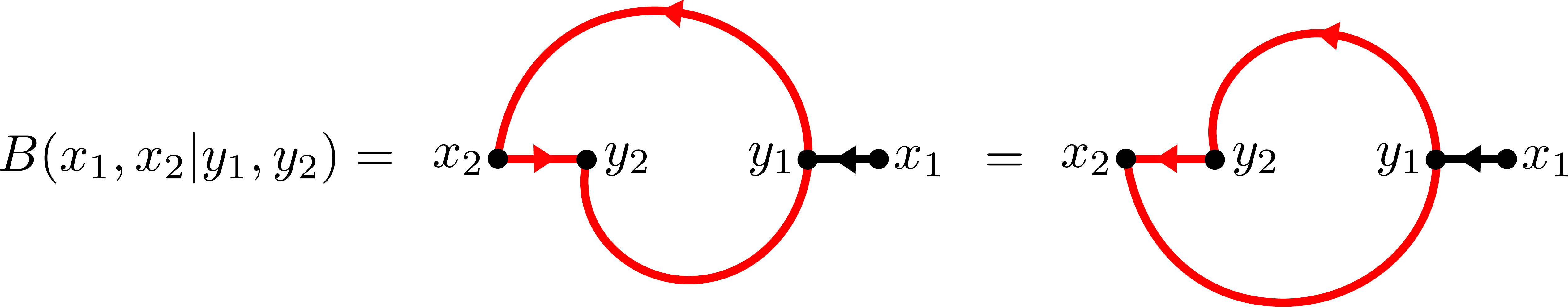}
\caption{\small The graph-building operator $\widehat B$ in (\ref{Bop}). The two drawings represent the two ways of adding a wheel to a diagram. The corresponding overall factor of two is part of our definition of the coupling $\xi_2^2$ in (\ref{fnlimit}).}
\label{fig:XXbar_GBO}
\end{figure}

The only interaction terms in the Lagrangian (\ref{l-int}) that contribute to this correlator in the limit (\ref{fnlimitgamma2}) are
\begin{equation}\label{intxx}
    \cL_\text{int}=Ng^2\,\tr(X^\dagger X^\dagger X X)+2 Ng^2e^{-\ii \gamma_3}\,\tr(X^\dagger Z^\dagger XZ) + \text{non-relevant for }G\, .
\end{equation}
\begin{figure}[t]
\centering
\includegraphics[width=0.8\textwidth]{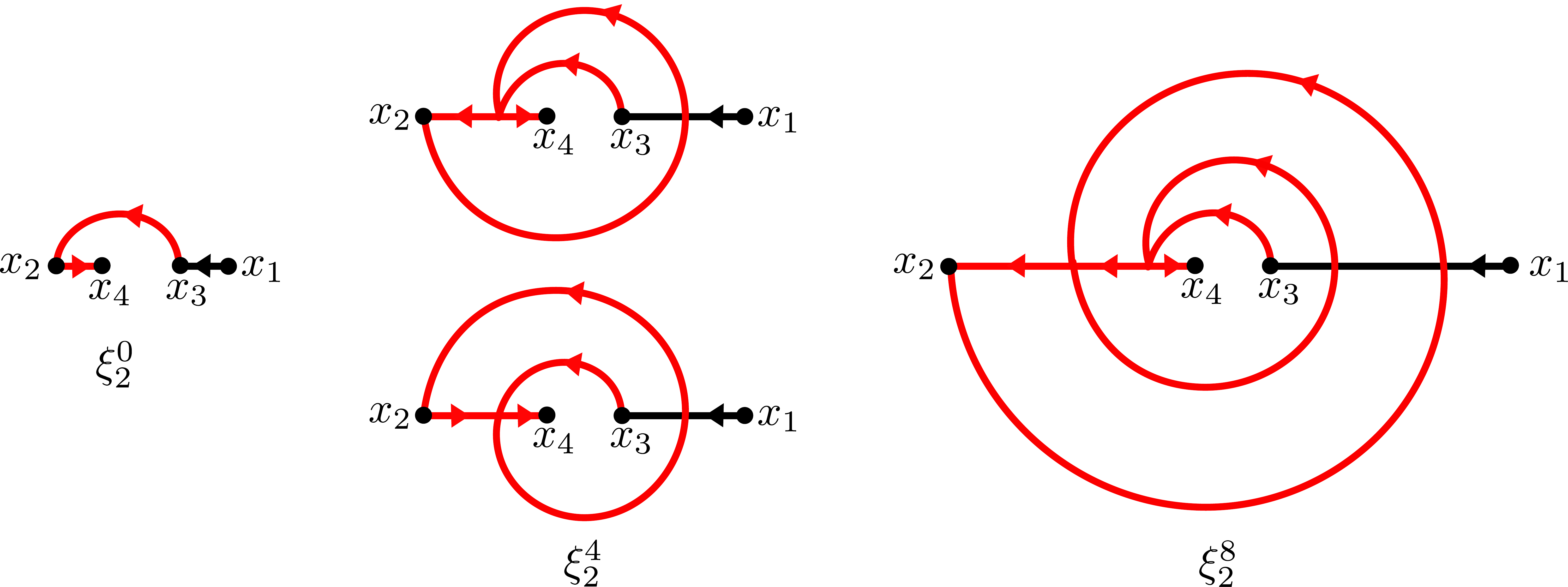}
\caption{\small The first few diagrams that contribute to the correlator (\ref{Gcorrelator}) in the double-scaling limit (\ref{fnlimitgamma2}). Here, the red lines denote $X$ propagators and black lines denote $Z$ propagators. The intersection of a red line and a black line represents the vertex $\tr(X^\dagger Z^\dagger XZ)$, and the intersection of two red lines represents the vertex $\tr(X^\dagger X^\dagger XX)$.}
\label{fig:xxbar_pertub}
\end{figure}
In figure \ref{fig:xxbar_pertub} we have plotted some of the first few diagrams contributing to the correlator (\ref{Gcorrelator}). At tree-level, (order $\xi_2^0$), the correlator is plotted in figure \ref{fig:xxbar_pertub}. and is given by the product of three propagators
\begin{equation} \label{g0-scalar}
    G^{(0)}(x_1 ,x_2|x_3 ,x_{4}) = \frac{(4\pi^2)^{-3}}{x_{13}^2 x_{23}^2 x_{24}^2}\,,
\end{equation}
where $x_{ij}=x_i-x_j$. 
At the next order in perturbation theory (order $\xi_2^4$), we have the two diagrams of figure \ref{fig:xxbar_pertub}. They give exactly the same contribution. This phenomenon repeats itself at each order in perturbation theory. Namely, at each order, one can add a ring to the diagrams of the previous order using either of the drawings in figure \ref{fig:XXbar_GBO}, both resulting in the same kernel. After taking this factor of $2$ into account in the definition of $\xi_2$ in \eqref{fnlimitgamma2}, we arrive at 
\begin{equation}
    G^{(2)}(x_1,x_2|x_3,x_4) = \int \frac{\dd^4 y_1 \dd^4 y_2}{\pi^4}\frac{G_1 ^{(0)}(y_1 ,y_2|x_3 ,x_{4})}{(x_1-y_1)^2 (x_2-y_1)^2 (x_2-y_2)^2 y_{12}^2} \, .
\end{equation}

\begin{figure}[t]
\centering
\includegraphics[scale=0.4]{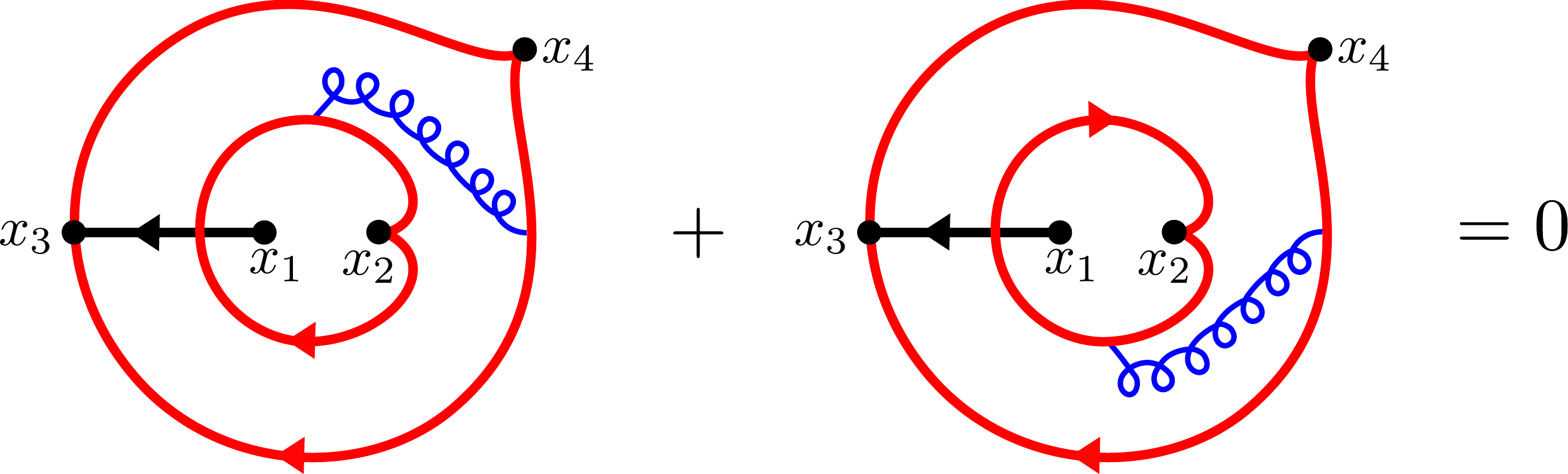}\caption{\small Cancellation of diagrams where a gluon appears.}\label{fig:XXbar_gluon_nomixing}
\end{figure}

At order $\xi_2^4$, we also have the two diagrams in figure \ref{fig:XXbar_gluon_nomixing}, where a gluon is exchanged between the $X$ scalars. These two contributions, however, cancel against each other and do not affect the correlator (\ref{Gcorrelator}). 
This cancellation persists at all orders in perturbation theory. Consequently, the only graphs that survive are of the type of those drawn in figure \ref{fig:xxbar_pertub}. They have an iterative structure that is generated by the graph-building operator $\widehat{B}$. It is defined by its action on the correlator (\ref{Gcorrelator}) as
\begin{equation}
    G^{(L+2)}(x_1,x_2|x_3,x_4) = \int \frac{\dd^4 y_1 \dd^4 y_2}{\pi^4}B(x_1,x_2|y_1,y_2)\,G^{(L)}(y_1 ,y_2|x_3 ,x_{4})\,,
\end{equation}
or in shorthand notation $G^{(L+2)}=\widehat B  G^{(L)}$, with the kernel
\beq\la{Bop}
B(x_1,x_2|y_1,y_2)\equiv\frac{1}{(x_1-y_1)^2 (x_2-y_1)^2 (x_2-y_2)^2 y_{12}^2}\,.
\eeq
The graph-building operator can be visualised diagrammatically as in figure \ref{fig:XXbar_GBO}.\footnote{Note that it is equal to the square of the graph-building operator for the one-magnon case of \cite{Gromov:2018hut}.}

The resummation of the perturbative expansion can therefore be represented as the 
geometrical series
\beq\la{geometricsum}
G= 
\frac{1}{1-\xi_2^4\widehat B}   G^{(0)}\,.
\eeq

If, for example, we had chosen to use a non-local operator with the other ordering of the fields in the trace, such as $\tr\big((Z^\dagger X^\dagger)(x_3) X(x_4)\big)$ instead of $\tr\big((Z^\dagger X)(x_3) X^\dagger(x_4)\big)$, then the only difference would have been in the definition of ${G}^{(0)}$, but $\widehat B$ would remain the same.

As we review in appendix \ref{app:Bethe-Salpeter},
the spectrum of local operators, $\Delta_{\ell,\bar \ell}(\xi_2)$, can be read from the locations of the poles in (\ref{geometricsum}). Namely, it is given by the solutions to the equation
\beq\la{Econdition}
\xi_2^4\,E(\Delta_*,\ell,\bar \ell) = 1\,,
\eeq
where $E(\Delta,\ell,\bar \ell)$ is the eigenvalue of the graph-building operator $\widehat{B}$ in the principal series representations of the conformal group $(\Delta,\ell,\bar \ell)$. The corresponding eigenfunctions are fixed by the conformal invariance of the graph-building operator (\ref{Bop}). They take the form 
of conformal three-point functions involving a scalar of dimension 1 and a scalar of dimension 2. The third operator, at $x_0$, can then only be a symmetric traceless tensor representation ($\bar\ell=\ell\equiv S$) of arbitrary rank. If we contract the index structure of that operator with a null vector $\zeta$, this eigenfunction reads
\begin{equation}
\Psi^{(\Delta,S,S)}_{x_0,\zeta}(x_1,x_2) = 
\frac{\left(\frac{\zeta\cdot x_{10}}{x_{10}^{2}}-\frac{\zeta\cdot x_{20}}{x_{20}^{2}}\right)^S}{|x_{12}|^{S-\Delta+3}|x_{10}|^{\Delta-S-1}|x_{20}|^{\Delta-S+1}}\, .
\end{equation}

To compute the eigenvalue $E(\Delta,S,S)$ we can either use the star-triangle identity \eqref{star-triangle1} to act with $\widehat B$ on $\Psi$, or directly act with the inverse 
\beq\la{BXm1}
\widehat{B}^{-1} = \frac{1}{16} x_{12}^2 \Box_{x_2} x_{12}^2 \Box_{x_1}\, .
\eeq
The corresponding eigenvalue is
\beq\la{Ess}
E(\Delta,S,S)=\frac{16}{(\Delta+S-1)^2((4-\Delta)+S-1)^2}\,.
\eeq

Equation (\ref{Econdition}) has two pairs of solutions related by the shadow transform, $\Delta\to4-\Delta$. The two solutions, which in the free theory have dimension above the unitarity bound, $\Delta_0\geqslant 2$, are 
\begin{equation}\label{eq: neat sol xxbar}
\Delta_{\pm}(S,\xi_2)=2+\sqrt{\left(S+1\right)^{2}\pm4{\xi}^{2}_2}\,.
\end{equation}

The weak coupling expansion of \eqref{eq: neat sol xxbar} in small ${\xi}^2_2$ reads
\begin{equation}\label{eq: xxbar weak}
\Delta_{\pm}(S,\xi_2)=3+S\pm\frac{2{\xi}^{2}_2}{(S+1)}-\frac{2{\xi}^{4}_2}{(S+1)^{3}}\pm\frac{4{\xi}^{6}_2}{(S+1)^{5}}+O\left({\xi}^{8}_2\right)\, .
\end{equation}
For $S=0$, these two dimensions are those of linear combinations of $\tr(ZXX^\dagger)$ and $\tr(ZX^\dagger X)$. An interesting feature of the weak coupling expansion in \eqref{eq: xxbar weak}
is that it is an expansion in ${\xi}^{2}_2$, while 
the graph-building operator appears in \eqref{Gcorrelator} with a factor of ${\xi}^{4}_2$. 
This has to do with the fact that the two operators $\tr(ZXX^\dagger)$ and $\tr(ZX^\dagger X)$ can mix with each other in perturbation theory. In appendix \ref{app:pert-xxbar}, we explicitly demonstrate how this happens at one-loop order in perturbation theory.

At strong coupling the dispersion relation (\ref{eq: neat sol xxbar}) becomes that of a holographic dual classical fishchain of length two \cite{Gromov:2019aku}
\beq
\Delta_\text{cl}^2=\ell_\text{cl}^2\pm4\xi_2^2+O(\xi_2)\,.
\eeq

Finally, following the procedure reviewed in appendix \ref{app:Bethe-Salpeter}, we have collected in appendix \ref{app:ZXX details} the other quantities required for a complete computation of the correlator \eqref{Gcorrelator}.

\subsection{The Operator $\tr(F_{\mu\nu}Z)$}
\label{sec:FZ}

The second example of a family of operators that becomes non-trivial in the limit (\ref{fnlimitgamma2}) is that of operators involving one $Z$-field and an $F_{\mu\nu}$, with any distribution of derivatives. To study these operators of abritrary spin, in analogy with (\ref{pointsplit}), we consider the non-local operators
\beq
\tr\big(Z(x)F^{\mu\nu}(y)\big)\,,\qquad \tr\big(Z^\dagger(x)F_{\mu\nu}(y)\big)\,.
\eeq
Their corresponding two-point function now takes the form
\beq\la{G3-gluon}
    G_{F,\rho\sigma}^{\mu\nu}(x_1,x_2|x_3,x_4)\equiv \langle \tr\big(Z(x_1) F^{\mu\nu}(x_2)\big)\, \tr\big(Z^\dagger(x_3) F_{\rho\sigma}(x_4)\big)\rangle\, . 
\eeq
As before, this correlator is gauge invariant in the limit (\ref{fnlimitgamma2}). We have found it useful to also consider the non-gauge invariant correlator
\beq\la{G3-gluon A}
G_{A,\nu}^{\mu}(x_1,x_2|x_3,x_4)\equiv\langle \tr\big(Z(x_1) A^{\mu}(x_2)\big)\, \tr\big(Z^\dagger(x_3) A_{\nu}(x_4)\big)\rangle\,.
\eeq
It is related to the correlator in (\ref{G3-gluon}) through
\beq\la{GAtoGF}
G_{F}^{\mu\nu,\rho\sigma}(x_1,x_2|x_3,x_4)=\d_{x_2}^{[\mu}\d_{x_4}^{[\rho}G_{A}^{\nu]\,\sigma]}(x_1,x_2|x_3,x_4)\,.
\eeq
where we have used that $g\rightarrow 0$ in the scaling limit (\ref{fnlimitgamma2}), and hence we can replace $F_{\mu\nu}$ by $\p_{[\mu} A_{\nu]}$.

The interaction terms in the Lagrangian \eqref{lagrangian} that contribute to this double scaled correlator are
\beq\la{AXint}
\cL_\text{int}=2Ng^2\tr\big(X^\dagger A_\mu X A^\mu+e^{-\ii \gamma_3}X^\dagger Z^\dagger XZ\big)-\ii Ng\,\tr\big([A^\mu,X]\p_\mu X^\dagger - \p_\mu X [X^\dagger,A^\mu]\big)+\dots\,.
\eeq

\begin{figure}
\centering{}
\includegraphics[scale=0.35]{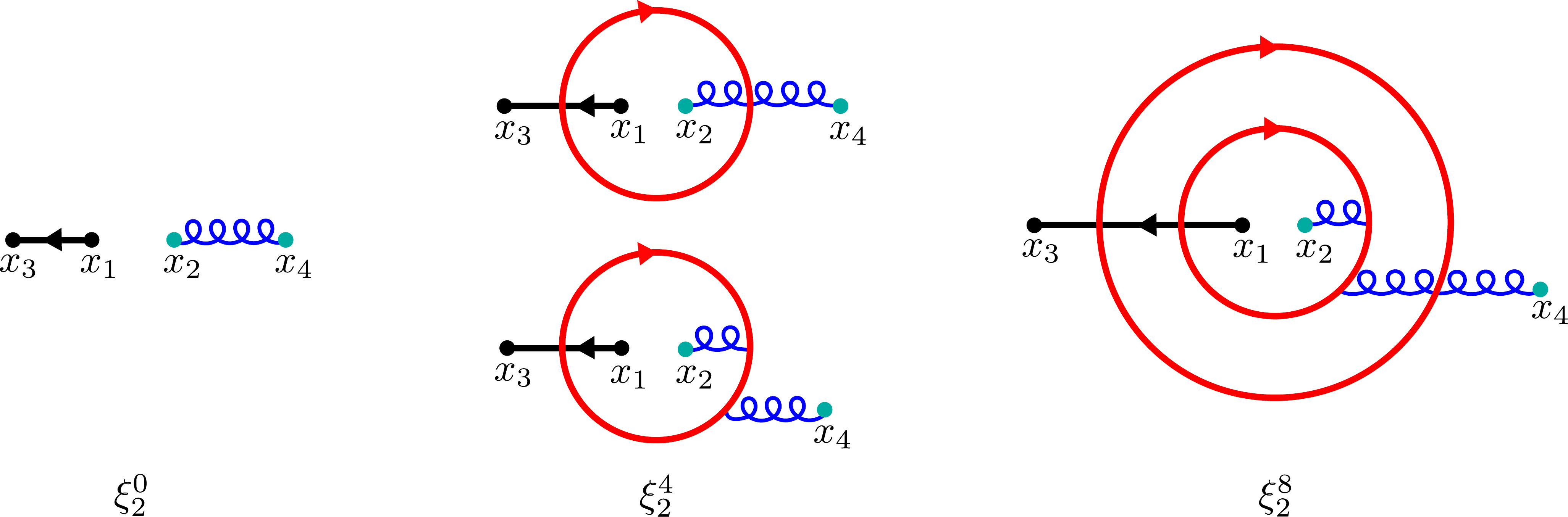}\caption{\small The first few diagrams in the perturbative expansion in $\xi_2$. The black line denotes the Z field propagator and the red line denotes the X field propagator. With each wheel, the gluon can either continue straight through or annihilate and immediately reappear. For the latter, the reappearing gluon can appear on either side of the annihilated gluon, but we have suppressed such diagrams for clarity.}
\label{fig:FZ_pertub}
\end{figure}
\begin{figure}
\centering
\includegraphics[scale=0.35]{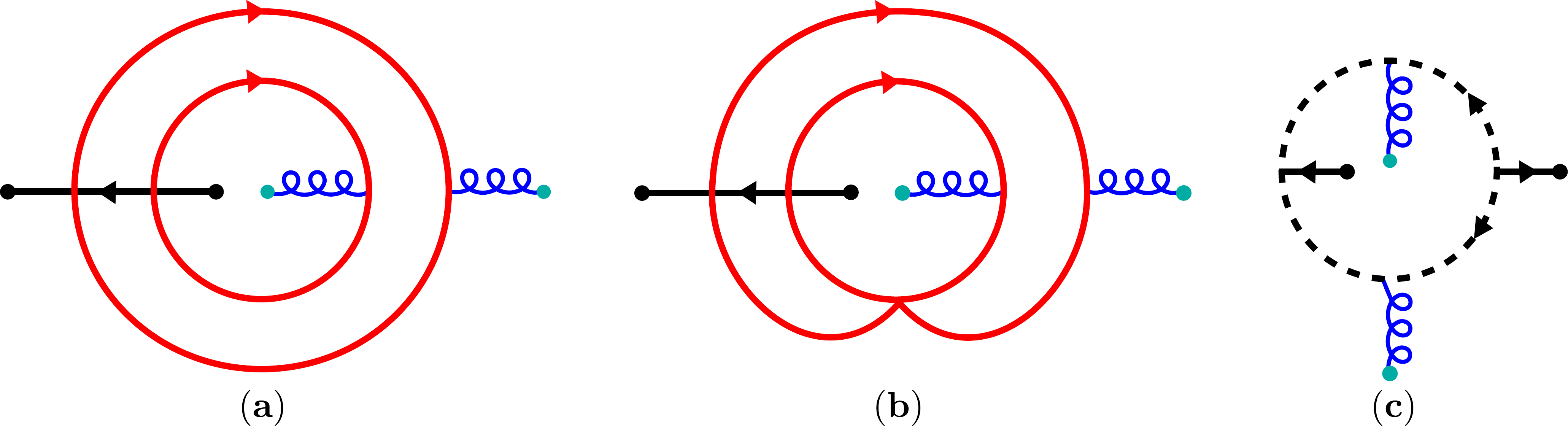}
\caption{\small Examples of diagrams which do not contribute in the double-scaling limit. Solid lines denote scalars and dashed lines denote fermions.}
\label{fig:three graphs}
\end{figure}

We have drawn some of the first few diagrams that contribute to the correlators (\ref{G3-gluon}) and (\ref{G3-gluon A}) in figure \ref{fig:FZ_pertub}. These diagrams have an iterative structure in which an $X$-wheel absorbs and emits the $Z$ scalar and the gluon. For this picture to hold true, it is important that an $X$-wheel cannot absorb the gluon without emitting one on its other side. Such a process would leave us with the trace of a single intermediate $Z$ field, see figure \ref{fig:three graphs}.a. This trace vanishes in the $SU(N)$ theory and is $1/N$-suppressed in the $U(N)$ theory.

Another type of process that is not supressed in the limit (\ref{fnlimitgamma2}) is that of a $\tr(XXX^\dagger X^\dagger)$ interaction between two $X$-wheels, see figure \ref{fig:three graphs}.b. These types of interactions come in pairs, one on each side of the incoming gluon. They cancel out due to the anti-symmetry of the $A\!\!-\!\!X\!\!-\!\!X^\dagger$ interaction vertex, (\ref{AXint}). All other processes, such as fermion wheels or ghost loops in figure \ref{fig:three graphs}.c, are suppressed in the limit (\ref{fnlimitgamma2}). 

In what follows we work in the $R_\alpha$ gauge, where the gluon propagator takes the form
\beq\la{gluonprop}
\Delta^\mu_\nu(x) ={1\over4\pi^2\alpha\, x^2}\[\frac{1}{2}(\alpha + 1)\delta^\mu_\nu + (\alpha - 1)\frac{x^\mu x_\nu}{x^2}\]\,.
\eeq 

The tree-level correlator is plotted in figure \ref{fig:FZ_pertub}.a and is given by the product of the gluon and scalar propagators
\beq\la{GA0}
G_{A,\mu\nu}^{(0)} = \frac{\Delta_{\mu\nu}(x_{42})}{4\pi^2 x_{13}^2}\, .
\eeq

\begin{figure}
\centering{}
\includegraphics[scale=0.4]{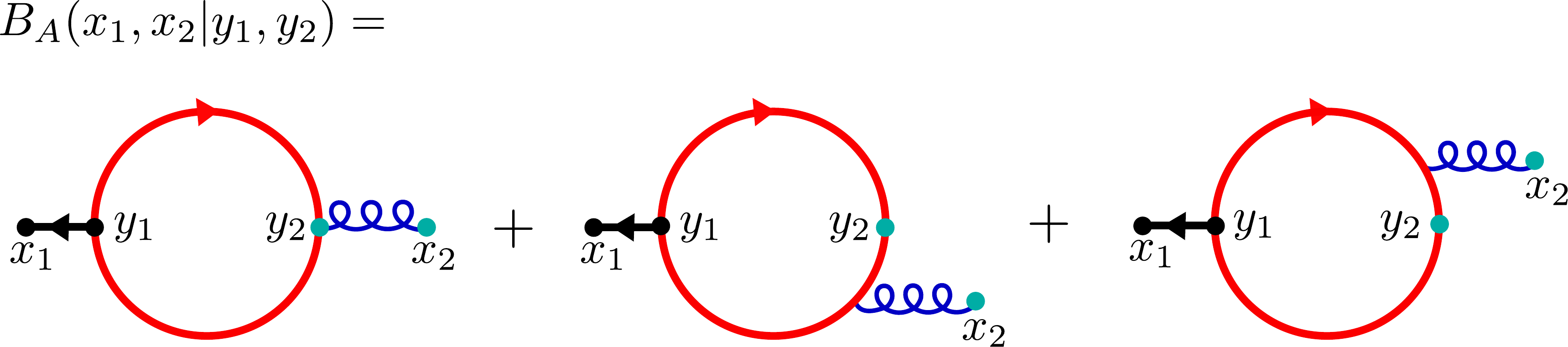}\caption{\small The graph-building operator $\widehat B_A$ in (\ref{BA}). The point $y_2$ marks the position of the gluon in the wave function $\Phi$.}
\label{fig:gluon_GBO}
\end{figure}

The higher loops are generated iteratively by the graph-building operator as
\begin{align}\la{gbo}
{G^{(L+2)}_A}^\mu_\nu(x_1,x_2|x_3,x_4) =\,&
\(B^{\mu\rho}_{A}  {G^{(L)}_A}_{\rho\nu}\)(x_1,x_2|x_3,x_4)\\
\equiv\,&
\int \frac{\dd^4 y_1 \dd^4 y_2}{\pi^4}B^{\mu\rho}_{A}(x_1,x_2|y_1,y_2)\,G^{(L)}_{A,\rho\nu}(y_1 ,y_2|x_3 ,x_{4})\,.\nn
\end{align}
Unlike the previous cases and the fishnet, this operator is not equal to a single diagram. Instead, it is given by the sum of the three diagrams in figure \ref{fig:gluon_GBO}
\beq\la{BA}
B^{\mu\nu}_{A}(x_1,x_2|y_1,y_2) = \frac{4\pi^2\Delta^{\mu\nu}(x_2-y_2)}{(x_1-y_1)^2y_{12}^4} +\int\! \dd^4 z\,\frac{\Delta^\mu_\sigma(x_2-z)}{(x_1 - y_1)^2}\Bigg[\frac{1}{(y_1-z)^2}\overleftrightarrow{\d}_z^\sigma \frac{1}{(y_2-z)^2}\overleftrightarrow{\d}_{y_2}^\nu\frac{1}{y_{12}^2}\Bigg]\,,
\eeq
where $f \overleftrightarrow {\partial}^\mu g \equiv f(\partial^\mu g) - (\partial^\mu f)g$. Here, the first term result from the quartic interaction in (\ref{AXint}), and the second term comes from the cubic ones.  

The graph-building operator $\widehat B_A$ depends on our gauge choice. This fact is manifest in the $\alpha$-dependence of the gluon propagator (\ref{gluonprop}) in (\ref{BA}). We now use it to derive a gauge-invariant graph-building operator $\widehat B_F$, that generates the perturbative expansion of $G_F$ as
\beq
{G_F^{(L+2)}}^{\mu\nu}_{\rho\sigma}(x_1,x_2|x_3,x_4) =
({B^{\mu\nu}_{F,\tau\chi}}  {G^{(L)}_F}^{\tau\chi}_{\rho\sigma})(x_1,x_2|x_3,x_4)\,,
\eeq
where $G^{(L)}_F$ is related to $G^{(L)}_A$ as in (\ref{GAtoGF}). 

To obtain $\widehat B_F$, we notice that the action of $\widehat B_A$ on an arbitrary vector function $\Phi^\mu$ takes the form 
\begin{equation}\la{Brelation}
    B_{A,\tau}^\nu  \Phi^\tau = 
    {H}_{\tau \rho}^{\nu} \left(\p_2^\tau\Phi^\rho - \p_2^\rho\Phi^\tau\right)\,,
\end{equation}
where the explicit form of ${H}_{\tau \upsilon}^{\nu}$ is given below. 
By plugging (\ref{gbo}) into (\ref{GAtoGF}) and using this relation we conclude that
\beq\la{HtoBF}
B^{\mu\nu}_{F,\rho\sigma} = \p_2^\mu H_{\rho\sigma}^{\nu} - \p_2^\nu H_{\rho\sigma}^{\mu}\,.
\eeq

Note that $\widehat B_F$ is indeed independent of the gauge-fixing parameter $\alpha$ in (\ref{gluonprop}). To show this we use that the dependence of $\widehat B_A$ on $x_2$ only enters through a gluon propagator connecting to $A(x_2)$. The dependence of this propagator on $\alpha$ drops out from $\d_\mu\Delta^{\nu\rho}-\d_\nu\Delta^{\mu\rho}$. Hence, when plugging (\ref{gbo}) into (\ref{GAtoGF}) the dependence of $\widehat B_A$ on $\alpha$ drops out. The $\alpha$-independence of $\widehat B_F$ then follows using (\ref{Brelation}) and (\ref{HtoBF}).

To prove (\ref{Brelation}), we first notice that the action of $\widehat B_A$ in (\ref{BA}) on an arbitrary vector function $\Phi$ takes the form
\begin{align}\la{H3check psi bis}
&\[\widehat{B}_A \Phi\right]^\mu(x_1,x_2)\\
&=4\int\frac{\dd^4y_1\, \dd^4 y_3}{\pi^2} \frac{\Delta^{\mu}_\tau(x_2-y_3)}{(x_1-y_1)^2 y_{13}^4} \Bigg[\Phi^{\tau}(y_1,y_3)+\p_{y_3}^\tau y_{13}^2 \int\frac{\dd^4 y_2}{4\pi^2}\,\Phi_\rho(y_1,y_2)\left(\frac{1}{y_{23}^2}\overleftrightarrow{\d}_{y_2}^\rho\frac{1}{y_{12}^2}
\right) \Bigg]\,,\nn
\end{align}
where the three terms correspond to the three diagrams in figure \ref{fig:gluon_GBO} respectively. Using the relation $\Box \,x^{-2} = -4\pi^2 \delta^{(4)}(x)$, the first term can be expressed as
\beq\la{ibp 1}
\Phi^{\tau}(y_1,y_3)=-\int\frac{\dd^4 y_2}{4\pi^2}\[{\Psi_\rho}^\tau(y_1,y_2)+\Phi_\rho(y_1,y_2) \p_{y_3}^\tau\]\p_{y_3}^\rho\frac{1}{y_{32}^2}\,,  
\eeq
where
\beq
\Psi^{\mu\nu}(x,y)\equiv\p_y^\mu\Phi^\nu(x,y)- \p_y^\nu\Phi^\mu(x,y)\,.
\eeq

For the other terms in (\ref{H3check psi bis}) we use that under the $y_2$-integration we have
\beq\la{secondterm}
\Phi_\rho(y_1,y_2)y_{13}^2\left[\frac{1}{y_{23}^2}\overleftrightarrow{\d}_{y_2}^\rho\frac{1}{y_{12}^2} 
\right]=2\Psi_{\rho\sigma}(y_1,y_2) \frac{y_{32}^\rho y_{12}^\sigma}{y_{32}^2\,y_{12}^2} 
+\Phi_\rho(y_1,y_2)\p_{y_2}^\rho\left[\frac{1}{y_{12}^2}-\frac{1 }{y_{32}^2}\right]\,.
\eeq
The term $\Phi_\rho(y_1,y_2)\p_{y_2}^\rho1/y_{12}^2$ on the right-hand side does not depend on $y_3$ and therefore drops out when we plug (\ref{secondterm}) into (\ref{H3check psi bis}). We are left with
\begin{align}\la{H3check psi}
\[\widehat{B}_A \Phi\right]^\mu(x_1,x_2) &=2\int\frac{\dd^4 y_1\dd^4 y_2 \dd^4 y_3}{\pi^4}\frac{\Delta^{\mu}_\tau(x_2-y_3) }{(x_1-y_1)^2y_{13}^4}
\Bigg[\frac{y_{32}^\rho}{y_{32}^4}\delta^{\tau\sigma}+ \p_{y_3}^{\tau}\frac{y_{32}^\rho\,y_{12}^\sigma}{y_{32}^2\,y_{12}^2}\Bigg]\Psi_{\rho\sigma}(y_1,y_2)\nn\\
&\equiv\[\widehat{H} \Psi\]^\mu(x_1,x_2)\, .
\end{align}
This concludes the proof of (\ref{Brelation}). 

The resumation of the perturbative expansion is again a geometrical series:
\beq\la{geometricseriesF}
G_F=\frac{1}{1-\xi_2^4\widehat B_F}   G^{(0)}_F\,.
\eeq
To obtain the spectrum, we need to diagonalise $\widehat B_F$. It is simpler to diagonalise the inverse $\widehat B_F^{-1}$ rather than $\widehat B_F$ itself because the former is a differential operator. 
For this aim, we first notice that the image of $\widehat B_F$ is always of the form $\p_2^\mu\Phi^\nu - \p_2^\nu\Phi^\mu$ for some $\Phi$ (that is only defined up to a gauge transformation), see (\ref{HtoBF}). Moreover, in (\ref{geometricseriesF}) $\widehat B_F$ only acts within the subspace of antisymmetric tensor functions of this type. Therefore, we should only invert $\widehat B_F$ inside this subspace.

Next, we observe that, for all antisymmetric tensor function $\Psi$, the vector function $\widehat{H} \Psi$ in \eqref{H3check psi} is divergence-less:
\beq\la{Hid}
{\p\over\p x_2^\mu}\[\widehat{H} \Psi\]^\mu=0\,.
\eeq
This non-trivial identity is straightforward to check, but we do not have a physical understanding of it. By combining (\ref{HtoBF}) with (\ref{Hid}) we conclude that
\begin{equation}
{\p\over\p x_2^\mu}\[\widehat{B}_F \Psi\]^{\mu\nu} = \Box_{x_2}\[\widehat{H} \Psi\]^\nu\, .
\end{equation}

Because the left-hand side is gauge invariant, so is the right-hand side. To proceed, we consider the right-hand side of this equation in Feynman gauge, 
$\Delta^\mu_\nu(x) = \delta^\mu_\nu/(4\pi^2 x^{2})$. The action of the Laplacian on $x_2$ then produces a $\delta$-function that localises the $y_3$-integral in (\ref{H3check psi}). To localise the $y_1$-integral we act with $\Box_{x_1}$. This leaves us with
\begin{align}
    x_{12}^4 \Box_{x_1} \p_{x_2}^\mu{\big[\widehat{B}_F \Psi\big]_\mu}^\nu&(x_1,x_2) = x_{12}^4 \Box_{x_1} \Box_{x_2}\big[\widehat{H} \Psi\big]^{\nu}(x_1,x_2)
    \\&=16 \Phi^{\nu}(x_1,x_2) + 4\p_{x_2}^\nu\[x_{12}^2 \int\frac{\dd^4 y_2}{\pi^2}\Phi_\rho(x_1,y_2)\Big(\frac{1}{(y_2-x_2)^2}\overleftrightarrow{\d}_{y_2}^\rho\frac{1}{(x_1-y_2)^2}
    \Big)\]\,,\nn
\end{align}
where $\Psi^{\mu\nu} = \p_2^\mu\Phi^\nu - \p_2^\nu\Phi^\mu$ for some $\Phi$. It therefore follows that
\begin{equation}\label{inverse H3}
    \p_{x_2}^{[\rho}x_{12}^4 \Box_{x_1} \p_{x_2}^\mu{\big[\widehat{B}_F \Psi\big]_\mu}^{\nu]}
    = 16\Psi^{\rho\nu}\, .
\end{equation}

\subsubsection{Spectrum}

We denote the eigenfunctions of $\widehat{B}_F$ by $\Psi^{\mu\nu}_E = \p_{x_2}^\mu\Phi^\nu_E - \p_{x_2}^\nu\Phi^\mu_E$, where $E$ is the eigenvalue and the vector-potential $\Phi_E^\mu$ is defined up to a total derivative. To represent these eigenfunctions it is useful to introduce an auxiliary polarisation vector $\theta$, whose components anticommute among themselves, $\{\theta^\mu,\theta^\nu\}=0$. In terms of these auxiliary variables, the functions $\Psi$ and $\Phi$ take the form
\beq
\Phi(\theta)=\theta^\mu\Phi_\mu\,,\qquad \Psi(\theta) = \frac{1}{2} \theta^\mu\theta^\nu\Psi_{\mu\nu}=(\theta\cdot\d_{x_2})\Phi(\theta)\,,
\eeq
and the eigenfunction equation becomes
\beq\la{BFm1}
\widehat B_F^{-1}\Psi_E={1\over16}(\theta\cdot\p_{x_2})\, x_{12}^4\,(\p_{x_2}\cdot \p_\theta)\, \Box_{x_1}\Psi_E = \frac{1}{E}\Psi_E\,.
\eeq
Equivalently, in terms of the vector-potential, this equation reads
\beq\la{Phieq}
 \theta\cdot\p_{x_2}\left[x_{12}^4\,(\p_{x_2}\cdot\p_\theta)(\theta\cdot\p_{x_2})\Box_{x_1}\Phi_E - \frac{16}{E}\Phi_E\right] = 0\,,\qquad\text{with}\qquad\Phi_E\simeq\Phi_E+(\theta\cdot\d_{x_2})f\,,
 \eeq
 for arbitrary function $f=f(x_1,x_2)$.

 As before, these functions are fixed by the conformal symmetry of the graph-building operator (\ref{inverse H3}). The vector-potential eigenfunction $\Phi_E^\mu$ is a conformal three-point functions between a dimension one primary scalar at $x_1$, a dimension one vector at $x_2$, and a primary operator at some arbitrary point $x_0$, that we take to transform in a representation $\rho=(\Delta,\ell,\bar\ell)$ of the conformal group
 \beq
 \Phi_E=\<\cO_{(\Delta,\ell,\bar\ell)}(x_0)\cO_{(1,0,0)}(x_1)\cO_{(1,1,1)}(x_2)\>+(\theta\cdot\d_{x_2})f\,.
 \eeq
Correspondingly, the eigenfunction $\Psi_E$ is a conformal three-point function involving the same operators at $x_0$ and $x_1$ and a dimension two self-dual or anti-self-dual  rank-2 anti-symmetric tensor at $x_2$.  

The simplest way of constructing these three point functions is using embedding coordinates. We relegate the details to appendix \ref{app:embedding and spectrum}, and we merely present here the results. 
We find that there are three families of such non-zero three point functions, with 
\beq
(\ell,\bar\ell)\in\{(S,S),(S-1,S+1),(S+1,S-1)\}\qquad\text{for}\qquad S\geqslant 1\,,
\eeq
where $S$ is the total spin. The first family of operators are symmetric traceless tensors of rank $S$. Their corresponding three-point function takes the form
\begin{equation}
    \Phi^{(\Delta,S,S)}_{x_0,\zeta}(x_1,x_2,\theta) =  \frac{\left(\frac{\theta\cdot x_{12}}{x_{12}^{2}}-\frac{\theta\cdot x_{02}}{x_{02}^{2}}\right) \left(\frac{\zeta\cdot x_{10}}{x_{10}^{2}}-\frac{\zeta\cdot x_{20}}{x_{20}^{2}}\right)^S}{|x_{12}|^{1+S-\Delta}| x_{10}|^{1+\Delta-S}|x_{20}|^{\Delta-S-1}}\, ,
\end{equation}
where $\zeta$ is a polarisation vector for the operator at $x_0$. By plugging this function into (\ref{Phieq}) we find the eigenvalue
\begin{equation}
E(\Delta,S,S) = \frac{16}{(\Delta-S-3)^2((4-\Delta)-S-3)^2}\, .
\end{equation}
The dimensions of the corresponding operators are determined by (\ref{Econdition}). The physical solutions, having $\Delta(0,S,S)\geqslant 2$, are
\begin{equation}\label{gluon-traceless-spec}
\Delta_{\pm}(\xi_2,S,S) = 2 + \sqrt{(S+1)^2\pm 4\xi_2^2}\, .
\end{equation}

Note that this is the same spectrum as the one we have found before in (\ref{eq: neat sol xxbar}). The operators are, however, different. For example, for $S=1$ the two non-trivial operators we have found here and the two in the previous section are different combinations of the operators\footnote{Here, we have used the equation of motion, $dF = J$, to trade field strength for a scalar bi-linear. We have also used integration by parts to remove the derivative from the $Z$-field.}
\beq
\tr(ZXD^\mu X^\dagger)\,,\qquad\tr(ZD^\mu X X^\dagger)\,,\qquad\tr(Z X^\dagger D^\mu X)\,,\qquad\text{and}\qquad\tr(ZD^\mu X^\dagger X)\,.
\eeq
We expect the degeneracy between (\ref{gluon-traceless-spec}) and (\ref{eq: neat sol xxbar}) to be lifted at the next order in the $g$ expansion around the large twist limit.

The other two families have $(\ell,\bar\ell)=(S-1,S+1)$ and $(\ell,\bar\ell)=(S+1,S-1)$. For $S=1$, they correspond to self dual and anti-self dual field strength. They can be combined into a single (reducible) representation $\rho=(\Delta,S-1,S+1)\oplus (\Delta,S+1,S-1)$. The index structure is encoded using two polarisation vectors, $\eta$ and $\zeta$, see appendix \ref{app:spectrum} for details. Two linearly independent wave functions are constructed in the appendix and are given by
\begin{equation}\label{eigenvectors 3 rho'}
    \Phi^{\rho,1}_{x_0,\zeta,\eta}(x_1,x_2,\theta) = \frac{x_{20}^{2}(\theta\cdot\eta) - 2(\theta\cdot x_{20})(\eta\cdot x_{20})}{|x_{12}|^{2+S-\Delta}|x_{10}|^{\Delta-S}|x_{20}|^{2+\Delta-S}}\,(\eta\cdot\p_\zeta)\left(\frac{\zeta\cdot x_{10}}{x_{10}^{2}}-\frac{\zeta\cdot x_{20}}{x_{20}^{2}}\right)^S\, ,
\end{equation}
and
\begin{equation}\label{eigenvectors 3 rho' bis}
\Phi^{\rho,2}_{x_0,\zeta,\eta}(x_1,x_2,\theta) = \epsilon_{\mu\nu\rho\sigma}\frac{(x_{20}^2x_{10}^\rho-x_{10}^2x_{20}^\rho)\theta^\sigma - 2(\theta\cdot x_{20})x_{10}^\rho x_{20}^\sigma}{|x_{12}|^{2+S-\Delta}|x_{10}|^{2+\Delta-S} |x_{20}|^{2+\Delta-S}}\eta^\nu(\eta\cdot\p_\zeta) \zeta^\mu
\left(\frac{\zeta\cdot x_{10}}{x_{10}^{2}}-\frac{\zeta\cdot x_{20}}{x_{20}^{2}}\right)^{S-1}\,.
\end{equation}
For $S=1$ these two functions are related by the duality map $*\cO_{\mu\nu}(x_0) = \frac{1}{2}\epsilon_{\mu\nu\rho\sigma} \cO^{\rho\sigma}(x_0)$. For $S>1$ they are related in an analogous way given in \eqref{dual transformation} and \eqref{dual transformation bis}, such that $\Phi^{\rho,1}_{x_0,\zeta,\eta}/S + \Phi^{\rho,2}_{x_0,\zeta,\eta}/(S+1)$ is self-dual, while $\Phi^{\rho,1}_{x_0,\zeta,\eta}/S - \Phi^{\rho,2}_{x_0,\zeta,\eta}/(S+1)$ is anti-self-dual.

These relations imply that the wave functions (\ref{eigenvectors 3 rho'}) and (\ref{eigenvectors 3 rho' bis}) correspond to two different representations of the same operator. 
Hence, they give rise to the same spectrum. In the appendix, we use (\ref{eigenvectors 3 rho'}) to obtain
\begin{equation}
E(\Delta,S\pm1,S\mp1) =\frac{16}{(\Delta-S-4)((4-\Delta)-S-4) (\Delta-S-2) ((4-\Delta)-S -2)}\, .
\end{equation}
The physical solutions of (\ref{Econdition}), namely with tree-level dimensions $\Delta_0\geqslant 2$, have twist two and four. They are given by\footnote{Note that under the replacement $\xi_2\to\xi_1$, the spectrum in (\ref{gluon-mix-spec}) exactly coincides with that of the fishnet operators $\tr(Z\d^SZ)$ and $\tr(Z\d^S\Box Z)$. In the fishnet case, the perturbative expansion of the twist-two dimension, $\Delta_-(\xi_1,-1,1)$, starts at order $\xi_1^2$ and is related to the presence of double-trace interactions. Here, $S\geqslant 1$, and no such issue arises.}
\begin{equation}\label{gluon-mix-spec}
\Delta_{\pm}(\xi_2,S+1,S-1) =\Delta_{\pm}(\xi_2,S-1,S+1) = 2 + \sqrt{1 + (S+1)^2 \pm 2\sqrt{(S+1)^2 + 4\xi_2^4}}\, . 
\end{equation}
At small coupling, the dimensions behave as
\begin{equation}
\Delta_{\pm} = \left(S+3 \pm 1\right) \pm \frac{2\, \xi_2^4}{(S+1) (S+1\pm 1)} - \frac{2\, \xi_2^8 (S+1 \pm(S+1\pm1)^2)}{(S+1)^3 (S+1 \pm 1)^3} + O\left(\xi_2^{12}\right)\,.
\end{equation}
When $S=1$, these solutions correspond to the operators $\tr(F_{\mu\nu} Z)$ and $\tr(F_{\mu\nu}\Box Z)$ respectively.

\section{Mixing Between Operators and Between Scaling Limits}\label{sec:mixing}

Consider the extension of the operators discussed in the previous section to operators with more $Z$-fields, such as $\tr(X X^\dagger Z^J)$ with $J>1$. These cases come with a new complication -- after adding more $Z$-lines to the diagrams in figure \ref{fig:XXbar_gluon_nomixing}, they no longer cancel. As a result, the graph-building operator mixes between wave functions with different number and type of fields, all having the same $R$-charge. In addition, diagrams where the $X$ and $X^\dagger$ annihilate or where the gluon in $\tr(F(x_1)Z(x_2)Z(x_3)\dots)$ is absorbed by the $X$-wheel no longer vanish. As a result, the graph-building operator also mixes between operators with different double-scaling limits (\ref{fnlimit}). Such a graph-building operator can be represented by a matrix. Importantly for us, this matrix is always finite. Different elements of this matrix have different dependence on the 't Hooft coupling and the twist.
By analysing it we can obtain both the spectrum of operators with different double-scaling limits, as well as the systematic corrections away from the large twist double-scaling limit. Depending on the operator we are interested in, we consider the same matrix-graph-building operator expanded around different double-scaling limits.

Our main focus in this section is the generalisation of the operators we have studied so far to operators with more $Z$-fields. Namely, the dimensions of the operators $\tr(FZ^J)$ and $\tr(XX^\dagger Z^J)$ with $J>1$. Their corresponding double-scaling limit is (\ref{fnlimit}) with $n=1+1/J$. In this limit, only the mixing between these two non-local operators and with the fishnet operator $\tr(Z^J)$ is relevant. Hence, the graph-building operator is a $3\times3$ matrix acting on the vector of wave functions
\beq\la{wavefunctions}
\vec\Psi = \begin{pmatrix}
\Psi_{\emptyset}\\ \Psi_{A} \\ \Psi_{X}
\end{pmatrix}\equiv\(\begin{array}{r}
\<\tr(Z(x_1)Z(x_2)\dots Z(x_J))\,\cO\>\\ \<\tr(A(x_0)Z(x_1)Z(x_2)\dots Z(x_J))\,\cO\>\\ \<\tr(XX^\dagger(x_0)Z(x_1)Z(x_2)\dots Z(x_J))\,\cO\>\end{array}\)\,.
\eeq
Here, like in (\ref{pointsplit}), we have chosen to use the operators with the fields $X$ and $X^\dagger$ placed at the same position. As before, we first consider the correlator with a gauge field insertion, $\Psi_A$, and then construct from it the correlator with a field strength insertion as
\beq\la{PsiAtoPsiF}
\Psi_F^{\mu\nu}=\d_{x_0}^{[\mu}\Psi_A^{\nu]}\,,\qquad\text{or}\qquad \Psi_F=\theta\cdot\p_{x_0}\Psi_A\,.
\eeq

The graph-building operator takes the following form
\beq\label{hres FZJ}
\widehat B = \begin{pmatrix}
g^{-2}{\mathcal{\widehat B}}_{\emptyset\emptyset}& g^{-1} {\mathcal{\widehat B}}_{\emptyset A} &  g^{-1} {\mathcal{\widehat B}}_{\emptyset X}\\ g^{-1} {\mathcal{\widehat B}}_{A\emptyset} &  {\mathcal{\widehat B}}_{A A} & {\mathcal{\widehat B}}_{A X}\\
g^{-1} {\mathcal{\widehat B}}_{X\emptyset} &  {\mathcal{\widehat B}}_{X A} & {\mathcal{\widehat B}}_{X X}
\end{pmatrix}\,,
\eeq
where ${\mathcal{\widehat B}}_{\emptyset\emptyset}$ is the fishnet graph-building operator. The corresponding eigenvalue equation reads
\begin{equation}\la{evequation}
    \begin{pmatrix}
    1& 0 & 0\\
    0 & \theta\cdot\p_0 & 0\\
    0 & 0 & 1
    \end{pmatrix} \left[\widehat B-E\,\boldsymbol{\mathbbm{1}}\right] \vec\Psi = 0\,.
\end{equation}
Finally, we read the spectrum by solving 
\beq\la{EtoDelta}
\xi^{2(J+1)}_{1+1/J}\,E(\Delta_*,\ell,\bar \ell)=1\,.
\eeq

We look for a solution to (\ref{evequation}) for which the eigenvalue $E$ stays finite when $g\rightarrow 0$ with fixed $\xi^2_{1+1/J}$. 
Due to the scaling of the ($\emptyset,j$) and ($i,\emptyset$) components of $\widehat{B}$ in (\ref{hres FZJ}) it follows that
\begin{equation}
    {\mathcal{\widehat B}}_{\emptyset\emptyset} \Psi_{\emptyset} = 0\, ,\qquad \text{and hence}\qquad  \Psi_{\emptyset} = 0\,.
\end{equation}
Similarly, it follows that the second and third components of $\Psi$ have to satisfy the reduced eigenvalue equation 
\begin{equation}\label{HPhi mixing}
     \begin{pmatrix}
     \theta\cdot\p_0 & 0\\
     0 & 1
     \end{pmatrix}\[{\mathfrak{B}}-E\,\boldsymbol{\mathbbm{1}}\] \begin{pmatrix}
\Psi_{A} \\ \Psi_X
\end{pmatrix}=0\,,
\end{equation}
where
\begin{equation}\label{HtildeHres}
    {\mathfrak{B}}_{ij} = {\mathcal{\widehat B}}_{ij} - {\mathcal{\widehat B}}_{i\emptyset}  {\mathcal{\widehat B}}_{\emptyset\emptyset}^{-1}  {\mathcal{\widehat B}}_{\emptyset j}\qquad \text{for}\qquad i,j\in\{A,X\}\, .
\end{equation}

\begin{figure}[t]
\centering
\includegraphics[width=0.9\textwidth]{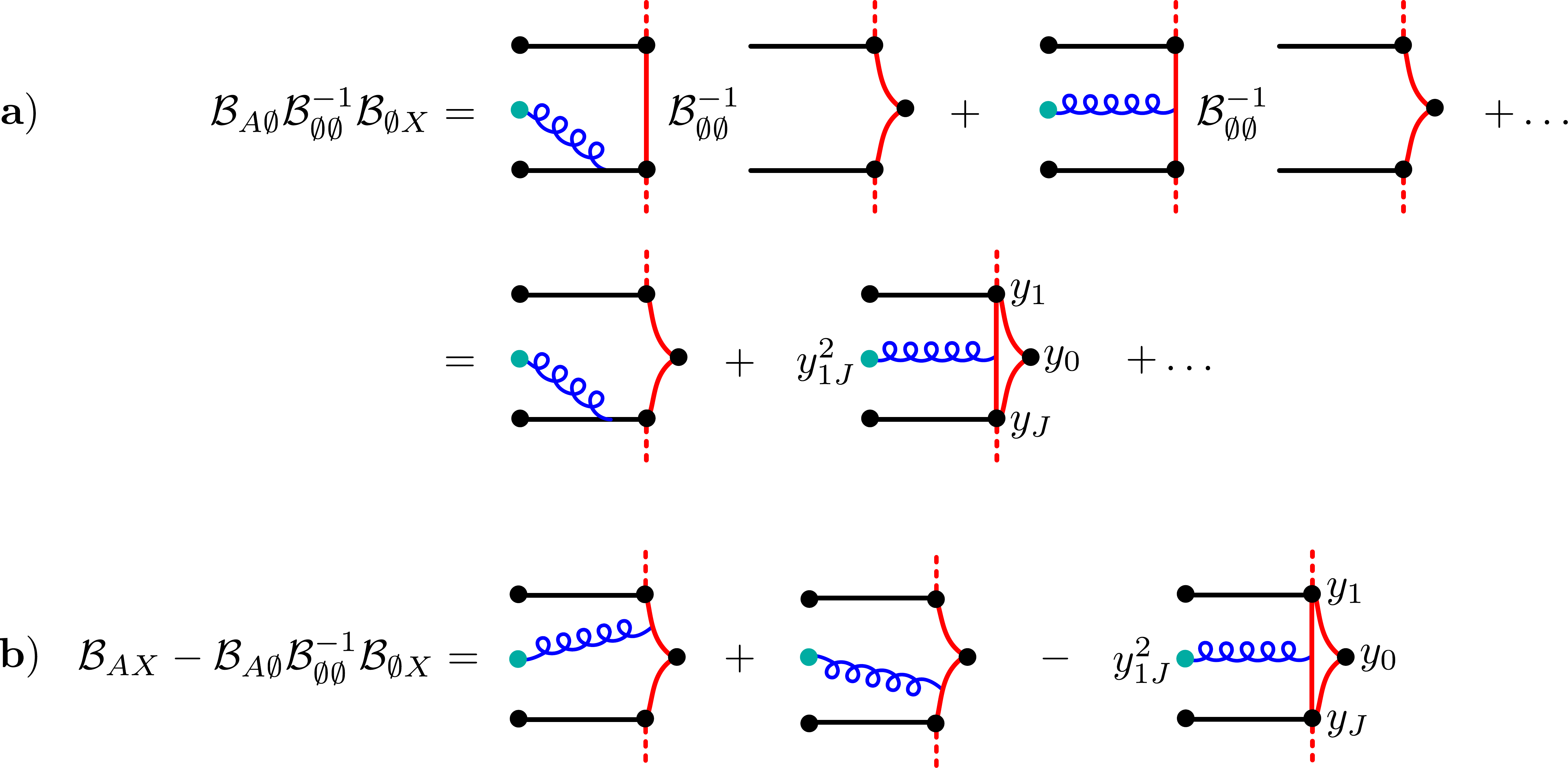}
\caption{\small {\bf a}) The combination ${\mathcal{\widehat B}}_{A\emptyset} {\mathcal{\widehat B}}_{\emptyset\emptyset}^{-1} {\mathcal{\widehat B}}_{\emptyset X}$ has the effect of joining diagrams from ${\mathcal{\widehat B}}_{A\emptyset}$ and ${\mathcal{\widehat B}}_{\emptyset X}$ where the gluon and the $XX^\dagger$ fields do not connect to the same $X$-propagator into a diagram of ${\mathcal{\widehat B}}_{AX}$. {\bf b}) As a result, the only diagrams that contributes to ${\mathcal{\widehat B}}_{AX} - {\mathcal{\widehat B}}_{A\emptyset} {\mathcal{\widehat B}}_{\emptyset\emptyset}^{-1} {\mathcal{\widehat B}}_{\emptyset X}$ are those with the interaction with the gluon and the $XX^\dagger$ fields is local on the chain of $Z$'s.}
\label{BBm1B}
\end{figure}

All the diagrams that contribute to $\widehat{B}$ in (\ref{hres FZJ}) have an $X$ wheel crossing all the $Z$ fields, that result in the $\xi^{2(J+1)}_{1+1/J}$ factor in (\ref{EtoDelta}). 
Among these diagrams, the only ones that contribute to ${\mathfrak{B}}_{ij}$ are those where a single $X$-propagator from that wheel absorbs and emits an $XX^\dagger$ or an $A$ field. All other diagrams, where an $XX^\dagger$ or an $A$ field is attached to different propagators, cancel out between the two terms in (\ref{HtildeHres}). This cancellation is demonstrated in figure \ref{BBm1B}.  
The remaining diagrams have the $XX^\dagger$ and the $A$ fields interact locally on the chain of $Z$'s. They are all drawn in figures \ref{BBm1B}.b, \ref{mathbbB}, and are evaluated in appendix \ref{app:mixing}. The result is
\begin{figure}
\centering
\includegraphics[width=0.8\textwidth]{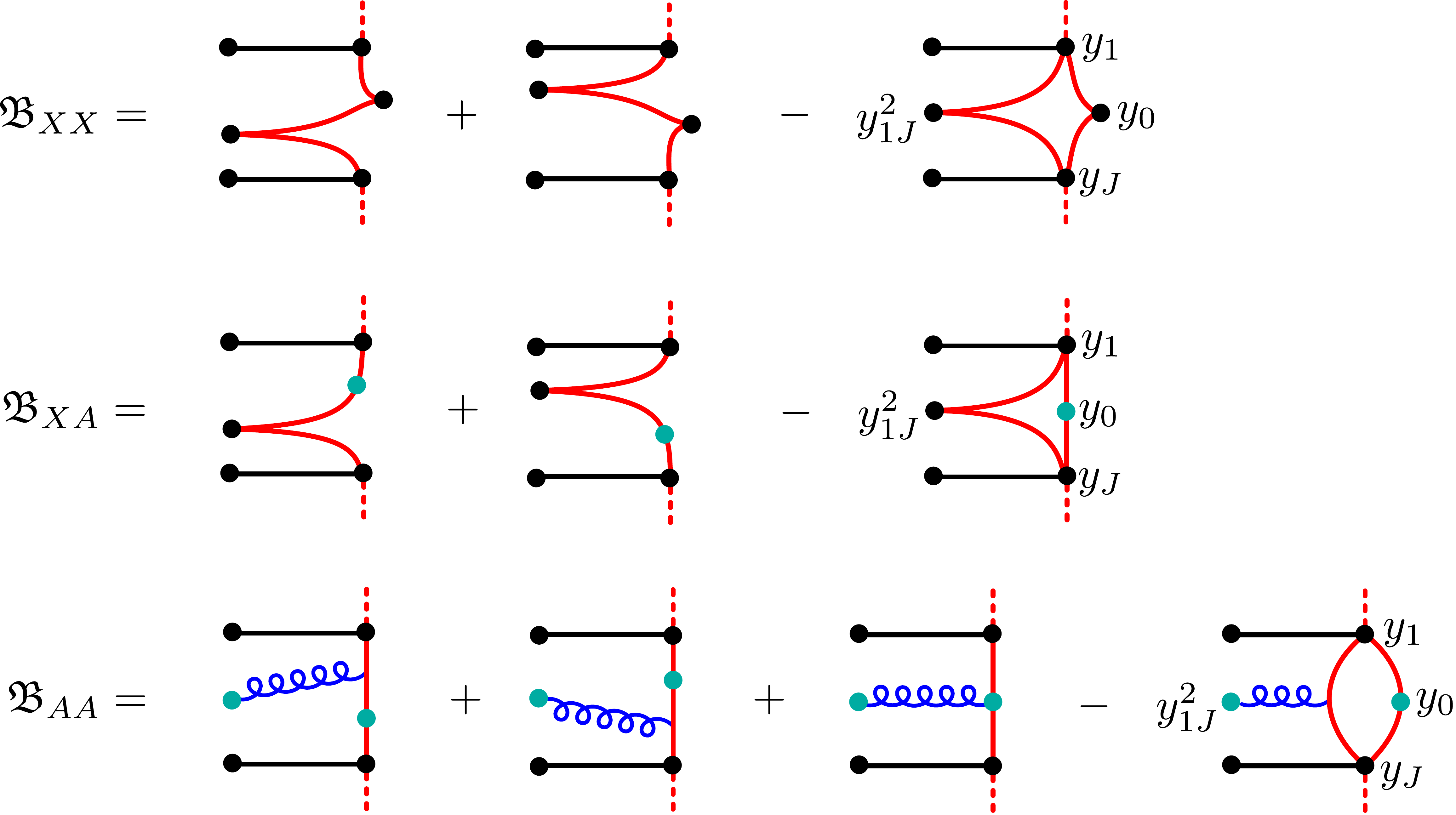}
\caption{\small The diagrams that contribute to the reduced graph-building operator ${\mathfrak{B}}$ in (\ref{HtildeHres}). The interactions with the gluon and the $XX^\dagger$ fields take place locally on the chain of $Z$'s. This is in contrast to the diagrams that contribute to $B$ in (\ref{hres}).}
\label{mathbbB}
\end{figure}
\begin{multline}
\left[{\mathfrak{B}}_{FF} \Psi_F\right](\theta,x_0,\dots,x_J) = \theta\cdot\p_0 \int\prod_{j=0}^J\frac{\dd^4 y_j}{\pi^2}\frac{\Psi_F^{\mu\nu}(y_0,\dots,y_J)}{\prod_{i=1}^J (x_i - y_i)^2 y_{i,i+1}^2  }\\
 \times
 \int{\dd^4 z\over\pi^2}
 {y_{J0}^2\over(x_0 - z)^2(y_1-z)^2(z-y_J)^2}\Bigg[\frac{(z - y_0)_\mu}{(z-y_0)^4} \theta_\nu
+ {1\over2}\theta\cdot \p_z\frac{(y_0 - z)_\mu}{(y_0 - z)^2}\left(\frac{y_{01,\nu}}{y_{01}^2} + \frac{y_{0J,\nu}}{y_{0J}^2}\right)\\
+\left(\frac{\theta\cdot (y_J - z)}{(y_J - z)^2} - \frac{\theta\cdot (y_1 - z)}{(y_1 - z)^2}\right) \bigg( \frac{(y_0 - z)_\mu y_{0J,\nu}}{(y_0 - z)^2 y_{0J}^2} + \frac{y_{0J,\mu} y_{01,\nu}}{y_{0J}^2 y_{01}^2}+ \frac{y_{01,\mu} (y_0 - z)_\nu}{y_{01}^2 (y_0 - z)^2}\bigg)\Bigg]\,,
\end{multline}
where $y_{J+1}\equiv y_0$ and we have used (\ref{PsiAtoPsiF}) to convert $\Psi_A$ into $\Psi_F$ as
\beq
\theta\cdot\p_{x_0}\,{\mathfrak{B}}_{AA} \Psi_A={\mathfrak{B}}_{FF} \Psi_F\,,\qquad \theta\cdot\p_{x_0}\,{\mathfrak{B}}_{AX} \Psi_X={\mathfrak{B}}_{FX} \Psi_X\,.
\eeq
Similarly, 
\begin{multline}
\left[{\mathfrak{B}}_{FX} \Psi_X\right](\theta,x_0,\dots,x_J) = \frac{1}{4}\theta\cdot\p_0\int\prod_{j=0}^J\frac{\dd^4 y_j}{\pi^2}\frac{\Psi_X(y_0,\dots,y_J)}{\prod_{i=0}^J (x_i - y_i)^2 y_{i,i+1}^2} \\
\times\int{\dd^4 z\over\pi^2}{(x_0-y_0)^2\over(x_0-z)^2}\Bigg[\frac{1}{2(z-y_1)^2 (z-y_J)^2}\theta\cdot \p_z\left(\frac{y_{10}^2 (z-y_J)^2}{(z-y_0)^2} - \frac{y_{J0}^2 (z-y_1)^2}{(z-y_0)^2}\right)\\
+\left(\frac{\theta\cdot (y_J - z)}{(y_J - z)^2} - \frac{\theta\cdot (y_1 - z)}{(y_1 - z)^2}\right) \bigg(\frac{y_{10}^2}{(z-y_1)^2 (z-y_0)^2} + \frac{y_{J0}^2}{(z-y_J)^2 (z-y_0)^2}
- \frac{y_{1J}^2}{(z-y_1)^2 (z-y_J)^2}\bigg)\Bigg]\, ,
\end{multline}
and
\begin{multline}
\left[{\mathfrak{B}}_{XF} \Psi_F\right](x_0,\dots,x_J) = 2\int\prod_{j=0}^J\frac{\dd^4 y_j}{\pi^2} \frac{\Psi_F^{\mu\nu}(y_0,\dots,y_J)}{\prod_{i=1}^J (x_i - y_i)^2y_{i,i+1}^2}{y_{J0}^2\over(x_0-y_1)^2 (x_0-y_J)^2}\\
\times\bigg[\frac{(y_0-x_0)_\mu y_{0J,\nu}}{(y_0-x_0)^2 y_{0J}^2} + \frac{y_{0J,\mu} y_{01,\nu}}{y_{0J}^2 y_{01}^2} +\frac{y_{01,\mu} (y_0-x_0)_\nu}{y_{01}^2 (y_0-x_0)^2}\bigg] \,.
\end{multline}
Finally,
\begin{multline}\label{HXX}
\left[{\mathfrak{B}}_{XX} \Psi_X\right](x_0,\dots,x_J) = \frac{1}{2}\int\prod_{j=0}^J \frac{\dd^4 y_j}{\pi^2}\frac{\Psi_X(y_0,y_1,\dots,y_J)}{\prod_{i=0}^J (x_i - y_i)^2  y_{i,i+1}^2} \\
\times\bigg[\frac{y_{10}^2}{(x_0 - y_1)^2}
+ \frac{y_{J0}^2}{(x_0 - y_J)^2}-\frac{y_{1J}^2(x_0 - y_0)^2}{(x_0 - y_J)^2 (x_0 - y_1)^2}\bigg]\, .
\end{multline}

Surprisingly, we find that ${\mathfrak{B}}$ has a simple inverse
\begin{align}\label{inverse B mixing}
{\mathfrak{B}}^{-1}=&(-4)^{-J-1}\begin{pmatrix}
(\theta\cdot\p_{x_0})\, x_{J0}^2 x_{10}^2\, (\p_{x_0}\cdot\p_\theta)& 2(\theta\cdot\p_{x_0})\left[x_{10}^2(\theta\cdot x_{J0})-x_{J0}^2(\theta\cdot x_{10})\right] \\[6pt]2\left[x_{J0}^2(x_{10}\cdot \p_\theta)-x_{10}^2(x_{J0}\cdot\p_\theta)\right](\p_{x_0}\cdot\p_\theta)& \p_{x_0,\mu}\, x_{J0}^2 x_{10}^2\, \p_{x_0}^{\mu}+8(x_{10}\cdot x_{J0})
\end{pmatrix}\\
&\times\prod_{i=1}^{J-1}x_{i,i+1}^2 \prod_{j=1}^J \Box_{x_j}\, .\nn
\end{align}

For $J=1$, this operator becomes diagonal with its elements exactly coinciding with (\ref{BFm1}) and (\ref{BXm1}) respectively. For $J>1$, this operator is conformally invariant, but conformal symmetry is no longer sufficient to fix its eigenfunctions. Instead, we now prove that it is part of the commuting charges of an integrable model.

\subsection{Integrability}\la{intsec}

Let us consider a conformal spin chain of length $J+1$ consisting of the reducible representation $\rho_0 = \left[\theta\cdot\p_{x} \left(1,1,1\right)\right] \oplus (2,0,0)$ at site $0$ and of (irreducible) scalar representations of dimension $1$ at all the other sites.\footnote{Note that $\theta\cdot\p_{x} \left(1,1,1\right)$ is included in the usual dimension-$2$ antisymmetric rank-$2$ tensor representation. The fact that it is obtained by taking an anti-symmetric derivative of the vector representation is an additional constraint. It implies that the tensor is annihilated by $\theta\cdot\p_{x}$, and is therefore conserved.} We also consider a six-dimensional auxiliary space carrying the defining representation $\mathbf{6}$ of the conformal group. In order to simplify some of the computations, we choose to work in the embedding space realisation of the principal series representations, see appendix \ref{embfor} for details. We denote by $Y^M$ the coordinates and by $\Theta^M$ the polarisation vectors in embedding space. The relevant Lax matrix at site $0$ takes the form
\begin{equation}\la{Lnew}
    L^{(\rho_0;\mathbf{6})}_Y(u) = u^2 - u\, q_{MN}^{(\rho_0)} \otimes e^{MN} + \mathcal{L}^{MN}_Y\otimes e_{MN}\, ,
\end{equation}
where $?e_M^N?$ is the $6\times6$ matrix with the $(M,N)$ element equal to one and the others are equal to zero. Here, $q_{MN}$ are the conformal generators. In the representation $\rho_0$, they take the form 
\begin{equation}
    q_{MN}^{(\rho_0)} = \begin{pmatrix}
Y_{M}\p_{Y^N} - Y_{N}\p_{Y^M} + \Theta_{M}\p_{\Theta^N} - \Theta_{N}\p_{\Theta^M} & 0 \\ 0 & Y_{M}\p_{Y^N} - Y_{N}\p_{Y^M}
\end{pmatrix}\,. 
\end{equation}
Finally, the operator $\mathcal{L}_Y$ is
\begin{equation}\la{Lnew2}
    \mathcal{L}^{MN}_Y = -\frac{1}{2}\begin{pmatrix}
(\Theta\cdot\p_Y)\, Y^MY^N\, (\p_Y\cdot\p_\Theta) & (\Theta\cdot\p_Y)\left[ Y^{M}\Theta^N - Y^{N}\Theta^M\right] \\
\left[Y^{N}\p_{\Theta}^{M} - Y^{M}\p_{\Theta}^{N}\right](\p_Y\cdot\p_\Theta) & \frac{1}{2}\left[Y^{M}\Box_Y Y^{N} + Y^{N} \Box_Y Y^{M}\right] + 2\eta^{MN}
\end{pmatrix}\, .
\end{equation}
The first column of these operators acts on weight-two functions $\Psi(\Theta,\lambda\,Y)=\lambda^{-2}\Theta^M\Theta^N \p_{Y^M}\Phi_{N}(Y)$, which also satisfy the transversality condition $Y\cdot\p_{\Theta}\Psi = 0$.\footnote{In order to simplify the computations, we also impose the relations $\Psi = \Theta\cdot\p_Y \Phi$ and $Y\cdot \p_{\Theta} \Psi = 0$ away from $Y^2 = \Theta\cdot Y = 0$. This is not a restriction; it simply means that we have fixed the gauge associated to $\Phi\rightarrow \Phi + (\Theta\cdot Y) F$. Otherwise, we would have to add $\frac{1}{2}(\p_Y^{M}\Theta^{N} + \p_Y^{M}\Theta^{N}) Y\cdot \p_{\Theta}$ to the $FF$ coefficient of $\mathcal{L}^{MN}_Y$ and replace the $XF$ coefficient by $\frac{1}{4}\left[Y^{N}\p_{\Theta}^{M} - Y^{M}\p_{\Theta}^{N}\right](Y\cdot\p_\Theta)\Box_Y$.} The second column acts on weight-two functions of $Y$ only.
Introducing a second copy of the auxiliary space, we have checked that
\begin{equation}\label{YBE}
    R_{12}^{(\mathbf{6},\mathbf{6})}(u-v) L^{(\rho_0;\mathbf{6})}_{Y,1}(u) L^{(\rho_0;\mathbf{6})}_{Y,2}(v) = L^{(\rho_0;\mathbf{6})}_{Y,2}(v) L^{(\rho_0;\mathbf{6})}_{Y,1}(u) R_{12}^{(\mathbf{6},\mathbf{6})}(u-v)\,,
\end{equation}
where the indices $1$ and $2$ indicate in which of the auxiliary spaces the operators are acting non-trivially, and the Zamolodchikovs' R-matrix is \cite{Zamolodchikov:1978xm}
\begin{equation}
    R_{12}^{(\mathbf{6},\mathbf{6})}(u) = u(u+2) + (u+2)\left(e_{1,MN}\otimes e_2^{NM}\right) - u \left( e_{1,MN}\otimes e_2^{MN}\right)\, .
\end{equation}

At sites $1,\ldots,J$ we have a dimension $1$ scalar. The corresponding fishnet Lax matrix, which satisfies the RLL-relation \eqref{YBE}, reads
\begin{equation}
    L^{(1,0,0;\mathbf{6})}_Y(u) = u^2 - u\left(Y^{M}\p_{Y}^{N} - Y^{N}\p_{Y}^{M}\right) \otimes e_{MN} - {1\over2}Y^{M}Y^{N}\Box_Y\otimes e_{MN}\,.
\end{equation}
It acts on weight-one functions $\Phi(\lambda\,Y)=\lambda^{-1}\Phi(Y)$. The transfer matrix 
\begin{equation}
    T_{\mathbf{6}}(u) \equiv \tr_{\mathbf{6}}\left(L^{(\rho_0;\mathbf{6})}_{Y_0}(u) L_{Y_1}^{(1,0,0;\mathbf{6})}(u) \cdots L_{Y_J}^{(1,0,0;\mathbf{6})}(u)\right)
\end{equation}
thus commutes with itself for arbitrary values of the spectral parameter,
\begin{equation}
\left[T_{\mathbf{6}}(u),T_{\mathbf{6}}(v)\right] = 0\,.
\end{equation}

Finally, we claim that
\begin{equation}\label{T6=B}
    T_{\mathbf{6}}(0) = (-1)^{J+1}\mathfrak{B}^{-1}\,.
\end{equation}
Hence, the inverse graph-building operator is part of the conserved charges of an integrable model. The proof of (\ref{T6=B}) uses the map to embedding coordinates that we review in appendix \ref{embfor}.\footnote{To that end, one can also use the fact that the diagonal elements of $\mathcal{L}_{Y,MN}$ in \eqref{Lnew2} are equal to $\frac{1}{2}?q^{(\rho_0)}_M^P? q^{(\rho_0)}_{PN} -  q^{(\rho_0)}_{MN}$. Eventually, it turns out that the reduction $\mathcal{L}_Y \rightarrow \mathcal{L}_x$ is equivalent to performing the replacement $(\Theta\cdot\p_Y, \p_Y\cdot\p_\Theta,\Box_Y)\rightarrow(\theta\cdot\p_x,\p_{x}\cdot\p_\theta,\Box_x)$, and plugging the following expressions in $L^{(\rho_0;\mathbf{6})}_Y(0)$:
\begin{equation}\la{coc1}
    Y^\mu = x^\mu\, , \quad Y^+ = 1\, ,\quad  Y^- = x^2\, ,\quad \Theta^{\mu} = \theta^{\mu} \, ,\quad \Theta^+ = 0\, ,\quad  \Theta^- = 2 x\cdot \theta\, ,
\end{equation}
and
\begin{equation}
    \p_{\Theta^\mu} = \p_{\theta^\mu}\, ,\quad \p_{\Theta^-} = 0\, ,\quad \p_{\Theta^+} = - x\cdot\p_{\theta}\, .
\end{equation}}

We point out that the Lax matrix \eqref{Lnew} is different from the na\"{i}ve guess associated to the reducibility of the representation $\rho_0$. This guess would correspond to a direct sum of the two standard Lax matrices constructed from the fusion procedure, recalled in appendix \ref{gencon}. Moreover, because this direct sum contains an epsilon-tensor, the Lax matrix \eqref{Lnew} does not seem to be related to it by a change of basis. This situation is in contrast to all other examples studied in this paper, where the standard fusion procedure produces the relevant integrable models. It would therefore be interesting to understand how the Lax matrix \eqref{Lnew} fits inside the known classification of integrable models.

\subsection{Higher-Order Corrections}\la{fncorrectionssec}

The \emph{same} graph-building operator (\ref{hres}) can also be used to compute perturbative corrections in $g$ to the fishnet dimensions
\beq\la{fncorrection}
\Delta_\text{f.n.}(\xi_1,g)= \Delta_\text{f.n.}^{(0)}(\xi_1)+g^2\Delta_\text{f.n.}^{(2)}(\xi_1) + O(g^3)\,.
\eeq
In order to compute $\Delta_\text{f.n.}^{(2)}$, we consider the limit (\ref{fnlimit}) with $n=1$ and expand around it in $g$. At order $g^2$ we must take into account elements of the graph-building operator that mix between the zeroth order fishnet wave function, $\Psi_\emptyset$ in (\ref{wavefunctions}), and $\Psi_A$, $\Psi_X$, as well as wave functions with other intermediate fields such as fermions. 
Using the relation $\xi_{1+1/J}^{2(J+1)} = g^2 \xi_{1}^{2J}$, the corresponding graph-building operator takes the form
\beq\label{hres}
 B_{\text{f.n.}} = \begin{pmatrix}
{\mathcal{ B}}_{\emptyset\emptyset} 
& g{\mathcal{ B}}_{\emptyset A} &  g{\mathcal{  B}}_{\emptyset X} & g{\mathcal{ B}}_{\emptyset,\text{other}}\\ g{\mathcal{  B}}_{A\emptyset} &  g^2{\mathcal{  B}}_{A A} & g^2{\mathcal{  B}}_{A X} & 0\\
g{\mathcal{  B}}_{X\emptyset} &  g^2{\mathcal{  B}}_{X A} & g^2{\mathcal{  B}}_{X X} & 0\\
g{\mathcal{  B}}_{\text{other},\emptyset} & 0 & 0 & g^2{\mathcal{  B}}_{\text{other},\text{other}}\\
\end{pmatrix}\,.
\eeq

Our main point in this subsection is that at the upper $3\times3$ corner we have the graph-building operator $g^2   B$ from (\ref{hres}). Similarly, the elements with ${\mathcal{B}}_\text{other}$ enter the computation of other type of operators in their corresponding scaling limit (\ref{fnlimit}) with $n>1$. As we explain below, using these elements, one can compute $\Delta_\text{f.n.}^{(2)}(\xi_1)$. Carrying out this computation explicitly is beyond the scope of this paper, and we leave it to a future work.  

We are interested in eigenvalues of $  B_\text{f.n.}$ of the form
\beq
E = E^{(0)} + g E^{(1)} + g^2 E^{(2)} + O(g^3)\,,
\eeq
where the coefficients $E^{(j)}$'s are independent of $\xi_1$. The corresponding correction to the scaling dimension in (\ref{fncorrection}) is obtained by solving the equation $\xi_1^{2J} E(\Delta_\text{f.n.},\ell,\bar\ell) = 1$.

At leading order we have the fishnet equation
\begin{equation}\label{fishnet equation}
{\mathcal{  B}}_{\emptyset\emptyset} \Psi_{\emptyset}^{(0)} = E^{(0)}\Psi_{\emptyset}^{(0)}\,.
\end{equation}
The following orders are determined iteratively from $\Psi_{\emptyset}^{(0)}$ in a standard perturbative fashion. At the next order, we obtain
\begin{equation}\la{Psi1}
    \Psi_i^{(1)} = \frac{1}{E^{(0)}} {\mathcal{  B}}_{i\emptyset} \Psi_{\emptyset}^{(0)} \qquad\text{for}\qquad i\in\{A,X,\text{other}\}\, ,
\end{equation}
and
\begin{equation}\label{fishnet equation first order}
    \left({\mathcal{B}}_{\emptyset\emptyset} - E^{(0)}\right)  \Psi_{\emptyset}^{(1)} = E^{(1)}\Psi_{\emptyset}^{(0)}\,.
\end{equation}
Because of \eqref{fishnet equation}, the only finite solution of this equation is $E^{(1)}=0$, and therefore $\Psi_{\emptyset}^{(1)}\propto \Psi_{\emptyset}^{(0)}$. This solution represents our freedom in choosing a normalisation for $\vec\Psi$. We choose it to be $g$-independent, so that $\Psi_{\emptyset}^{(1)}=0$.

At second order, there is only one non-trivial equation
\begin{equation}\label{fishnet equation second order}
    \left({\mathcal{  B}}_{\emptyset\emptyset} - E^{(0)}\right) \Psi_{\emptyset}^{(2)} = \Big(E^{(2)} 
    - \frac{1}{E^{(0)}} \left[{\mathcal{  B}}_{\emptyset A} {\mathcal{  B}}_{A\emptyset} + {\mathcal{  B}}_{\emptyset X} {\mathcal{  B}}_{X \emptyset} + {\mathcal{  B}}_{\emptyset,\text{other}} {\mathcal{  B}}_{\text{other},\emptyset}\right]\Big) \Psi_{\emptyset}^{(0)}\, .
\end{equation}
As in the previous order, the right-hand side needs to be orthogonal to $\Psi_{\emptyset}^{(0)}$ for $\Psi_{\emptyset}^{(2)}$ to be well defined. Hence,\footnote{To compute this ratio one has to introduce a UV regulator, see \eqref{ortho-relation}.} 
\begin{equation}\la{E2}
    E^{(2)} = \frac{\<\Psi_{\emptyset}^{(0)}| 
    {\mathcal{  B}}_{\emptyset A} {\mathcal{  B}}_{A\emptyset} + {\mathcal{  B}}_{\emptyset X} {\mathcal{  B}}_{X \emptyset} +{\mathcal{  B}}_{\emptyset,\text{other}} {\mathcal{  B}}_{\text{other},\emptyset}|\Psi_{\emptyset}^{(0)} \>}{E^{(0)}\<\Psi_{\emptyset}^{(0)}|\Psi_{\emptyset}^{(0)}\>}\, .
\end{equation}
Once this condition is satisfied, equation \eqref{fishnet equation second order} is trivially solved by expanding $\Psi_{\emptyset}^{(2)}$ over a complete basis of eigenstates of ${\mathcal{  B}}_{\emptyset\emptyset}$.

The spectrum is obtained as a solution to $\xi_1^{2J} E(\Delta_\text{f.n.},\ell,\bar\ell) = 1$. This gives the fishnet condition $\xi_1^{2J} E^{(0)}(\Delta^{(0)}_\text{f.n.},\ell,\bar\ell) = 1$ as well as the first correction to the anomalous dimension:
\begin{equation}
    \Delta^{(2)}_\text{f.n.} = -\left.\frac{E^{(2)} (\Delta,\ell,\bar\ell)}{\frac{\dd}{\dd \Delta} E^{(0)}(\Delta,\ell,\bar\ell)}\right|_{\Delta=\Delta^{(0)}_\text{f.n.}} \, .
\end{equation}

Note that beyond leading order, the basis of wave functions we work with is no longer gauge invariant, and neither are the matrix elements of the graph-building operator. However, the eigenvectors of the graph-building operator are gauge invariant. For example, the order $g$ mixing between the fishnet state and the state with one more gluon in (\ref{Psi1}) is responsible for converting the $Z$-derivatives into covariant ones. 

Note also that in (\ref{hres}) we did not include order $g^2$ correction to $\mathcal{B}_{\emptyset\emptyset}$. The absence of such corrections would imply that all the loop corrections to the twist 2 and twist 4 fishnet states (\ref{fncorrection}) can be obtained through the mixing structure in (\ref{E2}). Further research will be needed to determine whether or not this simple structure is valid.

\section{More Examples Without Mixing} \label{section-longeroperators}

By adjusting the twist, the construction of the previous section can be extended to any operator.\footnote{For a generic operator one has to use the twisting method of \cite{Cavaglia:2020hdb} instead of the $\gamma$-deformed action (\ref{l-int}).} We now present more examples of operators with different double-scaling limits and no mixing. The operators we study have one or two fermions and any number of $Z$-scalars. For operators with more than two fields in the trace, conformal symmetry is not sufficient to fix eigenfunctions of the corresponding graph-building operator, as before. Instead, we prove that the graph-building operator is integrable.

\subsection{The Operators $\tr(Z\psi_{4})$ and $\tr(Z\psi_{1}^\dagger)$} \label{sec:scalar-fermion}

A relatively simple family of operators that have a closed non-trivial double scaling limit 
are those made of one $Z$ field, one $\psi_4$ or $\psi_1^\dagger$ field, and at most one $X$ field. The corresponding limit is (\ref{fnlimit}) with $\theta=-\gamma_3$ and $n=4/3$. Namely, the limit we consider is 
\beq\la{eq: xxbar ds}
e^{-\ii\gamma_3}\rightarrow\infty\,,\qquad\text{with}\qquad\xi_{4/3}^2 = \frac{g^2\,e^{-\ii\frac{3}{4}\gamma_3}}{8\pi^2} \quad \text{fixed}\, .
\eeq

In this case, we find four non-trivial operators at any half-integer spin, one twist 2, one twist 4, and two of twist 3. In analogy with (\ref{Gcorrelator}), here it is sufficient to consider the correlator
\begin{equation}\label{g-2}
G_{2,\alpha\dot\alpha}(x_1,x_2|x_3,x_4) =\langle \tr\big(Z(x_1) \psi_{4\alpha}(x_2)\big)\,\tr\big(Z^\dagger(x_3) \psi_{4\dot\alpha}^\dagger(x_4)\big) \rangle\,,
\end{equation}
where $\alpha$ and $\dot\alpha$ are spinor indices.

The relevant interaction terms that contribute to this correlator in the limit (\ref{fnlimitgamma2}) are 
\beq\la{Zpsiint}
\cL_\text{int}=    \sqrt{2}N g e^{\frac{\ii}{2}\gamma_1 ^{-}}\,\tr\big(\psi^\dagger_4 X \psi^\dagger_1\big) -\sqrt{2}Ng e^{\frac{\ii}{2}\gamma_1^{-}}\,\tr\big(\psi_4 X ^{\dagger} \psi_{1}\big)+2 Ng^2e^{-\ii \gamma_3}\,\tr\big(X^\dagger Z^\dagger XZ\big)+\dots\,.
\end{equation}

\begin{figure}
\centering
\includegraphics[width=1\textwidth]{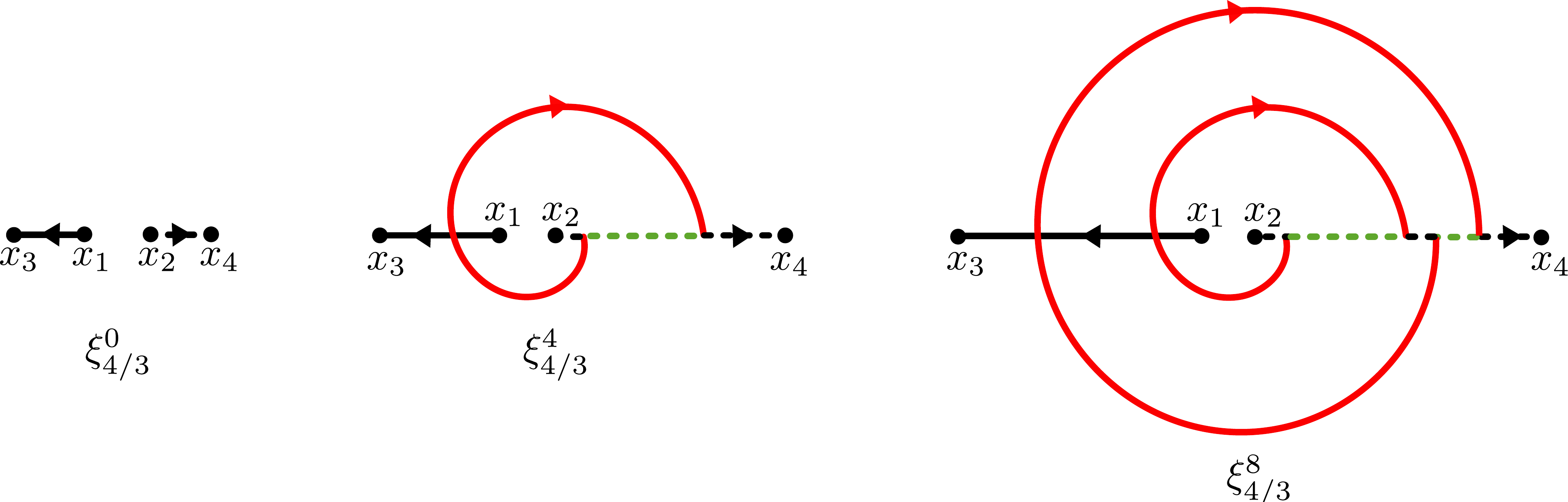}
\caption{\small The first few diagrams that contributes to the correlator (\ref{g-2}). 
Black dashed lines denote $\psi_4$ propagators and green dashed line denote $\psi_1$ ones.}\label{fig:Zpsi_pertu}
\end{figure}
          
The first few diagrams contributing to the correlator $G_{2,\alpha\dot\alpha}$ are plotted in figure \ref{fig:Zpsi_pertu}. At tree level, we have
\begin{equation} \label{g2-fermion}
    G_{2}^{(0)}(x_1,x_2|x_3,x_4) = \frac{\cancel{x}_{24}}{8\pi^4 x_{13}^2 x_{24}^4}\,,
\end{equation}
where $\cancel{x} = x^\mu \sigma_\mu$ and $\cancel{\bar x} = x^\mu \bar \sigma_\mu$ are $2\times 2$ matrices. As before, the diagrams have an iterative structure with $G^{(L+2)}=\xi_{4/3}^4\,  \widehat{B}  G^{(L)}$. The corresponding graph-building operator $\widehat{B}$ now also acts on the spinor index of $G$. 
\begin{figure}[h]
\centering
\includegraphics[width=0.55\textwidth]{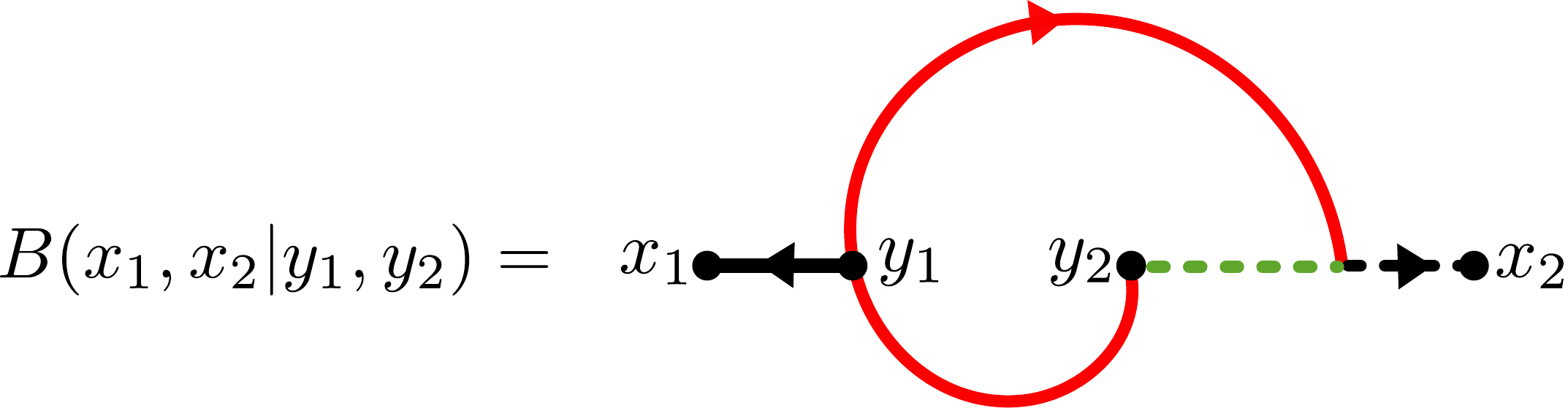}
\caption{\small The graph-building operator $B$ in (\ref{psiGBO}).}\label{fig:fermion_GBO}
\end{figure}
The kernel of this integral operator is plotted in figure \ref{fig:fermion_GBO}. It takes the form
\begin{align}\la{psiGBO}
B(x_1,x_2|y_1,y_2)\equiv&\int\frac{\dd^4 z}{\pi^2}\frac{- (\cancel{x}_2 - \cancel{z}) (\cancel{\bar z} - \cancel{\bar y}_2)}{(x_1-y_1)^2 (x_2 - z)^4 (z-y_2)^4 (z-y_1)^2 y_{12}^2} \\
=&\frac{(\cancel{y}_1-\cancel{x}_2) \cancel{\bar y}_{12}}{(x_1-y_1)^2 (x_2 - y_1)^2 (x_2-y_2)^2 y_{12}^4}\,,\nn
\end{align}
where in the second step we have preformed the integration using the star-triangle relation \eqref{star-triangle1}. 

As before, the eigenfunctions of the graph-building operator (\ref{psiGBO}) are fixed by its conformal symmetry. They take the form of conformal three-point functions involving a scalar of dimension 1 and a fermion of dimension $\frac{3}{2}$. The third operator can be in either the $(\Delta,S-1/2,S+1/2)$ or the $(\Delta,S+1/2,S-1/2)$ representation with half-integer spin $S$. Using \cite{Cuomo:2017wme} and the accompanying Mathematica notebook, we find that the eigenfunctions are given respectively by
\begin{equation}\label{eq: l,l+1 correaltor}
\Psi_{x_0,s,\bar s}^{(\Delta,S-{1\over2},S+{1\over2})}(x_1,x_2) = \frac{\cancel{x}_{20}\bar s\left[s \left(\frac{\cancel{x}_{10}}{x_{10}^{2}}-\frac{\cancel{x}_{20}}{x_{20}^{2}}\right) \bar s \right]^{S-{1\over2}}}{|x_{12}|^{S-\Delta+2}|x_{10}|^{\Delta -S}|x_{20}|^{\Delta-S+2}}\,,
\end{equation}
and
\begin{equation}\label{l+1,l correaltor}
\Psi_{x_0,s,\bar s}^{(\Delta,S+{1\over2},S-{1\over2})}(x_1,x_2) = \frac{\cancel{x}_{21} \cancel{\bar x}_{10} \epsilon s \left[s\left(\frac{\cancel{x}_{10}}{x_{10}^{2}}-\frac{\cancel{x}_{20}}{x_{20}^{2}}\right) \bar s\right]^{S-{1\over2}}}{|x_{12}|^{S-\Delta+3}|x_{10}|^{\Delta - S+1}|x_{20}|^{\Delta -S+1}} \,.
\end{equation}
Here, $s$ and $\bar s$ are the two spinor polarisations of the third operator is located at $x_0$.

The corresponding eigenvalues can be obtained by either acting with $\widehat{B}$ on the eigenfunctions (\ref{eq: l,l+1 correaltor}), (\ref{l+1,l correaltor}) and performing the integrations using the star triangle identity (\ref{star-triangle1}),\footnote{The first integral is of the form \eqref{star-triangle1} with $S=1$, $\zeta^\mu = s_0\sigma^\mu \bar{s}_0$, and $A_{\mu\nu} = \bar\sigma_\nu \sigma_\mu \bar s_0$. The second integral is of the form \eqref{star-triangle1} with $S=0$.} or by acting on them with the inverse
\begin{equation} \label{fermion-inverse}
\widehat{B}^{-1} = \frac{1}{16} x_{12}^2\, \cancel{\p}_{x_2}\, x_{12}^2\, \cancel{\bar\p}_{x_2}\Box_{x_1} = \frac{1}{16} x_{12}^2\, \cancel{x}_{21}\, \Box_{x_2}\, \cancel{\bar x}_{21} \Box_{x_1}\, .
\end{equation}
The eigenvalues we find are
\begin{align}
E(\Delta,S-1/2,S+1/2)=&\frac{16}{\left(\Delta + S- 1\right)^2 \left(\Delta - S- 2\right) \left(\Delta -S - 4\right)}\,,\\
E(\Delta,S+1/2,S-1/2)=&\frac{16}{\left(\Delta - S- 3\right)^2\left(\Delta +S-2\right)\left(\Delta + S\right)}\,.
\end{align}

Note that the shadow transform maps the $(\Delta,S-1/2,S+1/2)$ representation to the $(4-\Delta,S+1/2,S-1/2)$ one. Correspondingly, the eigenfunctions of the graph-building operators are interchanged under this transformation. 

There are two solutions to $\xi_{4/3}^4 E(\Delta,S-1/2,S+1/2) = 1$ with positive tree-level dimension, one with twist two and the other with twist four. They can be written in closed form, but these are not very illuminating. Instead, we only give the first few terms in their perturbative expansion
\begin{align}\label{Delta2 fermion}
\Delta_{2}(\xi_{4/3},S-1/2,S+1/2) & =2+S-\frac{2}{(S+1/2)^{2}}\xi_{4/3}^{4}+\frac{2(S-3/2)}{(S+1/2)^{5}}\xi_{4/3}^{8}+O(\xi_{4/3}^{12})\, ,\\
\Delta_{4}(\xi_{4/3},S-1/2,S+1/2) & =4+S+\frac{2}{(S+3/2)^{2}}\xi_{4/3}^{4}-\frac{2(S+7/2)}{(S+3/2)^{5}}\xi_{4/3}^{8}+O(\xi_{4/3}^{12})\, .
\label{Delta4 fermion}
\end{align}
For $S=1/2$ and $\xi_{4/3}=0$, they correspond to two sets of degenerate primary operators, $\tr(Z\psi_4)$ and $\tr(Z\Box \psi_4)+\dots$ or $\tr(Z\psi_1^\dagger)$ and $\tr(Z\Box\psi_1^\dagger)+\dots$ respectively.

Similarly, there are two solutions to $\xi_{4/3}^4 E (\Delta,S+1/2,S-1/2) = 1$ with $\Re(\Delta)>2$. Perturbatively, they take the form
\begin{equation}\label{Delta3 fermion}
    \Delta_{\frac{7}{2},\pm} =3+S \pm \frac{2}{\sqrt{(S+1/2)(S+3/2)}}\xi_{4/3}^{2} -\frac{2(S+1)}{(S+1/2)^{2}(S+3/2)^{2}}\xi_{4/3}^{4} + O(\xi_{4/3}^{6})\, .
\end{equation}
For $S=1/2$ and $\xi_{4/3}=0$, they correspond to linear combinations of $\tr(Z \psi_1^\dagger X)$ and $\tr(Z X \psi_1^\dagger)$ or $\tr(Z \psi_4 X)$ and $\tr(Z X \psi_4)$. As before, the perturbative expansion is in terms of $\xi_{4/3}^2$ (and not $\xi_{4/3}^4$) because of the mixing between the two sets, see appendix \ref{app:fermion_perturbation_theory}.

Finally, following the procedure reviewed in appendix \ref{app:Bethe-Salpeter}, we have collected in appendix \ref{app:fermion details} the other quantities required for the complete computation of the correlator \eqref{g-2}.

\subsection{The Operators $\tr(Z\psi_{2})$ and $\tr(Z\psi_{3}^\dagger)$}

Instead of the fermion $\psi_4$ and $\psi_1^\dagger$, we can consider operators that are composed from one $\psi_2$ or $\psi_3^\dagger$ field, (together with one $Z$ field and at most one $X$ field). 
The suitable double-scaling limit in this case is (\ref{fnlimit}) with $n=4$. 
The relevant interaction terms are now 
\begin{equation}
    \mathcal{L}_{\text{int}} =    \ii \sqrt{2}N g e^{\frac{\ii}{2}\gamma_1 ^{+}}\,\tr\big(\psi_3 X \psi_2 \big) +  \ii \sqrt{2}N g e^{\frac{\ii}{2}\gamma_1 ^{+}}\,\tr\big(  \psi^\dagger_3 X ^{\dagger} \psi^\dagger_{2} \big) +2 Ng^2e^{-\ii \gamma_3}\,\tr\big(X^\dagger Z^\dagger XZ\big) + \dots\,.
\end{equation}

The diagrams that survive in this limit look the same as the ones before, see figure \ref{fig:Zpsi_pertu}. The only difference is in the labeling of the fermions. Hence, all the results above also apply here, with $\xi_{4/3}\to\xi_4$.

\subsection{The Operators $\tr(\psi_4Z^J)$}

One can generalise the operator from section \ref{sec:scalar-fermion} and study  $\tr(\psi_4Z^J)$ with $J>1$ perturbatively under the limit of \eqref{fnlimit} with $n=\frac{2J+2}{2J+1}$.

We take the wave function in this case to be the correlator
\beq
\psi=\<\tr(\psi_4(x_0)Z(x_1)\dots Z(x_J))\,\cO\>\,.
\eeq
The inverse of the corresponding graph-building operator takes the form
\begin{equation} \label{n-scalar-fermion}
    \widehat{B}^{-1} =x^2_{0J}\cancel{\partial}_{x_0} \prod_{i=1}^Jx^2_{i,i-1} \cancel{\bar{\partial}}_{x_0} \prod_{j=1}^J\Box_{x_j}\,.
\end{equation}
When $J=1$, this naturally coincides with \eqref{fermion-inverse}, up to normalisation and relabeling of the points. 

In the integrability picture, we construct a spin chain with one fermion at position $x_0$ and $J$ scalars at positions $x_1,\dots,x_J$. The inverse of the graph-building operator (\ref{n-scalar-fermion}) is proportional to the corresponding transfer matrix in the $\bf 6$ representation, evaluated at $u=0$:
\begin{equation}
\widehat{B}^{-1}\ \propto\ \widehat{T}_{\mathbf{6}}\big(0;-1/4,0,..,0\big) = \tr_{\mathbf{6}}\(L_{x_0}^{\left(3/2,1,0;\boldsymbol{6}\right)}\left(-1/4\right) \cdot L_{x_1}^{\left(1,0,0;\boldsymbol{6}\right)} (0)\cdot\ldots\cdot L_{x_J}^{\left(1,0,0;\boldsymbol{6}\right)} (0) \)\,.
\end{equation}
The relevant Lax matrices are given in equations \eqref{Lax scalar1} and \eqref{Lax fermion} of appendix \ref{app:Lax matrices}.

\subsection{The Operator $\tr(\psi_4 \psi_1^\dagger Z)$}

There are also operators with more than one fermion that do not mix. Consider for example the operator $\tr(\psi_4 \psi_1^\dagger Z)$ in the double-scaling limit \eqref{fnlimit} with $n=\frac{3}{2}$. 

We take the wave function in this case to be the correlator
\beq\la{psi4psi1Z}
\psi=s_1^\alpha s_2^\beta\,\<\tr(\psi_{4,\alpha}(x_1)\psi_{1,\beta}^\dagger(x_2)Z(x_3))\,\cO\>\,.
\eeq
where $s_1$ and $s_2$ are the (independent) spinor polarisations. 

\begin{figure}
\centering
\includegraphics[width=0.8\textwidth]{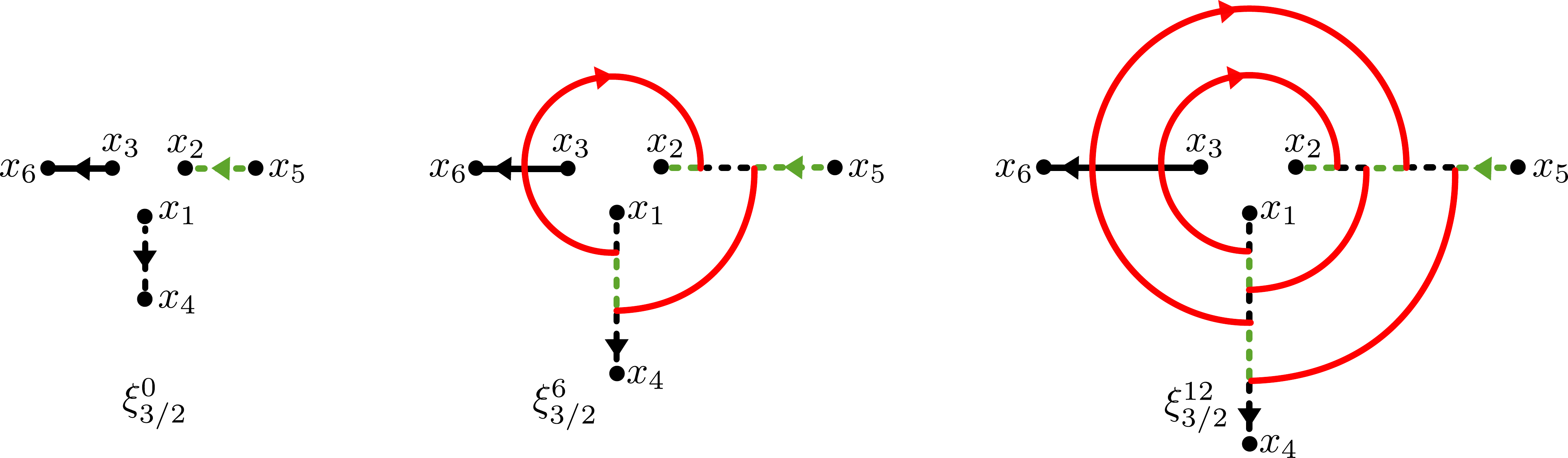}
\caption{\small The first few diagrams in the perturbative expansion that contribute to the correlator $\langle\tr(\psi_4(x_1)\psi_1^\dagger(x_2)Z(x_3))\,\tr(Z^\dagger(x_6)\psi_1(x_5)\psi_4^\dagger(x_4))\rangle$.}\label{fig:psi4psi1_perturb}
\end{figure}

\begin{figure}    
\centering
\includegraphics[width=0.55\textwidth]{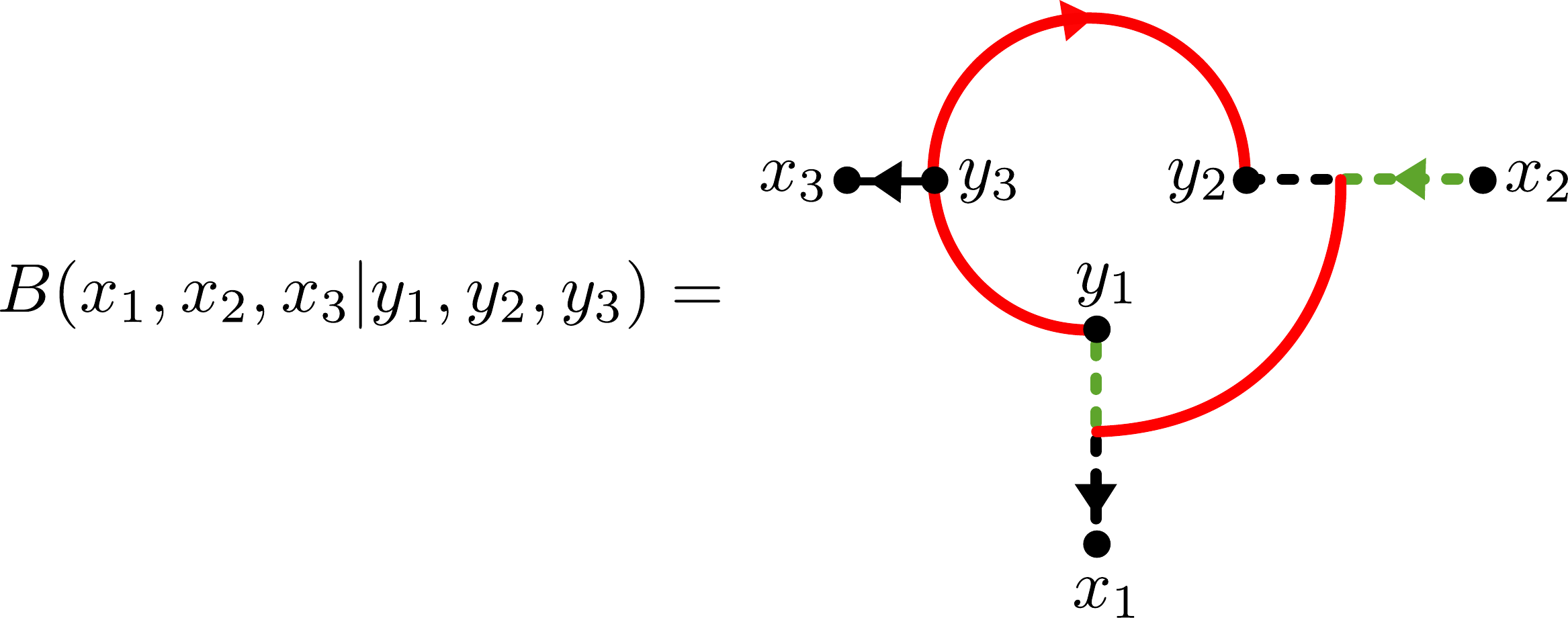}
\caption{\small The graph-building operator for the correlator $\psi$ in (\ref{psi4psi1Z}).}\label{fig:psi4psi1_GBO}
\end{figure}

The bulk diagrams contributing in perturbation theory appear in figure \ref{fig:psi4psi1_perturb}. 
The corresponding graph-building operator is plotted in figure \ref{fig:psi4psi1_GBO}. Its inverse reads
\beq\la{gbopsi2Z}
\widehat{B}^{-1} = x_{13}^2\, x_{23}^2\, s_1^\alpha\,\bar{s}_{2,\dot\alpha} \,\cancel{\p}_{x_1,\alpha \dot \beta}\, \bar{\cancel{\p}}_{x_2}^{\,\dot\alpha \beta} x_{12}^2\, \bar{\cancel{\p}}_{x_1}^{\,\dot\beta\gamma}\, \cancel{\p}_{x_2,\beta\dot \gamma}\, \p_{s^\gamma_1} \p_{\bar s_2}^{\dot\gamma}\, \Box_{x_3}\,.
\eeq

The associated integrable spin chain contains a fermion and an anti-fermion of dimension $\frac{3}{2}$, and a scalar of dimension $1$. The inverse of the graph-building operator (\ref{gbopsi2Z}) is proportional to the corresponding transfer matrix in the $\bf 6$ representation, evaluated at $u=0$:
\beq
\widehat{B}^{-1}\ \propto\ \widehat{T}_{\mathbf{6}}\left(0;-1/4,1/4,0\right) =\tr_{\mathbf{6}}\(L_{x_1}^{\left(3/2,1,0;\boldsymbol{6}\right)} \left(-1/4\right) \cdot L_{x_2}^{\left(3/2,0,1;\boldsymbol{6}\right)} \left(1/4\right)\cdot\text{\ensuremath{L_{x_3}^{\left(1,0,0;\boldsymbol{6}\right)}}}  \left(0\right)\)\,.
\eeq
The relevant Lax matrices are given in equations \eqref{Lax scalar1}, \eqref{Lax fermion}, and \eqref{Lax antifermion} of appendix \ref{app:Lax matrices}.

Similarly, one could consider operators with more scalars like $\tr(\psi_4 Z^m \psi_1^\dagger Z^n)$, or operators with more fermions of the same flavour like $\tr(\psi_4 \psi_4 Z^m \psi_4 Z^n)$.

\section{Holography and the Fishchain Model} \label{sec:holography}

One of the main motivations for this work is to extend our understanding of holography at large $N$. For the fishnet model, a first-principles derivation of a holographic dual was put forward in \cite{Gromov:2019aku,Gromov:2019bsj,Gromov:2019jfh}. Ideally, one would like to extend this derivation to ${\cal N}=4$ SYM theory. The dual ``fishchain model'' however, only captures a tiny set of operators of the mother ${\cal N}=4$ SYM theory. In this section we show that the fishchain model can be extended to include all the operators that we have studied here. They all have the same strong coupling classical description as that of the fishnet operators. They differ at the quantum level, and may include internal degrees of freedom at the fishchain sites. 

The basic idea of the holographic derivation is to identify the equation
\begin{equation}\la{BtoH}
   \mathcal{H}\,|\psi\rangle \equiv \(\xi^{-2M}\widehat{B}^{-1}- 1\)|\psi\rangle=0\,,
\end{equation}
with a time reparametrisation constraint for a chain of particles. This identification then results in a chain of particles with nearest-neighbour interactions moving in $AdS_5$.

In the fishchain model, quantum corrections are controlled by $\hbar=1/\xi$. Because $\d={\ii\over\hbar}\hat p$, at large $|\xi|$ the only terms in $\widehat{B}^{-1}$ that remain are the ones with two derivatives at each site. 
For all the cases we have studied these terms turn out to be the same as the fishnet ones. Therefore, at strong coupling they are all described by the classical fishchain model \cite{Gromov:2019aku}. 

For example, consider the inverse of the graph-building operator in \eqref{fermion-inverse} for the case of $\text{Tr}(Z\psi_{4\alpha})$. It can be written as
\begin{equation}
     \widehat{B}^{-1} = x^4_{12}\,\Box_{x_2}\,\Box_{x_1}+2x^2_{12}\,\cancel{x}_{21}\cancel{\bar{\partial}}_{x_2}\,\Box_{x_1}\;.
\end{equation}
Here the first term is the inverse of the fishnet graph-building operator while the second has only one derivative at site $2$. It is therefore suppressed at large $|\xi_{4/3}|$. Moreover, at this site we have a fermionic degree of freedom on which the second term acts non-trivially.

The quantisation of the classical fishchain model results in a chain of particles moving in $AdS_5$. These particles may now have internal degrees of freedom. The map between the boundary wave function $\Phi_\cO$ and the fishchain bulk wave function $\Psi_\cO$ is defined as\footnote{To include fermions, a different bulk-to-boundary propagator is required and also a different embedding formalism, see \cite{Nishida:2018opl}.}
\beq\la{maptoAdS}
\Psi\(\{Z_i\}_{i=0}^M;\{\Omega_i\}_{i=0}^M\)\equiv\int\prod_{j=1}^M\(D^4Y_j\, G^{\Delta_j,[p_j]}\(Z_j,\Omega_j|Y_j,\d_{\Theta_j}\)\)\Phi\(\{Y_j\}_{i=0}^M;\{\Theta_i\}_{i=0}^M\)\,,
\eeq
where the $Y_i$'s and the $\Theta_i$'s are the boundary embedding coordinates and polarisation vectors. Similarly, the $Z_i$'s and $\Omega_i$'s are the bulk AdS embedding coordinates and polarisation vectors. These coordinates are subject to the constraints $Z\cdot Z=-1$ and $\Omega\cdot Z=0$. The definition of the integration measure can be found in Section 5.1 of \cite{Gromov:2019bsj}. Finally, $G^{\Delta,[p]}$ is the $p$-form bulk-to-boundary
propagator. It is given by \cite{Tamaoka:2017jce}
\begin{equation}\la{btbprop}
G^{\Delta,[p]} \left(Z,\Omega|Y,\d_\Theta\right) = \frac{\left[\left(\d_\Theta\cdot\Omega\right) \left(Y\cdot Z\right) - \left(Z\cdot\d_\Theta\right) \left(Y\cdot\Omega\right)\right]^{p}}{-4\pi^2\left(Z\cdot Y\right)^{4-\Delta+p}}\,.
\end{equation}

The map (\ref{maptoAdS}) can be extended to the case with mixing that we have studied in the previous section. In that case, 
$M=J$ and the wave function is a direct sum of two spaces
\beq
\Phi\(\{Y_j\}_{i=0}^J;\Theta_0\)=\(\!\!\begin{array}{c}\Phi_F\(\{Y_j\}_{i=0}^J;\Theta_0\)\\ \Phi_X\(\{Y_j\}_{i=0}^J\)\end{array}\!\!\)\,,\qquad\Psi\(\{Z_j\}_{i=0}^J;\Omega_0\)=\(\!\!\begin{array}{c}\Psi_F\(\{Z_j\}_{i=0}^J;\Omega_0\)\\ \Psi_X\(\{Z_j\}_{i=0}^J\)\end{array}\!\!\)\,.
\eeq
At sites $i=1,\ldots,J$ we have $(\Delta_i,p_i)=(1,0)$. At site $i=0$ we have the direct sum of $(\Delta,p)=(2,2)\bigoplus(2,0)$. Correspondingly, in (\ref{maptoAdS}) we have
\beq
G^{\Delta_0,[p_0]}\,\Phi=\(\!\!\begin{array}{c}G^{2,[2]}\,\Phi_F\\ G^{2,[0]}\,\Phi_X\end{array}\!\!\)\,.
\eeq

It follows from this map that $(Z\cdot \partial_\Omega)\Psi_{F}(Z,\Omega)=0$. Moreover, the constraint $(Y\cdot\d_\Theta)\Phi_F=0$ implies that $(\d_Z\cdot \partial_\Omega)\Psi_{F}(Z,\Omega)=0$. Using (\ref{T6=B}), (\ref{BtoH}), and (\ref{maptoAdS}), the inverse of the graph-building operator 
maps into the following quantum fishchain Hamiltonian 
\beq
\mathcal{H}=\xi^{-2(J+1)}_{1+1/J}\,\tr_{\bf 6}\[\left(\!\!\begin{array}{cc}
{\mathbb L}_{FF}(Z_0,\Omega_0)& {\mathbb L}_{FX}(Z_0,\Omega_0)\\
{\mathbb L}_{XF}(Z_0,\Omega_0) & {\mathbb L}_{XX}(Z_0,\Omega_0)
\end{array}\!\!\right)\cdot\mathbb{L}\left(Z_1\right)\cdot\ldots\cdot\mathbb{L}\left(Z_J\right)\]-1\,,
\eeq
with 
\begin{align}
{\mathbb L}(Z) & = -\frac{1}{2}\left(Z^{M}Z^{N}\Box_{Z}+2Z^{(M}\partial_{Z}^{N)}-\partial_{Z}^{M}\partial_{Z}^{N}+2\eta^{MN}\right), \nn\\
{\mathbb L}_{FF}(Z,\Omega) & =-\frac{1}{2}\left(\left(\partial_{Z}\cdot\partial_{\Omega}\right)Z^{M}Z^{N}\left(\Omega\cdot\partial_{Z}\right)-\partial_{Z}^{M}\partial_{Z}^{N}\right),\nn\\
{\mathbb L}_{XX}(Z,\Omega) & =-\frac{1}{2}\left(\frac{1}{2}Z^{(M}\Box_{Z}Z^{N)}-\partial_{ Z}^{M}\partial_{ Z}^{N}+2\eta^{MN}\right), \\
{\mathbb L}_{FX}(Z,\Omega) & =\frac{1}{2}\left(\Omega\cdot\partial_{ Z}\right)\Omega^{[M}\partial_{ Z}^{N]},\nn\\
{\mathbb L}_{XF}(Z,\Omega) & =-\frac{1}{4}\partial^{[N}_{\Omega}\partial_{\Omega}^{M]}\,.\nn
\end{align}

\section{Conclusion and Discussion}

In this paper, we have argued that any single-trace operator in the planar limit of ${\cal N}=4$ SYM theory has a large twist double-scaling limit. In this operator-dependent limit, the loop corrections to the operator two-point function have an iterative structure. They can therefore be analysed to all loop order using the graph-building operator technique. Correspondingly, they all have a dual fishchain description. This description is discrete at any order in the expansion around the large twist limit. The picture that emerges is that of operators breaking into sectors with different double-scaling limits. At the level of the spectrum, these sectors decouple from each other. 

We have considered several types of examples. The case of mixing between operators with different double-scaling limits is both interesting and generic. The corresponding graph-building operator is a matrix, which can be used both to study the large twist operator's dimensions and to compute corrections away from this limit. Finally, we have proven the integrability of all the cases that we have considered. 

There are many future directions to pursue, some of which we list below. 
\begin{itemize}
\item In section \ref{intsec} we have found that the case with mixing is an integrable spin-chain model. It would be interesting to understand whether it fits inside the standard construction of integrable models or whether it is new.  
\item In section \ref{fncorrectionssec} we have described how certain corrections away from the large twist double-scaling limit arise from mixing between operators with a different double-scaling limit. It would be interesting to understand if this construction can be extended such that all the corrections are of this type. If true, it would imply that the elements of the corresponding graph-building operator do not receive loop corrections.
\item To solve the theory at large twist we also need to compute the structure constants. Three point functions between operators with the same double-scaling limit are expected to take a similar form to the ones in the fishnet theory. On top of these, it would be interesting to explore three point functions between operators with different double-scaling limits. 
\item A similar fishnet-like limit for $\mathcal{N}=2$ supersymmetric gauge theories was introduced in \cite{Pittelli:2019ceq}. It would be interesting to study this structure in such theories using similar generalised double-scaling limits.
\end{itemize}

\section*{Acknowledgements}
We thank S. Chester and D.-l. Zhong for invaluable discussions. We thank D.-l. Zhong for comments on the draft. AS, GF and EO are supported by the Israel Science Foundation (grant number 1197/20). GF is grateful to the Azrieli Foundation for the award of an Azrieli Fellowship.

\begin{appendix}

\section{Notation and Conventions}

In this appendix, we detail the conventions and notation we use throughout the paper.

\subsection{Metric and Sigma Matrices}
\label{app:convention}

\subsubsection{4 Dimensions}

We work in Euclidean signature and use Greek indices to denote the flat four dimensional coordinates as $x^\mu$, with $\mu=1,2,3,4$. 
The sigma matrices (or chiral projections of the gamma matrices) are
\begin{equation}
\sigma^\mu = \left(\ii\sigma_{x},\ii\sigma_{y},\ii\sigma_{z},\mathrm{I}_2\right) \quad \text{and} \quad \bar{\sigma}^\mu = \left(-\ii\sigma_{x}, -\ii\sigma_{y}, -\ii\sigma_{z}, \mathrm{I}_2\right)\, ,
\end{equation}
where $\sigma_x, \sigma_y,$ and $\sigma_z$ are the usual Pauli matrices while $\mathrm{I}_2$ is the $2\times 2$ identity matrix.

The sigma matrices satisfy
\begin{equation}
    \sigma^\mu \bar\sigma^\nu + \sigma^\nu \bar \sigma^\mu = \bar\sigma^\mu \sigma^\nu + \bar\sigma^\nu  \sigma^\mu = 2\,\delta^{\mu\nu} \mathrm{I}_2\, ,
\end{equation}
\begin{equation}
    \sigma^\mu = (\bar \sigma^\mu)^\dagger\, ,\qquad \epsilon \sigma^\mu \epsilon = - ^t\!\bar \sigma^\mu\, ,
\end{equation}
where $\epsilon = \begin{pmatrix}
0 & 1\\
-1 & 0\\
\end{pmatrix}$ and $^t\!\bar \sigma^\mu$ is the transpose of $\bar \sigma^\mu$. Moreover,
\begin{equation}
    \sigma^\mu\otimes \bar \sigma_\mu = 2\,\mathbb{P}\, ,\quad \sigma^\mu\otimes \sigma_\mu = \bar \sigma^\mu\otimes \bar \sigma_\mu = 2\left(\mathrm{I}_4 - \mathbb{P}\right)\, ,
\end{equation}
where $\mathrm{I}_4$ is the $4\times 4$ identity matrix and $\mathbb{P}$ is the permutation operator on $\mathbb{C}^2\otimes \mathbb{C}^2$. In explicit index notations, the previous relations read
\begin{equation}
    (\sigma^\mu)_{\alpha \dot\alpha} (\bar \sigma_\mu)^{\dot\beta \beta} = 2\,\delta^{\beta}_\alpha \delta^{\dot \beta}_{\dot \alpha}\, ,\quad (\sigma^\mu)_{\alpha \dot\alpha} (\sigma_\mu)_{\beta \dot\beta} = 2\,\epsilon_{\alpha\beta} \epsilon_{\dot\alpha \dot\beta}\, ,\quad (\bar\sigma^\mu)^{\dot\alpha \alpha} (\bar\sigma_\mu)^{\dot\beta \beta} = 2\,\epsilon^{\alpha\beta} \epsilon^{\dot\alpha \dot\beta}\, ,
\end{equation}
with $\epsilon_{\alpha\beta}, \epsilon_{\dot\alpha \dot\beta}, \epsilon^{\alpha\beta}, \epsilon^{\dot\alpha \dot\beta}$ all antisymmetric in their indices and such that $\epsilon_{12} = \epsilon_{\dot 1 \dot 2} = \epsilon^{12} = \epsilon^{\dot 1 \dot 2} = 1$.

The matrices
\begin{equation}
    \sigma_{\mu\nu} = \frac{\sigma_\mu \bar\sigma_\nu - \sigma_\nu \bar\sigma_\mu}{4}\quad\text{and}\quad  \bar\sigma_{\mu\nu} = \frac{\bar\sigma_\mu \sigma_\nu - \bar\sigma_\nu \sigma_\mu}{4}
\end{equation}
are the generators of the two inequivalent two-dimensional representations of $\mathfrak{so}(4,\mathbb{C})$.

\subsubsection{6 Dimensions}

We use upper-case Latin indices $M,N,P$, etc. to denote indices ranging from $1$ to $6$. The indices are lowered using the tensor $\eta_{MN} = \operatorname{Diag}(1,1,1,1,1,-1)_{MN}$. For an arbitrary tensor $F^{\cdots MNP\cdots}$ it is also convenient to define
\begin{equation}
    F^{\cdots M \pm P\cdots} = F^{\cdots M 6 P\cdots} \pm F^{\cdots M 5 P\cdots}\, .
\end{equation}
A contraction of two tensors can then be written
\begin{equation}
    A^M B_M = A^M B^N \eta_{MN} = A^\mu B_\mu + A^5 B^5 - A^6 B^6 = A^\mu B_\mu -\frac{A^+ B^- + A^- B^+}{2}\, .
\end{equation}

\subsection{$\gamma$-Deformed $\mathcal{N}=4$ SYM}
\label{app:gammadeformed}

The $\gamma$-deformed theory is a family of deformations of the 
interaction terms in the ${\cal N}=4$ SYM Euclidean Lagrangian. It is parameterised by $\gamma_{i=1,2,3}$, and is defined as
\beq\la{lagrangian}
\mathcal{L}= - N\,\text{Tr}\left[\frac{1}{4}F_{\mu\nu}F^{\mu\nu} + 
 D^{\mu}\phi_{i}^{\dagger} D_{\mu}\phi^{i} + \psi^\dagger \overline{\cancel{D}}\psi\right] + \mathcal{L}_{int}\, ,
\eeq
with
\begin{align}\label{l-int}
\mathcal{L}_{\text{int}}=\ \  g^2N\, \tr&\Big[2\, e^{-\ii\epsilon^{ijk}\gamma_{k}} \phi_{i}^{\dagger} \phi_{j}^{\dagger} \phi^{i} \phi^{j} - \frac{1}{2}\{\phi_{i}^{\dagger},\phi^{i}\}\{\phi_{j}^{\dagger},\phi^{j}\}\Big]\nn\\
+\sqrt{2} g N\,\tr&\Big[e^{\frac{\ii}{2}\gamma_{j}^{-}} \(\psi^\dagger_{4}\phi^{j}\psi^\dagger_{j}-\psi^{4}\phi_{j}^{\dagger}\psi^{j}\)
+ e^{-\frac{\ii}{2}\gamma_{j}^{-}}\(\psi^{j}\phi_{j}^{\dagger}\psi^{4}- \psi^\dagger_{j}\phi^{j} \psi^\dagger_{4} \)\Big]\\
+\sqrt{2} g N\,\tr&\Big[ \ii e^{\frac{\ii}{2}\epsilon_{imk}\gamma_{m}^{+}} \(\epsilon_{ijk}\psi^{i}\phi^{j}\psi^{k}+ \epsilon^{ijk} \psi^\dagger_{i}\phi_{j}^{\dagger}\psi^\dagger_{k}\)\Big]\,.\nn
\end{align}
Here, 
\begin{equation}
\gamma_1^{\pm}=-\frac{\gamma_3 \pm\gamma_2 }{2},\qquad\gamma_2^{\pm}=-\frac{\gamma_1 \pm\gamma_3 }{2},\qquad\gamma_3^{\pm}=-\frac{\gamma_2 \pm\gamma_1 }{2}\,,
\end{equation}
and we have suppressed the spinor indices, assuming the contractions $\epsilon^{\beta\alpha} \psi^i_{\alpha} \psi^j_{\beta}$ and $\epsilon^{\dot\alpha\dot\beta} \psi^\dagger_{i,\dot{\alpha}} \psi^\dagger_{j,\dot{\beta}}$ for instance. The covariant derivative and the field strength are 
\begin{equation}
    D_{\mu}=\partial_{\mu}+\ii g[A_{\mu},\cdot]
\end{equation}
and
\begin{equation}
    F_{\mu\nu} = - \frac{\ii}{g}[D_{\mu},D_{\nu}] = \partial_{\mu}A_{\nu} - \partial_{\nu}A_{\mu} + \ii g [A_{\mu},A_{\nu}]\, ,
\end{equation}
And we use $\overline{\cancel{D}} = D_{\mu}\bar{\sigma}^\mu$. In \eqref{lagrangian}, moreover, summation over doubly and triply repeated indices is assumed. Throughout this paper we use either Feynman gauge, or more generally, the $R_\alpha$ gauge. The Feynman rules associated to the propagators are recalled in appendix \ref{feynman-rules}. In other parts of this paper, we also use notation where the three complex scalars are $X=\phi_1$, $Z=\phi_2$, and $Y=\phi_3$.

In terms of the three $\gamma$-parameters, the double-scaling limit in all the examples we consider are of the type (\ref{fnlimit}) with $\theta=-\gamma_3$, and $\gamma_{1} = \gamma_{2} = 0$. In particular, the fishnet limit is that limit with $n=1$. Namely,\footnote{Note that in (\ref{fnlimitgamma}) $\gamma_3$ is complex. Hence, the $\gamma$-deformed CFT is not unitary. Instead of twisting the theory, as in (\ref{l-int}), one can take the viewpoint where we stay with ${\cal N}=4$ SYM theory and twist the operators, see \cite{Cavaglia:2020hdb}.}
\beq\la{fnlimitgamma}
\gamma_{1} = \gamma_{2} = 0\,,\qquad e^{-\ii\gamma_3}\to\infty\,,\qquad\text{with}\quad  \xi_1^2\equiv {g^2\,e^{-\ii\gamma_3}\over8\pi^2}\quad\text{fixed}\,.
\eeq

The twists we consider can be obtained by replacing the product of fields in the Lagrangian with a star-product (before integrating out the auxiliary fields), \cite{Fokken:2013aea}. This star-product is defined as
\begin{equation}
    A\star B=e^{\frac{\ii}{2} q_A\wedge q_B}AB
\end{equation}
where
\begin{equation}
    q_{A} \wedge q_{B}=\left(q_{A}\right)^{\mathrm{T}} C q_{B}, \quad C=\left(\begin{array}{ccc}
0 & -\gamma_3  & \gamma_2  \\
\gamma_3  & 0 & -\gamma_1  \\
-\gamma_2  & \gamma_1  & 0
\end{array}\right)\;,
\end{equation}
and the $q_A$'s are the $SU(4)$ R-symmetry Cartan charges. Explicitly, the charges are given by
\begin{equation}
    \def\arraystretch{1.3}
    \begin{array}{c|cccc|c|ccc}
 & \psi_{\alpha}^{1} & \psi_{\alpha}^{2} & \psi_{\alpha}^{3} & \psi_{\alpha}^{4}  & A_{\mu} & \phi_1  & \phi_2  & \phi_3  \\
\hline q_{B}^{1} & +\frac{1}{2} & -\frac{1}{2} & -\frac{1}{2} & +\frac{1}{2}  & 0 & 1 & 0 & 0 \\
q_{B}^{2} & -\frac{1}{2} & +\frac{1}{2} & -\frac{1}{2} & +\frac{1}{2} & 0 & 0 & 1 & 0 \\
q_{B}^{3} & -\frac{1}{2} & -\frac{1}{2} & +\frac{1}{2} & +\frac{1}{2} & 0 & 0 & 0 & 1
\end{array}
\end{equation}
Note that the Lagrangian \eqref{lagrangian} needs to be supplemented  with ghost fields. They are not relevant in the examples we study.

\subsection{Feynman Rules} \label{feynman-rules}

The free propagators in the $R_\alpha$ gauge are given by
\beqa
\langle (\phi_i^\dagger)_{ab}(x) \phi^j_{cd}(y) \rangle_0 &=& \frac{1}{4\pi^2 N_c (x-y)^2} \delta_{i}^j\left(\delta_{ad} \delta_{bc} - \frac{\delta_{ab} \delta_{cd}}{N_c}\right)\,,\\
\langle A^\mu_{ab}(x) A^{\nu}_{cd}(y) \rangle_0&=& \frac{\frac{\alpha+1}{2\alpha}\delta^{\mu\nu} + \frac{\alpha - 1}{\alpha}\frac{(x-y)^\mu (x-y)^\nu}{(x-y)^2}}{4\pi^2 N_c (x-y)^2} \left(\delta_{ad} \delta_{bc} - \frac{\delta_{ab} \delta_{cd}}{N_c}\right)\, ,\label{gluon propagator}\\
\langle \psi^A_{\alpha cd}(x) (\psi_{\dot\alpha B}^\dagger)_{ab}(y) \rangle_0& =& \frac{(\cancel{x}-\cancel{y})_{\alpha \dot\alpha}}{2\pi^2 N_c (x-y)^4} \delta^{A}_B\left(\delta_{ad} \delta_{bc} - \frac{\delta_{ab} \delta_{cd}}{N_c}\right)\, ,
\eeqa
where $\cancel{x} = x_\mu\sigma^\mu$, and, in this equation only, the upper-case indices $A$ and $B$ range from 1 to 4.

\section{Star-Triangle Relations} \label{star-triangle}

We collect here some integral identities, useful for the eigenvalue computations of section \ref{sec:ZXX} and section \ref{sec:scalar-fermion}. Such identities appeared for the first time in \cite{Chicherin:2012yn} (see also \cite{Kazakov:2018gcy}). In this section, $\ell$ and $S$ are non-negative integers, $\zeta$ is a null vector ($\zeta^\mu \zeta_\mu = 0$), and $A$ is a 2-tensor such that
\begin{equation}
    A_{\mu\nu} + A_{\nu\mu} = \frac{\delta_{\mu\nu}}{2} ?A^\rho_\rho?\, ,\quad ?A^{\mu}_\nu? ?A_{\mu\rho}? = ?A_{\mu\nu}? ?A_{\rho}^{\nu}?= 0\, .
\end{equation}

Provided that $\zeta^\mu A_{\mu\nu} = 0$ and $\alpha+\beta+\gamma = 4 + S$, the following integral identities hold true
\begin{multline}\label{star-triangle1}
    \int \frac{\left[(x_1 - y)^\mu A_{\mu\nu} (y-x_2)^\nu\right]^S \left[\zeta\cdot \left(\frac{y-x_1}{(y-x_1)^2} - \frac{x_{21}}{x_{21}^2}\right)\right]^\ell}{(x_1 - y)^{2\alpha} (x_2 - y)^{2\beta} (x_3 - y)^{2\gamma}} \frac{\dd^4 y}{\pi^2}\\
    = \frac{\Gamma(2+S-\alpha) \Gamma(2+S+\ell-\beta) \Gamma(2-\gamma)}{\Gamma(\alpha+\ell) \Gamma(\beta) \Gamma(\gamma)} \frac{\left[x_{13}^\mu A_{\mu\nu} x_{32}^\nu\right]^S \left[\zeta\cdot \left(\frac{x_{31}}{x_{31}^2} - \frac{x_{21}}{x_{21}^2}\right)\right]^\ell}{x_{12}^{2(2-\gamma)} x_{23}^{2(2+S-\alpha)} x_{31}^{2(2+S-\beta)} }\, ,
\end{multline}
and
\begin{multline}\label{star-triangle2}
    \int \frac{\left[(x_1 - y)^\mu A_{\mu\nu} (y-x_2)^\nu\right]^S \left[\zeta\cdot \left(\frac{y-x_1}{(y-x_1)^2} - \frac{x_{31}}{x_{31}^2}\right)\right]^\ell}{(x_1 - y)^{2\alpha} (x_2 - y)^{2\beta} (x_3 - y)^{2\gamma}} \frac{\dd^4 y}{\pi^2}\\
    = \frac{\Gamma(2+S-\alpha) \Gamma(2+S-\beta) \Gamma(2+\ell-\gamma)}{\Gamma(\alpha+\ell) \Gamma(\beta) \Gamma(\gamma)} \frac{\left[x_{13}^\mu A_{\mu\nu} x_{32}^\nu\right]^S \left[\zeta\cdot \left(\frac{x_{21}}{x_{21}^2} - \frac{x_{31}}{x_{31}^2}\right)\right]^\ell}{x_{12}^{2(2-\gamma)} x_{23}^{2(2+S-\alpha)} x_{31}^{2(2+S-\beta)} }\, .
\end{multline}

If $\zeta_\rho\p_3^\rho A_{\mu\nu} = 0$, $\alpha+\beta+\gamma = 4 + S$, and there exists another null vector $\eta$ such that
\begin{equation}
    A_{\mu\nu} x_{32}^{\nu} = x_{23}^2 \eta_\mu\, ,\qquad \zeta^\mu \eta_\mu = 0\, ,
\end{equation}
then 
\begin{multline}\label{star-triangle3}
    \int \frac{\left[(x_1 - y)^\mu A_{\mu\nu} (y-x_2)^\nu\right]^S \left[\zeta\cdot \left(\frac{y-x_3}{(y-x_3)^2} - \frac{x_{23}}{x_{23}^2}\right)\right]^\ell}{(x_1 - y)^{2\alpha} (x_2 - y)^{2\beta} (x_3 - y)^{2\gamma}} \frac{\dd^4 y}{\pi^2}\\
    = \frac{\Gamma(2+S-\alpha) \Gamma(2+S+\ell-\beta) \Gamma(2-\gamma)}{\Gamma(\alpha) \Gamma(\beta) \Gamma(\gamma+\ell)} \frac{\left[x_{13}\cdot \eta\right]^S \left[\zeta\cdot \left(\frac{x_{13}}{x_{13}^2} - \frac{x_{23}}{x_{23}^2}\right)\right]^\ell}{x_{12}^{2(2-\gamma)} x_{23}^{2(2-\alpha)} x_{31}^{2(2+S-\beta)} }\,.
\end{multline}

\section{Correlators} \label{section-setup}

In this appendix, we compute some of the  correlators of short non-local operators studied in the main text.

\subsection{The Correlator of Two Length-Two Non-local Operators}
\label{app:Bethe-Salpeter}

We begin with a review of the method for computing the two-point function between non-local operators made from two fields. We loosely call it a ``four-point function''. The procedure has been widely developed in the literature, see for example \cite{Gromov:2018hut} for the case of scalars. Since we will also be interested in fermionic representations, we keep the discussion very general; we merely assume that the correlator has an iterative structure encoded in a graph-building operator. In line with this, we will heavily rely on the general results about harmonic analysis on the conformal group from \cite{Karateev:2018oml}.

Consider some four-point function
\begin{equation} \label{4-point}
    G(x_1,x_2|x_3,x_4)\equiv\langle \text{Tr}(\mathcal{O}_1(x_1)\mathcal{O}_2(x_2)) \text{Tr}(\mathcal{O}_3(x_3)\mathcal{O}_4(x_4)) \rangle \, ,
\end{equation}
where $\mathcal{O}$ is any combination of the fields in the theory and their derivatives.
We assume that all of the diagrams contributing to $G$ have an iterative structure. Namely, that we can write the correlator as the geometric series 
\begin{equation}\label{eq:iterative_ladder_structure}
\widehat{G}= \sum_{L=0}^\infty \chi^L \widehat B^{L}  \widehat{G}^{(0)} = \frac{1}{1-\chi \widehat B}  \widehat{G}^{(0)}\;,
\end{equation}
where $\chi$ is some coupling constant, $\widehat B$ is the graph-building operator, and $\widehat{G}^{(0)}$ is the first contribution. All these operators are integral operators whose kernels we denote by $B(x_1,x_2|y_1,y_2)$, $G^{(0)}(x_1,x_2|y_1,y_2)$, etc. In order to simplify the expressions, we include factors of $\pi^{-2}$ in the integration measure. The action of $\widehat B$ on an arbitrary function $\Phi$ is thus given by
\begin{equation}
    \widehat{B}  \Phi(x_1,x_2)=\int\frac{\dd^4 y_1\dd^4 y_2}{\pi^4}  B(x_1,x_2|y_1,y_2)\Phi(y_1,y_2) \,.
\end{equation}
And, for example, the kernel of $\widehat B^2=\widehat B \widehat B$ is
\begin{equation}
     B^2(x_1,x_2|y_1,y_2) = \int  B(x_1,x_2|z_1,z_2) B(z_1,z_2|y_1,y_2) \frac{\dd^4 z_1\dd^4 z_2}{\pi^4}\, .
\end{equation}
Finding a complete basis of eigenvectors of $\widehat B$ will allow us to compute the four-point function \eqref{eq:iterative_ladder_structure}. We now turn to the diagonalisation of $\widehat B$.

In what follows, we assume that $\widehat B$ is conformally invariant. Namely, we assume that $\widehat B$ commutes with the action of two principal series representations of the conformal algebra. One $\rho_1 = (\Delta_1,\ell_1,\bar \ell_1)$ acts on $x_1$, and the other $\rho_2= (\Delta_2,\ell_2,\bar \ell_2)$ acts on $x_2$. The diagonalisation is thus reduced to decomposing the tensor product into irreducible representations. Consequently, the eigenvectors of $\widehat B$ take the form of conformal three-point functions between the operators at $x_1$, $x_2$, and a third operator in a representation $\rho$ at $x_0$, $\Psi_{x_0}^{\rho}$. 
For brevity, we do not display the $\mathfrak{so}(4)$ indices of $\Psi_{x_0}^{\rho}$ associated with each representation. The eigenvalue equation takes the form
\begin{equation}\label{eq:eigenvalue_eq}
\widehat{B} \Psi^\rho_{x_0}(x_1,x_2)= E(\rho)\, \Psi^\rho_{x_0} \left(x_1 , x_{2}\right)\,.
\end{equation}
In this equation, for instance, there is an implicit contraction of the indices associated with representations $\rho_1$ and $\rho_2$ between $\widehat B$ and $\Psi^\rho_{x_0}$, whereas the indices associated with $\rho$ remain arbitrary. In order to have a complete basis of eigenvectors, it will be enough to consider $\rho = (\Delta,\ell,\bar \ell)$ such that $\Delta=2 + \ii \nu$ with $\nu$ real and positive. We introduce the left eigenvector $\widetilde{\Psi}_{x_0}^{\rho}$ of $\widehat B$ associated to the eigenvalue $E(\rho)$. It has the form of a conformal three-point function of operators transforming according to the representations $\tilde{\rho}\equiv (4-\Delta,\ell,\bar\ell)=(2-\ii\nu,\ell,\bar\ell)$, $\tilde{\rho}_1$, and $\tilde{\rho}_2$. The orthogonality relation will read
\begin{equation}\label{ortho-relation}
    \int\frac{\dd^{4} x_1 \dd^{4} x_2}{\pi^4} \widetilde{\Psi}_{x_{0}}^{\rho}(x_1,x_2) \Psi_{x'_0}^{\rho'}(x_1,x_2)  = \frac{2\pi^3}{\mu(\rho)} \delta_{\rho,\rho'} \delta^{(4)}(x_0 - x'_0) \operatorname{Id}_{\ell,\bar\ell}\, ,
\end{equation}
where $\Delta = 2 + \ii\nu$ and $\Delta' = 2 + \ii\nu'$ with $\nu, \nu'>0$, the indices for $\rho_1$ and $\rho_2$ have been contracted in the left-hand side, and
\begin{equation}
    \delta_{\rho,\rho'} = \delta(\nu-\nu') \delta_{\ell,\ell'} \delta_{\bar\ell,\bar\ell'}\, .
\end{equation}
Here, $\operatorname{Id}_{\ell,\bar\ell}$ denotes the unit operator in the representation $(\ell,\bar\ell)$ of $\mathfrak{so}(4)$. In practice, we will always contract the indices associated to this space with auxiliary polarisation vectors $s,\bar s,s',\bar s'$ so that $\operatorname{Id}_{\ell,\bar\ell}$ would be replaced with $(s\cdot s')^\ell (\bar s\cdot\bar s')^{\bar \ell}$. The coefficient $\mu(\rho)$ can be found in \cite{Karateev:2018oml}, and the completeness relation reads \cite{Dobrev:1977qv}\footnote{We only consider representations of the complementary series such that $|2-\Delta_1| + |2 - \Delta_2|\leqslant 2$. We can thus use the results of section 10.D of \cite{Dobrev:1977qv}.}
\begin{multline}
   \int\limits_0^{\infty}\frac{\dd \nu}{2\pi}\sum_{\substack{\ell,\bar\ell \\ \rho\subset \rho_1\otimes \rho_2}} \mu(\rho) \int \frac{\dd^4 x_0}{\pi^2}\,\Psi_{x_{0}}^{(2 + \ii\nu,\ell_1,\bar \ell_1)}(x_1,x_2) \widetilde{\Psi}_{x_{0}}^{(2 + \ii\nu,\ell_2,\bar \ell_2)}(y_1,y_2)\\
   = \pi^4 \delta^{(4)}(x_1 - y_1) \delta^{(4)}(x_2 - y_2)\operatorname{Id}_{\ell_1,\bar\ell_1} \operatorname{Id}_{\ell_2,\bar\ell_2}\, .
\end{multline}
Inserting this completeness relation in the expression for the four-point function yields
\begin{equation}\label{g-rep}
    G(x_1 ,x_2|x_3, x_4) = \int\limits_0^{\infty}\frac{\dd \nu}{2\pi}\sum_{\substack{\ell,\bar\ell \\ \rho\subset \rho_1\otimes \rho_2}} \frac{\mu(\rho)}{1-\chi E(\rho)}\int \frac{\dd^4 x_0}{\pi^2}\, \Psi_{x_{0}}^{(2 + \ii\nu,\ell,\bar \ell)}(x_1,x_2) \left[\widetilde{\Psi}_{x_{0}}^{(2 + \ii\nu,\ell,\bar \ell)} \widehat{G}^{(0)}\right](x_3,x_4) \, .
\end{equation}
The function $\widetilde{\Psi}_{x_{0}}^{(2 + \ii\nu,\ell,\bar \ell)} \widehat{G}^{(0)}$ will also be of the form of conformal three-point function---up to some power of $x_{34}^2$. The integral over the auxiliary point $x_{0}$ can therefore be expressed in terms of four-dimensional conformal blocks $\mathfrak{G}$ as \cite{Karateev:2018oml}\footnote{Here, the conformal blocks are defined as in subsection 2.7.1 of \cite{Karateev:2018oml}.}  
\begin{multline}\label{conformal-block}
  \int\frac{\dd^4 x_0}{\pi^2} \Psi_{x_{0}}^{(2 + \ii\nu,\ell,\bar \ell)}(x_1,x_2) \left[\widetilde{\Psi}_{x_{0}}^{(2 + \ii\nu,\ell,\bar \ell)} \widehat{G}^{(0)}\right](x_3,x_4) \\
  = \widetilde{G}_0(x_3,x_4) C(\rho) \left[S(\rho_3,\rho_4;\tilde\rho)\mathfrak{G}(x_1,x_2|x_3,x_4) + S(\rho_1,\rho_2;\rho)\widetilde{\mathfrak{G}}(x_1,x_2|x_3,x_4)\right]\, .
\end{multline}
In the equation above, $\widetilde{G}_0(x_3,x_4) C(\rho)$ comes from $\widetilde{\Psi}_{x_{0}}^{(2 + \ii \nu,\ell,\bar \ell)} \widehat{G}^{(0)}$, whereas the terms between brackets result from the integral over $x_0$.
The second term in the brackets 
can be absorbed into that of the first one upon extension of the integration domain to the whole real line:
\begin{equation}\label{blocks}
    G(x_1 ,x_2|x_3, x_4) = \widetilde{G}_0(x_3,x_4)\int\limits_{-\infty}^{\infty}\frac{\dd\nu}{2\pi} \sum_{\substack{\ell,\bar\ell \\ \rho\subset \rho_1\otimes \rho_2}} \frac{C(\rho)\mu(\rho) S(\rho_3,\rho_4;\tilde\rho)}{1-\chi E(\rho)} \mathfrak{G}(x_1,x_2|x_3,x_4) \, .
\end{equation}

The block $\mathfrak{G}$ decays exponentially in the $\nu$ lower half-plane, so we can deform the integration contour in that direction and write the integral as the sum over the residues:
\begin{equation}
G =\widetilde{G}_0\sum_{\rho^{*}} P\left(\rho^{*}\right)\mathfrak{G}_{\mathcal{O}_{*}}\, ,
\end{equation}
where 
\begin{equation}\label{coff-blocks}
P(\rho^*) =-\text{Res}_{\Delta=\Delta^{*}}\left[\frac{C(\rho)\mu(\rho) S(\rho_3,\rho_4;\tilde\rho)}{1-\chi E(\rho)}\right],
\end{equation}
are the conformal block coefficients. The locations of the associated physical poles are obtained by solving
\begin{equation}\label{spectrum-solve}
    \chi\, E(\Delta^*,\ell,\bar \ell) = 1\, .
\end{equation}

Note that when deforming the contour away from the principal series, there are also poles arising from the conformal block $\mathfrak{G}$ and the numerator. These are expected to cancel against each other, but we did not attempt to verify it.

\subsection{The Operator $\tr(XX^\dagger Z)$}
\label{app:ZXX details}

We give here the results for the correlator \eqref{Gcorrelator} of subsection \ref{sec:ZXX}. The left eigenvectors of $\widehat{B}$ are
\begin{equation}
\widetilde{\Psi}^{(\Delta,S,S)}_{x_0,\zeta}(x_1,x_2) = \frac{\left(\frac{\zeta\cdot x_{10}}{x_{10}^{2}}-\frac{\zeta\cdot x_{20}}{x_{20}^{2}}\right)^{S}}{|x_{12}|^{1+S+\Delta} |x_{10}|^{5 - \Delta - S} |x_{20}|^{3 - \Delta -S}}\, .
\end{equation}
In particular, they satisfy
\begin{equation}
\left[\widetilde{\Psi}^{(\Delta,S,S)}_{x_0,\zeta} \widehat{G}^{(0)}\right] (x_3,x_4) = \frac{16\, x_{34}^{2}}{(4 \pi^{2})^3 (3+S-\Delta)^2(-1+S+\Delta)^2} \widetilde{\Psi}^{(\Delta,S,S)}_{x_0,\zeta}(x_3,x_4)\, ,
\end{equation}
where $\widehat{G}^{(0)}$ is given by \eqref{g0-scalar}. Thus, we conclude that in (\ref{conformal-block}) we have 
\beq
C(\Delta,S) = \frac{16}{(4 \pi^{2})^3 (3+S-\Delta)^2(-1+S+\Delta)^2}\qquad\text{and}\qquad\widetilde{G}_0(x_3,x_4) = x_{34}^2\,.
\eeq

The kinematic factors are
\begin{align}
    \mu (\Delta,S,S ) = 2^{S-1}(S+1) (\Delta -2)^2 (\Delta - S - 3) (\Delta +S-1)\,,
\end{align}
and 
\begin{align}
    S((3,0,0),(2,0,0);(4-\Delta,S,S)) = \frac{\pi ^2  \Gamma (S-\Delta+3) \Gamma \left(\frac{1}{2} (S+\Delta -1)\right) \Gamma \left(\frac{1}{2} (S+\Delta +1)\right)}{(2-\Delta) \Gamma (S+\Delta ) \Gamma \left(\frac{1}{2} (S-\Delta  +3)\right) \Gamma \left(\frac{1}{2} (S-\Delta +5)\right)}\, .
\end{align}

Putting everything together, we obtain
\begin{align}
    P(\Delta_\pm,S,S)=\mp \frac{(S+1) 2^{S-2\Delta_{\pm}} \Gamma \left(\frac{S+4-\Delta_{\pm}}{2}\right) \Gamma\left(\frac{S+1+\Delta_{\pm}}{2}\right)}{4\pi^4\, \xi^{2}_2\, \Gamma \left(\frac{S+3-\Delta_{\pm}}{2}\right) \Gamma \left(\frac{S+\Delta_{\pm}}{2}\right)}\, ,
\end{align}
where $\Delta_\pm$ is given in \eqref{eq: neat sol xxbar}.

\subsection{The Operator $\tr(Z\psi_4)$}
\label{app:fermion details}

We give here the results for the correlator \eqref{g-2} of subsection \ref{sec:scalar-fermion}. The first family of left eigenvectors of $\widehat{B}$ is
\begin{equation}
\widetilde{\Psi}^{(\Delta,S-\frac{1}{2},S+\frac{1}{2})}_{x_0,s,\bar{s}}(x_1,x_2) =  \frac{\bar{s}\, \cancel{\bar{x}}_{02} \left[s\left(\frac{\cancel{x}_{10}}{x_{10}^{2}}-\frac{\cancel{x}_{20}}{x_{20}^{2}}\right) \bar s_{0} \right]^{S-\frac{1}{2}}}{|x_{12}|^{\Delta+S+1} |x_{10}|^{-\Delta-S+5} |x_{20}|^{-\Delta-S+5}}\, .
\end{equation}
In particular, using the  formulae from appendix \ref{star-triangle}, they satisfy
\begin{multline}
    \left[\widetilde{\Psi}^{(\Delta,S-\frac{1}{2},S+\frac{1}{2})}_{x_0,s,\bar s}  \widehat{G}_2^{(0)}\right] (x_3,x_4) = \frac{1}{\pi^{4} \left(\Delta+S-1\right) \left(\Delta-S-2\right) \left(\Delta-S-4\right)}\\
    \times \frac{\bar{s}\,\cancel{\bar{x}}_{03}\,\cancel{x}_{34} \left[s\left(\frac{\cancel{x}_{30}}{x_{30}^{2}}-\frac{\cancel{x}_{40}}{x_{40}^{2}}\right) \bar s \right]^{S-\frac{1}{2}}}{|x_{34}|^{\Delta+S-1} |x_{30}|^{-\Delta-S+5} |x_{40}|^{-\Delta-S+5}}\, ,
\end{multline}
\beq
\left[\widetilde{\Psi}^{(\Delta,S-\frac{1}{2},S+\frac{1}{2})}_{x_0,s,\bar s}  \widehat{G}_2^{(0)}\right] (x_3,x_4) = \frac{x_{34}^2\,\widetilde{\Psi}^{(\Delta,S-\frac{1}{2},S+\frac{1}{2})}_{x_0,s,\bar{s}}(x_3,x_4)\cdot\cancel{x}_{04}\cancel{x}_{03}\cancel{x}_{34}}{\pi^{4} \left(\Delta+S-1\right) \left(\Delta-S-2\right) \left(\Delta-S-4\right)x_{04}^2}\, ,
\eeq
where $\widehat{G}_2^{(0)}$ is given in \eqref{g2-fermion}. Since $\widehat{G}_2^{(0)}$ is not proportional to $\widehat{B}$, we are not getting back the same eigenvector. Instead, the resulting function is the shadow transform with respect to point $x_1$ and point $x_2$. From the previous equation, we read off
\beq
C(\Delta,S-\tfrac{1}{2},S+\tfrac{1}{2}) = \frac{1}{\pi^{4} \left(\Delta+S-1\right) \left(\Delta-S-2\right) \left(\Delta-S-4\right)}\quad\text{and}\quad\widetilde{G}_0(x_3,x_4)=1\,.
\eeq

The kinematic factors are
\begin{align}
    \mu(\Delta,S-\tfrac{1}{2},S+\tfrac{1}{2}) = \frac{1}{2}(S+\tfrac{1}{2}) \left(\Delta - \tfrac{5}{2}\right) \left(\Delta - \tfrac{3}{2}\right) \left(\Delta - S - 3\right) \left(\Delta + S - 1\right)\, ,
\end{align}
and
\begin{align}
   S\left((1,0,0),\left(\tfrac{3}{2},0,1\right);(4-\Delta,S-\tfrac{1}{2},S+\tfrac{1}{2})\right) = \frac{\ii \pi ^2 \Gamma \left(S-\Delta +3\right) \Gamma \left(\frac{1}{2} \left(S+\Delta -1\right)\right) \Gamma \left(\frac{1}{2} \left(S+\Delta +1\right)\right)} {\left(\Delta - \frac{5}{2}\right) \Gamma \left(\frac{1}{2} \left(S-\Delta +4\right)\right)^2 \Gamma \left(S+\Delta \right)}\, .
\end{align}

Putting everything together, we obtain
\begin{multline}
P\left(\Delta,S-\tfrac{1}{2},S+\tfrac{1}{2}\right)= \frac{- \ii (S+\frac{1}{2}) \left(\Delta - \frac{3}{2}\right) \left(\Delta+S-1\right)^{3} \left(\Delta-S-4\right)} {\xi^4_{4/3}2^{2(\Delta+1)}\pi^{2}\left(\Delta^2 - \left(S+5\right)\Delta + 2S + \frac{11}{2}\right)}\\
    \times\frac{\Gamma\left(\frac{1}{2}\left(S-\Delta+5\right)\right) \Gamma\left(\frac{1}{2}\left(S+\Delta-1\right)\right)}{\Gamma\left(\frac{1}{2}\left(S-\Delta+2\right)\right) \Gamma\left(\frac{1}{2}\left(S+\Delta\right)\right) }\, ,
\end{multline}
where $\Delta=\Delta_2$ or $\Delta=\Delta_4$, are the dimensions given in \eqref{Delta2 fermion} and \eqref{Delta4 fermion}.

The second family of left eigenvectors is
\begin{equation}
\widetilde{\Psi}_{x_{0},s,\bar{s}}^{\left(\Delta,S+\frac{1}{2},S-\frac{1}{2}\right)}\left(x_{1},x_{2}\right) = \frac{s\,\cancel{x}_{01}\,\cancel{\bar{x}}_{12} \left[s\left(\frac{\cancel{x}_{10}}{x_{10}^{2}}-\frac{\cancel{x}_{20}}{x_{20}^{2}}\right) \bar s\right]^{S-\frac{1}{2}}}{|x_{12}|^{\Delta+S+2} |x_{10}|^{-\Delta-S+6} |x_{20}|^{-\Delta-S+4}}\,.
\end{equation}
They satisfy the relation
\begin{multline}
\left[\widetilde{\Psi}_{x_{0},s,\bar s}^{\left(\Delta,S+\frac{1}{2},S-\frac{1}{2}\right)}\widehat{G}_{2}^{\left(0\right)}\right]\left(x_{3},x_{4}\right) = \frac{1}{\pi^{4} \left(-\Delta+S+3\right) \left(\Delta+S-2\right) \left(\Delta+S\right)}\\ 
    \times \frac{s\,\cancel{x}_{40} \left[s\left(\frac{\cancel{x}_{30}}{x_{30}^{2}}-\frac{\cancel{x}_{40}}{x_{40}^{2}}\right) \bar s\right]^{S-\frac{1}{2}}}{|x_{34}|^{\Delta+S-2} |x_{30}|^{-\Delta-S+4} |x_{40}|^{-\Delta-S+6}}\, .
\end{multline}
The resulting function is again the shadow transform with respect to $x_1$ and $x_2$. We have 
\begin{align}
C(\Delta,S+\tfrac{1}{2},S-\tfrac{1}{2})=\frac{1}{\pi^{4} \left(-\Delta+S+3\right) \left(\Delta+S-2\right) \left(\Delta+S\right)}\qquad\text{and}\qquad\widetilde{G}_0(x_3,x_4)=1\, .
\end{align}

Regarding the kinematic factors, one has $\mu(\Delta,S+\frac{1}{2},S-\frac{1}{2}) = \mu(\Delta,S-\frac{1}{2},S+\frac{1}{2})$, but
\begin{align}
   S\left((1,0,0),\big(\tfrac{3}{2},0,1\big);\big(4-\Delta,S+\tfrac{1}{2},S-\tfrac{1}{2}\big)\right) = \frac{\ii \pi ^2 \Gamma \left(S-\Delta +3\right) \Gamma \left(\frac{1}{2} \left(S+\Delta \right)\right)^2}{\left(\Delta - \frac{5}{2}\right) \Gamma \left(\frac{1}{2} \left(S-\Delta +3\right)\right) \Gamma \left(\frac{1}{2} \left(S-\Delta +5\right)\right) \Gamma \left(S+\Delta \right)}\, .
\end{align}

Putting everything together, we obtain
\begin{multline}
    P\left(\Delta_{\frac{7}{2},\pm},S+\tfrac{1}{2},S-\tfrac{1}{2}\right)= \frac{-\ii (S+\frac{1}{2}) 2^{-2\Delta_{\frac{7}{2},\pm}-2} \left(\Delta_{\frac{7}{2},\pm} - \frac{3}{2}\right) \left(\Delta_{\frac{7}{2},\pm}-S-3\right)^{3} \left(\Delta_{\frac{7}{2},\pm}+S-2\right)} {\xi^4_{4/3}\pi^{2}\left(\Delta_{\frac{7}{2},\pm}^2 + \Delta_{\frac{7}{2},\pm}\left(S-3\right) - 2S + \frac{3}{2}\right)}\\
    \times\frac{\Gamma\left(\frac{1}{2}\left(S-\Delta_{\frac{7}{2},\pm}+4\right)\right) \Gamma\left(\frac{1}{2}\left(S+\Delta_{\frac{7}{2},\pm}+2\right)\right)}{\Gamma\left(\frac{1}{2}\left(S-\Delta_{\frac{7}{2},\pm}+5\right)\right) \Gamma\left(\frac{1}{2}\left(S+\Delta_{\frac{7}{2},\pm}-1\right)\right)}\, ,
\end{multline}
where $\Delta_{\frac{7}{2},\pm}$ is given in \eqref{Delta3 fermion}.

\section{Integrability}

In this appendix we review the construction of integrable spin chains with four-dimensional conformal symmetry. We then apply it to demonstrate the integrability of the graph-building operators for all the cases without mixing.

\subsection{General Construction}
\label{gencon}

Integrable spin chains based on the four-dimensional conformal algebra $\mathfrak{so}(5,1)$ can be built for an arbitrary representation $\rho$ at each site of the chain. This is done using the Lax matrices
\begin{equation}
L^{(\rho;\mathbf{4})}(u) = u\operatorname{Id} - \frac{1}{2} q_{MN}^{(\rho)}\otimes \Sigma^{MN}\,.
\end{equation}
This matrix acts on $\mathcal{V}_\rho\otimes \mathbb{C}^4$, with $\mathcal{V}_\rho$ being the physical space in the representation $\rho$, and $\mathbb{C}^4$ is the auxiliary space transforming in the ${\bf 4}$ of $\mathfrak{so}(5,1)$. This matrix is constructed such that it satisfies the Yang--Baxter equation
\begin{equation}
R^{(\mathbf{4};\mathbf{4})}(u-v) L^{(\rho;\mathbf{4})}_1(u) L^{(\rho;\mathbf{4})}_2(v) = L^{(\rho;\mathbf{4})}_2(v) L^{(\rho;\mathbf{4})}_1(u) R^{(\mathbf{4};\mathbf{4})}(u-v)\, .\label{eq: YBeq}
\end{equation}
Here, $\Sigma_{MN} = q^{(\mathbf{4})}_{MN}$ are the $\sigma$-matrices in 6D. They realise one of the two four-dimensional irreducible representations of the conformal algebra. The generators $q_{MN}^{(\rho)}$ satisfy the commutation relations
\begin{equation}
\left[q_{MN}^{(\rho)},q_{KL}^{(\rho)}\right] = \eta_{NK} q_{ML}^{(\rho)} - \eta_{MK} q_{NL}^{(\rho)} - \eta_{NL} q_{MK}^{(\rho)} + \eta_{ML} q_{NK}^{(\rho)}\,.
\end{equation}
The R-matrix is given by
\begin{equation}
R^{(\mathbf{4};\mathbf{4})}(u) = L^{(\mathbf{4};\mathbf{4})}\left(u+\frac{1}{4}\right) = u\operatorname{Id} + \mathbb{P}\, ,
\end{equation}
where $\mathbb{P}$ is the permutation operator on $\mathbb{C}^4\otimes \mathbb{C}^4$. Since $R^{(\mathbf{4};\mathbf{4})}(-1)$ projects onto the irreducible representation $\mathbf{6}\subset \mathbf{4}\otimes\mathbf{4}$, one can use the fusion procedure to compute the Lax matrices acting on $\mathcal{V}_\rho\otimes \mathbb{C}^6$, \cite{Kulish:1981gi,Kulish:1981bi}. We checked, using Mathematica, that they are given by\footnote{The sign in front of the term involving $\epsilon_{MNABCD}$ is related to our convention for the $\sigma$-matrices, which satisfy $\epsilon_{MNABCD} \Sigma^{AB} \Sigma^{CD} = -12\Sigma_{MN}$.}
\begin{multline}\label{Lrho6}
L^{(\rho;\mathbf{6})}(u) = \Bigg[u^2\eta_{MN} - u q_{MN}^{(\rho)}
+ \frac{1}{2}\bigg( ?q^{(\rho),}_M^P? q^{(\rho)}_{PN} - 2?q^{(\rho)}_{MN}?\\ - \frac{C_\rho + 2}{4}\eta_{MN} - \frac{1}{8} ?\epsilon_{MN}^{ABCD}? q^{(\rho)}_{AB} q^{(\rho)}_{CD}\bigg)\Bigg]\otimes ?e^{MN}?\,.
\end{multline}
Here, $?e_M^N?$ is the $6\times 6$ matrix with a single non-zero coefficient, equal to one, at the intersection of row $M$ and column $N$, and $q^{(\mathbf{6})}_{MN} = e_{MN} - e_{NM}$, the indices being lowered using the metric $\eta_{MN}$, see appendix \ref{app:convention}. The quadratic Casimir reads
\begin{equation}
C_\rho = q^{(\rho),MN}q^{(\rho)}_{NM}\, .
\end{equation}
Finally, for the completely antisymmetric tensor in (\ref{Lrho6}) we use the convention where $\epsilon_{123456} = +1$.

As a consequence of the Yang--Baxter equation (\ref{eq: YBeq}), the transfer matrices (or T-operators)
\begin{equation}\label{t-operator}
T_{\mathbf{6}}\left(u;\theta_1,\dots,\theta_N\right)\equiv \tr_{\mathbf{6}} \left(L_{N}^{\left(\rho_N;\mathbf{6}\right)}\left(u-\theta_{N}\right)L_{N-1}^{\left(\rho_{N-1};\mathbf{6}\right)}\left(u-\theta_{N-1}\right)\cdots L_1 ^{\left(\rho_1;\mathbf{6}\right)}\left(u-\theta_1 \right)\right),\end{equation}
commute with each other for different values of the spectral parameter $u$, and fixed values of the inhomogeneities $\theta_1,\dots,\theta_N$.

For each correlator in the sections below, we construct the associated Lax matrices and show that $T_{\mathbf{6}}\left(0;\theta_1,\dots,\theta_N\right)$ coincides with the inverse of the graph-building operator---for an appropriate choice of the inhomogeneities.

\subsection{Principal Series Representations}
\label{Principal Series}

We will be interested in representations belonging to the principal series of $\mathrm{Spin}(5,1)$, (which is a double cover of the connected component of the identity in the conformal group). These representations are labeled by a complex number $\Delta$, the conformal dimension, and two non-negative integers $\ell$ and $\bar\ell$, the spins. They may be realised as acting on functions $f$ on $\mathbb{R}^4\otimes\mathbb{C}^2\otimes \mathbb{C}^2$, such that $f(x,s,\bar{s})$ is homogeneous of degree $\ell$ in $s\in\mathbb{C}^2$ and, separately, of degree $\bar\ell$ in $\bar s\in\mathbb{C}^2$. The generators of the conformal algebra take the following form
\begin{itemize}
    \item rotations $q^{(\Delta, \ell,\bar\ell)}_{\mu \nu} = x_{\mu}\p_{x^\nu} - x_{\nu}\p_{x^\mu} + s^\alpha \left(\sigma_{\mu\nu}\right)?\!_\alpha^\beta? \p_{s^\beta}  + \bar s_{\dot\alpha} \left(\bar \sigma_{\mu\nu}\right)?\!^{\dot\alpha}_{\dot\beta}? \p^{\dot\beta}_{\bar s}\, ,$
    \item translations $q^{(\Delta, \ell,\bar\ell),{\mu +}} = -\p_x^{\mu}\, ,$
    \item dilation $q^{(\Delta, \ell,\bar\ell)}_{56} = -\Delta -x\cdot\p_x\, ,$
    \item special conformal transformations
    
    $q^{(\Delta, \ell,\bar\ell),{\mu -}} = 2x^\mu\left(\Delta + x\cdot \p_x \right) - x^2\p_x^{\mu} + 2x_\nu\left(s^\alpha \left(\sigma^{\mu\nu}\right)?\!_\alpha^\beta? \p_{s^\beta}  + \bar s_{\dot\alpha} \left(\bar \sigma^{\mu\nu}\right)?\!^{\dot\alpha}_{\dot\beta}? \p^{\dot\beta}_{\bar s}\right)\, .$
\end{itemize}
Here we have introduced $\sigma_{\mu\nu} = \frac{\sigma_\mu\bar\sigma_\nu - \sigma_\nu\bar\sigma_\mu}{4}$ and $\bar\sigma_{\mu\nu} = \frac{\bar\sigma_\mu \sigma_\nu - \bar\sigma_\nu \sigma_\mu}{4}$. Introducing the additional notations
\begin{equation}\label{physical space xM}
    x^+ = 1\, ,\quad  x^- = x^2\, ,\quad \p_{x,-} = 0\, ,\quad \p_{x,+} = -\Delta - x\cdot\p_x\, ,
\end{equation}
\begin{equation}\label{physical space sigmaM}
    \sigma^+ = \bar\sigma^+ = 0\, , \quad\sigma^- = 2\,\cancel{x}, \quad \text{and}\quad  \bar\sigma^- = 2\,\cancel{\bar x}\, ,
\end{equation}
we can write all the generators in the compact form
\begin{equation}
    q^{(\Delta, \ell,\bar\ell)}_{MN} = x_{M}\p_{x,N} - x_{N}\p_{x,M} + s^\alpha \left(\sigma_{MN}\right)?\!_\alpha^\beta? \p_{s^\beta}  + \bar s_{\dot\alpha} \left(\bar \sigma_{MN}\right)?\!^{\dot\alpha}_{\dot\beta}? \p^{\dot\beta}_{\bar s}\, .
\end{equation}

In the following, we shall be interested in representations for which either $\ell$ or $\bar\ell$ is zero. In this case, the L-matrices \eqref{Lrho6} become
\begin{multline}\label{Laxl0}
    L^{(\Delta,\ell,0;\mathbf{6})}_{MN}(u) = \left(u^2 - \frac{(\Delta-1)^2}{4}\right)\eta_{MN} - \left(u+\frac{\ell}{4}\right)q^{(\Delta,\ell,0)}_{MN}\\
    + \frac{1}{2}\left[-x_M x_N\Box_x - x_N s\sigma_{M}\cancel{\bar{\p}}_x\p_{s} + \left(1-\Delta+\frac{\ell}{2}\right)\left(x_M\p_{x,N} + x_N\p_{x,M} + s\sigma_{MN} \p_{s} + 2\ell V_M x_N\right)\right]\, ,
\end{multline}
and
\begin{multline}\label{Lax0lbar}
    L^{(\Delta,0,\bar\ell;\mathbf{6})}_{MN}(u) = \left(u^2 - \frac{(\Delta-1)^2}{4}\right)\eta_{MN} - \left(u-\frac{\bar\ell}{4}\right)q^{(\Delta,0,\bar\ell)}_{MN}\\
    + \frac{1}{2}\left[-x_M x_N\Box_x - x_M \bar s \bar{\sigma}_{N}\cancel{\p}_x \p_{\bar s} + \left(1-\Delta+\frac{\bar \ell}{2}\right)\left(x_M\p_{x,N} + x_N\p_{x,M} - \bar s\bar\sigma_{MN} \p_{\bar s} + 2\bar\ell x_M V_N\right)\right]\, ,
\end{multline}
where $V$ is a constant vector: $V^\mu = V^+ = 0$, and $V^- = 1$. These computations were also done using Mathematica.

\subsection{Lax Matrices for Operators Without Gluons}
\label{app:Lax matrices}

We collect here the various Lax matrices required to reproduce the inverse graph-building operators \eqref{BXm1} for $\tr(XX^\dagger Z)$, \eqref{n-scalar-fermion} for $\tr(\psi_4 Z^J)$, and \eqref{gbopsi2Z} for $\tr(\psi_4 \psi^\dagger Z)$. Applying the formulae \eqref{Laxl0} and \eqref{Lax0lbar}, we find
\begin{align}
L_{MN}^{\left(1,0,0;\boldsymbol{6}\right)} \left(0\right) &= -\frac{1}{2}x_{M}x_{N}\Box_{x}\, ,\label{Lax scalar1}\\
L_{MN}^{\left(2,0,0;\boldsymbol{6}\right)} \left(0\right) &= -\frac{1}{4}\eta_{MN}-\frac{1}{2}\left[x_{M}x_{N}\Box_{x}+x_{M}\p_{x,N}+x_{N}\p_{x,M}\right]\, ,\label{Lax scalar2}\\
L_{MN}^{\left(\frac{3}{2},1,0;\boldsymbol{6}\right)} \left(-\frac{1}{4}\right) &= -\frac{1}{2}\left[x_{M}x_{N}\Box_{x} + x_N s\sigma_{M}\cancel{\bar{\p}}_x\p_{s}\right]\, ,\label{Lax fermion}\\
L_{MN}^{\left(\frac{3}{2},0,1;\boldsymbol{6}\right)} \left(\frac{1}{4}\right) &= -\frac{1}{2}\left[x_{M}x_{N}\Box_{x} + x_M \bar s \bar{\sigma}_{N}\cancel{\p}_x \p_{\bar s}\right]\label{Lax antifermion}\, .
\end{align}
We recall that they are all written in four-dimensional space using the notation introduced in \eqref{physical space xM} and \eqref{physical space sigmaM}.

The inverse graph-building operator \eqref{BXm1} for instance is given by
\begin{equation}
    \widehat{B}^{-1} = \tr_{\mathbf{6}}(L_{x_1}^{\left(1,0,0;\boldsymbol{6}\right)} \left(0\right) L_{x_2}^{\left(2,0,0;\boldsymbol{6}\right)} \left(0\right))\, .
\end{equation}

\subsection{Integrability for $\tr(F_{\mu\nu}Z)$}

In the case of the operator $\tr\big(Z(x_1) F^{\mu\nu}(x_2)\big)$ in (\ref{G3-gluon}) we have an antisymmetric rank-2 tensor of dimension 2 at $x_2$. This representation is reducible, given by the direct sum $(2,2,0)\oplus(2,0,2)$. These representations correspond to the self-dual and anti-self-dual parts of the tensor, respectively.\footnote{Note that because the graph-building operator we got is not expressed in terms of spinors, we cannot directly apply the general results of section \ref{Principal Series}.}

In order to lighten the notations, we introduce an auxiliary polarisation vector $\theta$, whose components anticommute among themselves: $\{\theta^\mu,\theta^\nu\}=0$. We are thus interested in functions homogeneous of degree $2$ in $\theta$, $\psi(x,\theta) = \theta^\mu \theta^\nu \psi_{\mu\nu}(x)$. In this representation, the generators of the conformal group take the form
\begin{align}
q_{\mu\nu} &= x_\mu\p_{x^\nu} - x_\nu\p_{x^\mu} + \theta_\mu\p_{\theta^\nu} - \theta_\nu\p_{\theta^\mu}\, ,\label{generators asym1}\\
q^{\mu +} &= -\p^\mu\, ,\\
q_{56} &= - \Delta - x\cdot\p_x\, ,\\
q^{\mu -} &= 2\,x_\mu\left(\Delta + x\cdot\p_x\right) - x^2\p_x^\mu + 2\left(\theta^\mu (x\cdot\p_{\theta}) - (\theta\cdot x) \,\p^{\mu}_\theta\right)\, .\label{generators asym4}
\end{align}
One can then apply \eqref{Lrho6} to obtain the Lax matrices at each site of a chain of length $2$ containing a scalar of dimension 1 (site 1) and an antisymmetric rank-2 tensor of dimension 2 (site 2). The corresponding transfer matrix \eqref{t-operator} becomes
\begin{equation}
    \widehat{T}_{\mathbf{6}}(0;0,0) = \frac{1}{16}\left[(\theta\cdot\p_{x_2}) \, x_{12}^4\, (\p_{x_2}\cdot \p_{\theta}) + (\p_{x_2}\cdot \p_{\theta})\, x_{12}^4\, (\theta\cdot\p_{x_2})\right]\Box_{x_1}\, .
\end{equation}
Clearly, if $\Psi = \theta\cdot\p_2\, \Phi$ for some function $\Phi(x_1,x_2,\theta) = \theta^\nu \Phi_\nu(x_1,x_2)$, then this operator simplifies to
\begin{equation}
    \widehat{T}_{\mathbf{6}}(0;0,0) \Psi = \frac{1}{16} (\theta\cdot\p_{x_2})\, x_{12}^4\, (\p_{x_2}\cdot \p_{\theta}) \Box_{x_1}\Psi\, .
\end{equation}
This is indeed the inverse of (the restriction of) $\widehat{B}_F$, see \eqref{inverse H3}.

\section{One-Loop Checks of the Spectrum}
\label{app:perturbative}

In this appendix we preform a few checks of the spectrum at one-loop order in perturbation theory.

\subsection{The operators $\tr(ZXX^\dagger)$ and $\tr(ZX^\dagger X)$} \label{app:pert-xxbar}

We would like to check the spectrum of $\tr(ZXX^\dagger)$, given in \eqref{eq: neat sol xxbar}, at small $\xi_2$. Expanding \eqref{eq: neat sol xxbar} at $S=0$ we obtain
\begin{equation}\label{eq:pert_spectrum}
\Delta_\pm(0,\xi_2)=3\pm 2\xi_2^2
+O(\xi_2^4)\, .
\end{equation}
Surprisingly the leading term is $ \xi_2^2$ instead of $ \xi_2^4$, even though the graph-building operator is of order $\xi_2^4$. This is an outcome of the fact that we have mixing between the operators\footnote{In the original fishnet model, this was due to the presence of double-trace interactions in the action.}
\beq
\mathcal O_1=\tr(Z X X^\dagger)\qquad\text{and}\qquad\mathcal O_2=\tr(Z X^\dagger X)\,.
\eeq
Let us see explicitly how this spectrum emerges at one-loop order in the perturbative expansion. 

\begin{figure}[t]
\centering
\includegraphics[scale=0.4]{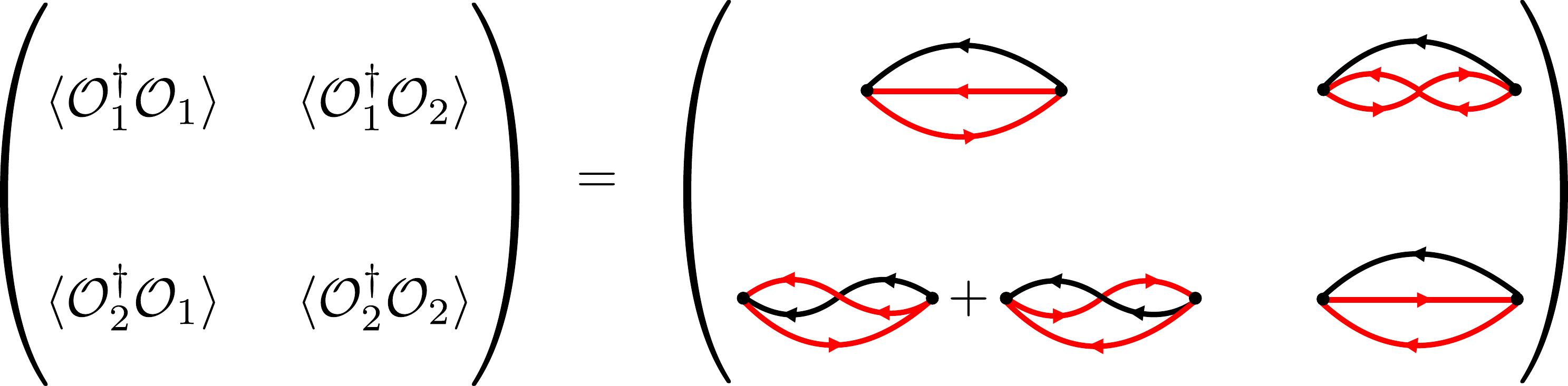}\caption{\small Diagrams contributing to the two-point functions of $\cO_1=\tr(Z X^\dagger X)$ and $\cO_2=\tr(Z X X^\dagger)$. Black lines denotes $Z$ propagator and red lines denote $X$ propagator. The interactions are the same as in \eqref{intxx}.}
\label{fig:perturb_mixing}
\end{figure}

In the case at hand, the dilatation operator is a $2\times 2$ matrix, acting on the operators $\mathcal O_1$ and $\mathcal O_2$. The diagrams that contribute to the two point functions of these operators at leading order in $g^2$ and $\xi_2^2$ are drawn in figure \ref{fig:perturb_mixing}. 
We found that they take the form
\begin{equation}\label{eq:2_pt_matrix}
\begin{pmatrix}\<\cO^\dagger_1\cO_1\>&\<\cO^\dagger_1\cO_2\>\\\<\cO^\dagger_2\cO_1\>&\<\cO^\dagger_2\cO_2\>\end{pmatrix}=
\frac{1}{(4\pi^2)^3 x^6}\begin{pmatrix}1 & g^2 K  \\ 256 \pi^4\frac{\xi_2^4}{g^2} K&1\end{pmatrix}\(1+O(g^2,\xi_2^2)\)\, ,
\end{equation}
where $x$ is the separation between the two operators and $K$ is the integral

\beq \label{integralK}
K=\int{\dd^4 y\over(4\pi^2)^2}{x^4\over y^4(x-y)^4}={1\over4\pi^2}\log(x/\epsilon)+\text{finite}\,,
\eeq
where $\epsilon$ is some UV cutoff length scale. The logarithmic divergence of $K$ comes from the region of integration where $y\to0,x$. 
The corresponding eigenvalues of the matrix (\ref{eq:2_pt_matrix}) are therefore
\begin{equation}\la{Kpm}
    K_\pm =\frac{1\mp4\xi_2^2\( \log (x/\epsilon)+\text{finite}\)}{(4\pi^2)^3x^6}\,.
\end{equation}

From (\ref{Kpm}) we read the anomalous dimensions
\begin{equation}
    \gamma_\pm=\pm 2 \xi_2^2+O( \xi_2^4)\,,
\end{equation}
in agreement with \eqref{eq:pert_spectrum}.

\subsection{The operators $\tr(Z^\dagger X^\dagger \psi_{1\alpha})$ and $\tr(Z^\dagger \psi_{1\alpha}X^\dagger)$}\label{app:fermion_perturbation_theory}

\begin{figure}[h]
\centering
\includegraphics[scale=0.4]{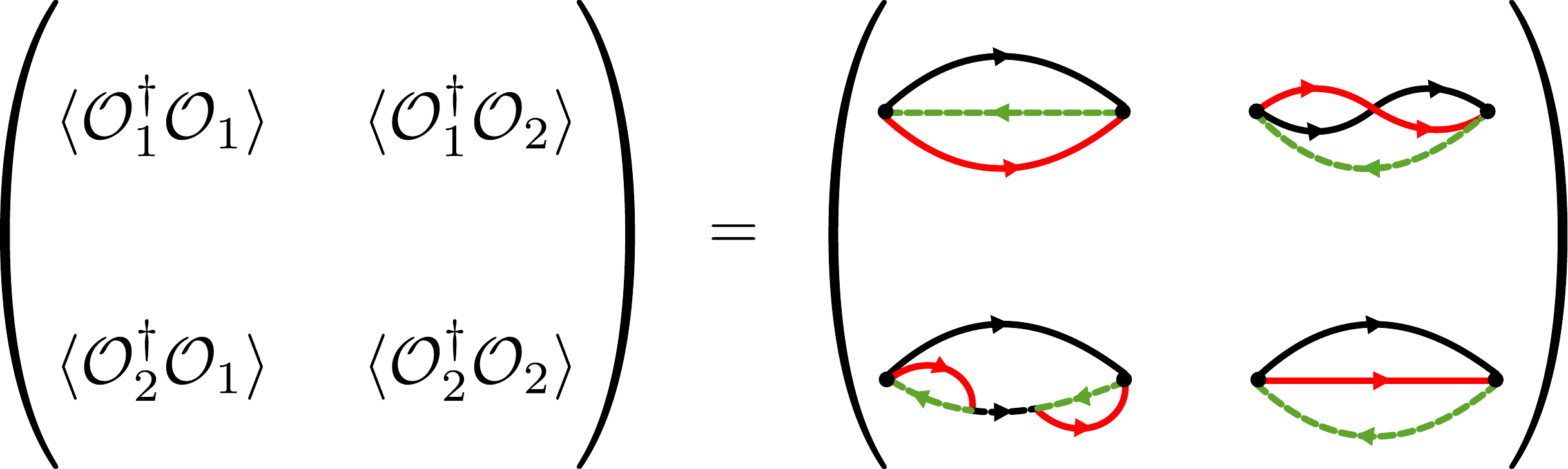}\caption{\small Diagrams contributing to the two-point functions of $\mathcal{O}_{1}=\tr(Z^\dagger\psi_{1\alpha}X^\dagger)$ and $\mathcal{O}_{2}=\tr(Z^\dagger X^\dagger\psi_{1\alpha})$. The black lines represent $Z$ propagator, the red lines are $X$ propagator, and the dashed black lines are fermion propagators of $\psi_1$ (green) and $\psi_4$ (black). The interactions are given in \eqref{Zpsiint}.}
\label{fig:fermion_perturb_mixing}
\end{figure}

Similarly to the case considered above, we compute the dimension of the spin one-half operators
\beq
\mathcal{O}_{1}=\tr(Z^\dagger\psi_{1\alpha}X^\dagger)\qquad\text{and}\qquad\mathcal{O}_{2}=\tr(Z^\dagger X^\dagger\psi_{1\alpha})\,,
\eeq
at one loop order in $\xi_{4/3}$. The diagrams that contribute to these two point functions are drawn in figure \ref{fig:fermion_perturb_mixing}. 
They take the form 
\beq
\begin{pmatrix}\label{fmatrix}\<\cO^\dagger_1\cO_1\>&\<\cO^\dagger_1\cO_2\>\\\<\cO^\dagger_2\cO_1\>&\<\cO^\dagger_2\cO_2\>\end{pmatrix}=\frac{\cancel{x}}{32\pi^6(x^2)^4}
\begin{pmatrix}
 1 & g^{-2/3}\xi_{4/3}^{8/3} \,Y_1  \\  \,g^{2/3} \xi_{4/3}^{4/3}\,Y_2 & 1
\end{pmatrix}\(1+O(g^2,\xi_{4/3}^2)\)\, ,
\eeq

where 
\beq
Y_1=32\pi^{8/3}\,K\,,\qquad Y_2=\frac{1}{2\pi^{2/3}}\int \frac{\dd^4z \,\dd^4y}{\pi^4}\frac{x^4\,\cancel{\bar x}\,\cancel{z}(\cancel{\bar z}-\cancel{\bar y})(\cancel{y}-\cancel{x})}{z^6 (y-z)^4 (y-x)^6}\,.
\eeq
Working in dimensional regularisation, with $d=4- \epsilon$, we find
\beqa
Y_1&=&8 \pi ^{2/3}\left(\frac{1}{\epsilon}+\log(\mu \, x)+3\log(x)+2\gamma+2\log(\pi)+O\left(\epsilon\right)\right)\,,\\
Y_2&=& \frac{1}{\pi ^{2/3}}\left(\frac{1}{\epsilon}+\log(\mu \, x)+3\log(x)+2\gamma-\frac{1}{4}+2\log(\pi)+O\left(\epsilon\right)\right)\,,
\eeqa
where the $d$-dimensional and the 4-dimensional couplings are related through $g_d^2=g_4^2 \mu^\epsilon$. The $3\log(x)$ term results from the $\epsilon$-expansion of the tree-level factor, dressing the $1/\epsilon$ piece. It cancels out against the wave function renormalisation of the operator and hence, it does not contribute to the anomalous dimension. The resulting eigenvalues of the renormalised matrix of two-point functions are given by
\begin{equation}
Y_\pm=
\frac{ \cancel{x}}{32\pi^6(x^2)^4}\(1\mp 2 \sqrt{2} \xi_{4/3}^2 (\log(\mu \,x)+\text{finite})\)\,.
\end{equation}
The corresponding one-loop anomalous dimensions read  
\begin{equation}
\gamma_\pm=\pm\sqrt{2}\xi_{4/3}^2 + O(\xi_{4/3}^4)\,,
\end{equation}
in agreement with \eqref{Delta3 fermion} at $S={1\over2}$.

\section{Embedding Formalism and the Spectrum of $\tr(F_{\mu\nu}Z)$}
\label{app:embedding and spectrum}

Some of the computations of this article are done using the so-called embedding formalism. We review this formalism here in section \ref{embfor}, and we use it in section \ref{app:spectrum} to compute the eigenvalues of the graph-building operator $\widehat{B}_F$ of section \ref{sec:FZ}.

\subsection{Embedding Formalism}
\label{embfor}

We introduce here a realisation of the principal series representations of the conformal group as homogeneous functions on the null cone in six dimensions, with Lorentzian metric $\eta_{MN}$. The degree of homogeneity is $-\Delta$. We will denote by $x,\theta,\zeta$, etc. the four-dimensional vectors and by $Y,\Theta,Z$, etc. the six-dimensional ones. The use of the embedding formalism is quite standard, see (\cite{Costa:2011mg} and references therein. The idea of working in six dimensions, where conformal transformations are linearly realised, dating back to \cite{Dirac:1936fq}. We shall focus here on the scalar and rank-2 antisymmetric tensors representations, (see \cite{Costa:2014rya} for more general cases).

For scalar representations, the map between the two realisations sends a $f:\mathbb{R}^4\rightarrow \mathbb{C}$ of dimension $\Delta$ to
\begin{equation}
    \overline f(Y)\equiv\frac{1}{(Y^+)^\Delta} f\left(\frac{Y^\mu}{Y^+}\right)\, .
\end{equation}
The inverse sends a homogeneous function $F$ on the null cone $\{Y\in\mathbb{R}^6; Y^M Y_M = 0\}$ to
\begin{equation}
    \overline F(x) = F(\bar Y)\, ,
\end{equation}
where
\begin{equation}
    \bar{Y}^\mu = x^\mu\, ,\quad \bar{Y}^+ = 1\, ,\quad \bar{Y}^- = x^2\, .
\end{equation}
Clearly, $\overline{\overline f} = f$ and $\overline{\overline F} = F$, (on the null cone). It is also straightforward to check that the generators $q_{MN}^{(\Delta,0,0)}$ given in subsection \ref{Principal Series} are mapped to 
\begin{equation}
    \mathfrak{q}^{(\Delta,0,0)}_{MN} = Y_M \p_{Y^N} - Y_N \p_{Y^M}\,,
\end{equation}
in embedding space.

For rank-2 antisymmetric representations, there is an additional dependence on an auxiliary polarisation vector $\theta$ (such that $\{\theta^\mu,\theta^\nu\} = 0$) in the usual realisation. In the embedding formalism, the functions $\Psi$ depend on $Y$ and $\Theta$ and satisfy the following requirements:
\begin{itemize}
    \item they are defined on $Y^M Y_M = Y^M \Theta_M  = 0$,
    \item they are homogeneous of degree $-\Delta$ in $Y$ and of degree $2$ in $\Theta$,
    \item they are transverse: $\Psi(Y,\Theta + \alpha Y) = \Psi(Y,\Theta)$ for arbitrary $\alpha$ anticommuting with $\Theta$.
\end{itemize}
A function $\psi$ is now sent to
\begin{equation}\label{Theta-lift}
    \overline \psi(Y,\Theta)\equiv \frac{1}{(Y^+)^\Delta}\psi\left(\frac{Y^\mu}{Y^+}, \Theta^\mu - \Theta^+ \frac{Y^\mu}{Y^+} \right)\, .
\end{equation}
The inverse map sends $\Psi$ to
\begin{equation}
    \overline{\Psi}(x,\theta) = \Psi(\bar Y,\bar \Theta)\, ,
\end{equation}
where
\begin{equation}\label{projection Theta}
    \bar{\Theta}^\mu = \theta^\mu\, ,\quad \bar{\Theta}^+ = 0\, ,\quad \bar{\Theta}^- = 2\, \theta\cdot x\, .
\end{equation}
The generators of conformal transformations, given by \eqref{generators asym1}-\eqref{generators asym4} in physical space, are mapped to
\begin{equation} \label{AS-charge-density}
    \mathfrak{q}_{MN} = Y_M \p_{Y^N} - Y_N \p_{Y^M} + \Theta_M\p_{\Theta^N} - \Theta_N\p_{\Theta^M}\,,
\end{equation}
in embedding space. Moreover, when $\Delta = 2$, the analogue of $\psi = \theta\cdot \p_x \phi$ is $\Psi = \Theta\cdot \p_Y\, \Phi$ with $\Phi$ homogeneous of degree $-1$ in $Y$.

\subsection{Spectrum}
\label{app:spectrum}
To compute the spectrum of $\widehat{B}_F$ in (\ref{BFm1}), we first 
translate \eqref{Phieq} to embedding space. We start with the transfer matrix. The main advantage of this new realisation is that the computations are much simpler. For instance, the Lax matrix \eqref{Lrho6} for a scalar of dimension $1$ (corresponding to the first site of the chain) is easily verified to be
\begin{equation}
    L_Y^{(1,0,0;\mathbf{6})}(u) = \left[u^2\eta^{MN} - u\left(Y^M\p_{Y}^N - Y^M\p_{Y}^N\right)-\frac{1}{2}Y^M Y^N\Box_{Y}\right]\otimes e_{MN}\, .
\end{equation}
For rank-2 antisymmetric representations of dimension 2 (second site), one has
\begin{equation} 
    \mathfrak{q}_{MP} \mathfrak{q}\,?\!^P_N? - 2\, \mathfrak{q}_{MN} = -\Theta\cdot \p_{Y}\, Y_{M} Y_{N}\, \p_{Y}\cdot\p_{\Theta} - \p_{Y}\cdot\p_{\Theta}\, Y_{M} Y_{N}\, \Theta\cdot \p_{Y}\, .
\end{equation}
Consequently, the transfer matrix \eqref{t-operator} at $u=0$ is
\begin{equation}
    \widehat{T}_{\mathbf{6}} (0;0,0) = \frac{1}{4} \left[\Theta\cdot \p_{Y_2}\, (Y_1\cdot Y_2)^2 \, \p_{Y_2}\cdot\p_{\Theta} + \p_2\cdot\p_{\Theta}\, (Y_1\cdot Y_2)^2\, \Theta\cdot \p_{Y_2}\right]\Box_{Y_1}\, .
\end{equation}
The eigenvalue equation $\widehat{T}_{\mathbf{6}} (0;0,0)\Psi = \frac{1}{E}\Psi$ for $\Psi = \Theta\cdot \p_{Y_2} \Phi$ becomes
\begin{equation}\label{eigenvalue Phi}
    \Theta\cdot \p_2\left[(Y_1\cdot Y_2)^2 \, \p_{Y_2}\cdot\p_{\Theta}\, \Theta\cdot \p_{Y_2}\Box_{Y_1} \Phi -\frac{4}{E}\Phi\right] = 0\, .
\end{equation}

As usual, the solutions $\Phi$ should have the form of three-point functions. One of the operators is a scalar of dimension 1 and another is a vector of dimension 1. The only possibility for the third operator is either a symmetric traceless tensor or some mixed symmetry tensor.

\paragraph{Symmetric traceless tensors} We use a four-dimensional auxiliary polarisation vector $\zeta$ to parameterise the tensor structure.  
Because the tensor is traceless, it suffices to consider null polarisation vectors $\zeta^2 = 0$. 
We denote by $Z$ the polarisation vector in embedding space. 
It is defind such that its projection to four dimensions flat space, is similar to \eqref{projection Theta}, reads \begin{equation}\label{projection Zeta}
    \bar{Z}^\mu = \zeta^\mu\, ,\quad \bar{Z}^+ = 0\, ,\quad \bar{Z}^- = 2\, \zeta\cdot x_0\, .
\end{equation}
The three-point function $\Phi$ is therefore a function of $Z$ and $Y_0$, as well as $Y_1$, $Y_2$ and $\Theta$ described above. 
It is defined on the space $Z^2 = Z\cdot Y_0 = Y_i^2 = 0$ and is transverse. Namely, $\Phi(Z+\alpha Y_0) = \Phi(Z)$ for arbitrary $\alpha\in\mathbb{R}$. Moreover, it is an homogeneous function of degree $S$ in $Z$ and of degree $-\Delta$ in $Y_0$.

For $S = 0$, the only possibility is
\begin{equation}
    \Phi\propto \Theta\cdot \p_{Y_2}\left(\frac{1}{(Y_1\cdot Y_2)^{\frac{1-\Delta}{2}} (Y_1\cdot Y_0)^{\frac{1+\Delta}{2}} (Y_2\cdot Y_0)^{\frac{\Delta-1}{2}}}\right)\, ,
\end{equation}
which trivially satisfies \eqref{eigenvalue Phi}, and gives $\Psi = 0$. We can thus only consider $S\geqslant 1$. In that case, there are \textit{a priori} two possible structures:
\begin{equation}
\Phi_1 = \frac{J_{\Theta,Z} K_Z^{S-1}}{(Y_1\cdot Y_2)^{\frac{1+S-\Delta}{2}} (Y_1\cdot Y_0)^{\frac{\Delta+S-1}{2}} (Y_2\cdot Y_0)^{\frac{\Delta+S+1}{2}}}
\end{equation}
and
\begin{equation}
\Phi_2 = \frac{K_{\Theta} K_Z^{S}}{(Y_1\cdot Y_2)^{\frac{3+S-\Delta}{2}} (Y_1\cdot Y_0)^{\frac{\Delta+S+1}{2}} (Y_2\cdot Y_0)^{\frac{\Delta+S+1}{2}}}\, ,
\end{equation}
where
\begin{equation}\label{transverse structures J}
J_{\Theta,Z} = (\Theta\cdot Z) (Y_2\cdot Y_0) - (\Theta \cdot Y_0) (Z\cdot Y_2)\, ,\quad J_{\Theta,H} = (\Theta\cdot H) (Y_2\cdot Y_0) - (\Theta \cdot Y_0) (H\cdot Y_2)\, ,
\end{equation}
and
\begin{equation}\label{transverse structures K}
K_{Z} = (Z\cdot Y_1) (Y_2\cdot Y_0) - (Z \cdot Y_2) (Y_1\cdot Y_0)\, ,\quad K_{\Theta} = (\Theta\cdot Y_1) (Y_2\cdot Y_0) - (\Theta \cdot Y_0) (Y_1\cdot Y_2)\, .
\end{equation}

However, these two structures are related via
\begin{equation}\label{gauge equivalence C1C2}
\Theta \cdot\p_{Y_2}\left(\frac{K_Z^{S}}{(Y_1\cdot Y_2)^{\frac{1+S-\Delta}{2}} (Y_1\cdot Y_0)^{\frac{1+S+\Delta}{2}} (Y_2\cdot Y_0)^{\frac{\Delta+S-1}{2}}}\right) = - S\Phi_1 + \frac{\Delta - 1 - S}{2}\Phi_2\, ,
\end{equation}
which means that they give rise to the same $\Psi$, and are therefore equivalent. We now plug $\Phi_1$ in equation \eqref{eigenvalue Phi}. Firstly,
\begin{equation}
\Box_{Y_1}\Phi_1 \approx -\frac{(\Delta - S - 1)(\Delta + S - 1)}{2} \frac{J_{\Theta,Z} K_Z^{S-1}}{(Y_1\cdot Y_2)^{\frac{3+S-\Delta}{2}} (Y_1\cdot Y_0)^{\frac{1+S+\Delta}{2}} (Y_2\cdot Y_0)^{\frac{-1+S+\Delta}{2}}}
\end{equation}
where $\approx$ means equal up to terms proportional to $Y_2^2, Y_0^2, Z\cdot Y_0$, etc., which we can consistently ignore as they project to zero in physical space and are stable under the various differential operators we apply. Then,
\begin{equation}
(Y_1\cdot Y_2)^2 \, \p_{Y_2}\cdot\p_{\Theta}\, \Theta\cdot \p_{Y_2}\Box_{Y_1} \Phi_1 \approx \frac{(\Delta - S - 1) (\Delta + S - 1) (\Delta - S - 3)}{4}\left[(\Delta - 3)\Phi_1 + \frac{\Delta - S - 5}{2} \Phi_2\right]\, .
\end{equation}
Using \eqref{gauge equivalence C1C2}, one verifies that equation \eqref{eigenvalue Phi} is satisfied and that the eigenvalue is
\begin{equation}
E(\Delta,S,S) = \frac{16}{(\Delta + S - 1)^2 (\Delta - S -3)^2}\, .
\end{equation}

\paragraph*{Mixed symmetry tensors}
In this case, the tensors we are interested in are of the form $f^{\mu\nu\mu_1\cdots\mu_\ell}$ with the following symmetry properties: they are antisymmetric in $(\mu,\nu)$, completely symmetric in $(\mu_1,\dots,\mu_\ell)$, and they also satisfy, for all $k\in\{1,\dots,\ell\}$, 
\begin{equation}\label{additional symmetry}
f^{\mu\nu\mu_1\cdots\mu_\ell} = f^{\mu_k\nu\mu_1\cdots\mu\cdots\mu_\ell} +  f^{\mu\mu_k\mu_1\cdots\nu\cdots\mu_\ell}\,,
\end{equation}
where on the right-hand side, $\mu$ and $\nu$ appear at the $k$'th position. Additionally, the tensors should be completely traceless. In order to get rid of the index structure, we consider $f(\zeta,\eta) = f^{\mu\nu\mu_1\cdots\mu_\ell} \eta_\mu\eta_\nu\zeta_{\mu_1}\dots\zeta_{\mu_\ell}$ for two arbitrary vectors satisfying
\begin{equation}
\zeta^2 = \eta^2 = \zeta\cdot\eta = 0\,,\qquad [\zeta_\mu,\zeta_\nu] = \{\eta_\mu,\eta_\nu\} = 0\, .
\end{equation}
It is important to notice that not all homogeneous polynomials, of degree $2$ in $\eta$ and $\ell$ in $\zeta$, come from a tensor with the symmetry we want. Indeed, we have not taken into account the constraints coming from \eqref{additional symmetry}. In particular, there should exist a polynomial $\tilde f$ such that
\begin{equation}
f = \eta\cdot\p_{\zeta} \tilde f\, .
\end{equation}
One could take $\tilde f(\zeta,\eta) = \frac{2}{\ell+2}f^{\mu\nu\mu_1\cdots\mu_\ell} \zeta_\mu\eta_\nu\zeta_{\mu_1}\dots\zeta_{\mu_\ell}$ for instance. In embedding space, the auxiliary vectors are promoted to $Z$ and $H$ satisfying the same properties and, additionally, we restrict ourselves to $Z\cdot Y_0 = H\cdot Y_0 = 0$. The three-point function $\Phi$ is homogeneous of degree $2$ in $H$, of degree $\ell$ in $Z$, and transverse in both variables
\begin{equation}
\Phi(H+\alpha Y_0,Z) = \Phi (H,Z+Y_0) = \Phi(H,Z)\, ,
\end{equation}
where $\alpha$ anti-commute with $H$. Moreover, there should exist some function $\widetilde{\Phi}$ such that
\begin{equation}\label{constraint mixed symmetry}
\Phi = H\cdot\p_{Z} \widetilde{\Phi} \, .
\end{equation}
As is now usual, the polarisation vectors are projected to \eqref{projection Zeta} and
\begin{equation}
    \bar{H}^\mu = \eta^\mu\, ,\qquad \bar{H}^+ = 0\, ,\qquad \bar{H}^- = 2\, \eta\cdot x_0\, .
\end{equation}
Furthermore, one may easily verify that \eqref{constraint mixed symmetry} implies
\begin{equation}
    \overline{\Phi} = \eta\cdot\p_{\zeta} \overline{\widetilde{\Phi}}\, .
\end{equation}

We now examine the possible structures. The only one which does not involve $\varepsilon^{ABCDEF}$ is
\begin{multline}
\Phi'_1 = \frac{K_H J_{\Theta,H} K_Z^{\ell}}{(Y_1\cdot Y_2)^{\frac{3+\ell-\Delta}{2}} (Y_1\cdot Y_0)^{\frac{1+\ell+\Delta}{2}} (Y_2\cdot Y_0)^{\frac{3+\ell+\Delta}{2}}}\\
= \frac{H\cdot\p_Z}{\ell+1} \left[\frac{J_{\Theta,H} K_Z^{\ell+1}}{(Y_1\cdot Y_2)^{\frac{3+\ell-\Delta}{2}} (Y_1\cdot Y_0)^{\frac{1+\ell+\Delta}{2}} (Y_2\cdot Y_0)^{\frac{3+\ell+\Delta}{2}}}\right]\, ,
\end{multline}
where $J_{\Theta,H}$ and $K_Z$ are defined in \eqref{transverse structures J} and \eqref{transverse structures K} above, and
\begin{equation}
K_{H} = (H\cdot Y_1) (Y_2\cdot Y_0) - (H \cdot Y_2) (Y_1\cdot Y_0)\, .
\end{equation}
Similarly, there is only one structure involving $\varepsilon^{ABCDEF}$
\begin{multline}
\Phi'_2 = \frac{\varepsilon(Y_0,Y_1,Y_2,H K_Z + \ell Z K_H,H,\Theta) K_Z^{\ell-1}}{(Y_1\cdot Y_2)^{\frac{3+\ell-\Delta}{2}} (Y_1\cdot Y_0)^{\frac{1+\ell+\Delta}{2}} (Y_2\cdot Y_0)^{\frac{1+\ell+\Delta}{2}}}\\
= H\cdot\p_Z \left[\frac{\varepsilon(Y_0,Y_1,Y_2,Z,H,\Theta) K_Z^{\ell}}{(Y_1\cdot Y_2)^{\frac{3+\ell-\Delta}{2}} (Y_1\cdot Y_0)^{\frac{1+\ell+\Delta}{2}} (Y_2\cdot Y_0)^{\frac{1+\ell+\Delta}{2}}}\right]\, ,
\end{multline}
with
\begin{equation}
\varepsilon(Y_0,Y_1,Y_2,Z,H,\Theta) = \varepsilon_{ABCDEF} Y_{0}^A Y_{1}^B Y_{2}^C Z^D H^E \Theta^F\, .
\end{equation}
Since $\p_2\cdot\p_{\Theta}\Phi'_2 = 0$ and
\begin{equation}
    (Y_1\cdot Y_2)^2 \Box_{Y_2} \Box_{Y_1}\Phi'_2 \approx \frac{(\Delta + \ell + 1) (\Delta + \ell - 1) (\Delta - \ell - 3) (\Delta - \ell - 5)}{4}\Phi'_2\, ,
\end{equation}
we find that equation \eqref{eigenvalue Phi} holds true, with eigenvalue
\begin{equation}
E(\Delta,\ell+2,\ell) = \frac{16}{(\Delta + \ell + 1) (\Delta + \ell - 1) (\Delta - \ell - 3) (\Delta - \ell - 5)}\, .
\end{equation}

A similar, though more cumbersome, computation, would show that $\Phi'_1$ shares the same eigenvalue. Instead of doing this, we will show that their projections to four dimensional flat space, \eqref{eigenvectors 3 rho'} and \eqref{eigenvectors 3 rho' bis}, are related as
\begin{equation}\label{dual transformation}
    \Phi^{\rho,2}_{x_0,\zeta,\eta} = \frac{1}{S} \left[\epsilon(\zeta,\eta,\p_{\zeta},\p_{\eta}) - \frac{1}{2} \epsilon(\eta,\eta,\p_{\eta},\p_{\eta})\right]\Phi^{\rho,1}_{x_0,\zeta,\eta}\,,
\end{equation}
and
\begin{equation}\label{dual transformation bis}
    \Phi^{\rho,1}_{x_0,\zeta,\eta} = \frac{S}{(S+1)^2} \left[\epsilon(\zeta,\eta,\p_{\zeta},\p_{\eta}) - \frac{1}{2} \epsilon(\eta,\eta,\p_{\eta},\p_{\eta})\right]\Phi^{\rho,2}_{x_0,\zeta,\eta}\, .
\end{equation}
We recall that $\Phi^{\rho,i}_{x_0,\zeta,\eta} = (\eta\cdot\p_{\zeta}) \widetilde{\Phi}^{\rho,i}_{x_0,\zeta,\eta}$ with the tilde function being linear in $\eta$. Moreover,
\begin{equation}
    \left[\epsilon(\zeta,\eta,\p_{\zeta},\p_{\eta}) - \frac{1}{2} \epsilon(\eta,\eta,\p_{\eta},\p_{\eta}), \eta\cdot\p_{\zeta}\right] = 0\, .
\end{equation}
Using this, it is straight forward to check \eqref{dual transformation}. Now that we have proven \eqref{dual transformation}, we insert it into the second equation, which is thus equivalent to
\begin{equation}
    \Phi^{\rho,1}_{x_0,\zeta,\eta} = \frac{1}{(S+1)^2} (\eta\cdot\p_{\zeta}) \left[\epsilon(\zeta,\eta,\p_{\zeta},\p_{\eta})\right]^2 \widetilde{\Phi}^{\rho,1}_{x_0,\zeta,\eta}\, .
\end{equation}
Since $\widetilde{\Phi}^{\rho,1}_{x_0,\zeta,\eta}$ is linear in $\eta$, we can replace $\left[\epsilon(\zeta,\eta,\p_{\zeta},\p_{\eta})\right]^2$ with $\epsilon^{\alpha\beta\gamma\delta} \eta_\alpha \zeta_\beta  \p_{\zeta^\gamma} ?\epsilon^{\mu\nu\rho}_\delta? \zeta_\mu \p_{\eta^\nu} \p_{\zeta^\rho}$ so that the equation becomes
\begin{align*}
    \Phi^{\rho,1}_{x_0,\zeta,\eta} & = \frac{1}{(S+1)^2} (\eta\cdot\p_{\zeta}) \left[ \epsilon^{\alpha\beta\gamma\delta} \eta_\alpha \zeta_\beta  \p_{\zeta^\gamma}  ?\epsilon^{\mu\nu\rho}_\delta? \zeta_\mu \p_{\eta^\nu} \p_{\zeta^\rho} \right] \widetilde{\Phi}^{\rho,1}_{x_0,\zeta,\eta}\\
    & = \frac{1}{S+1} (\eta\cdot\p_{\zeta}) \left[(\eta\cdot \p_{\eta}) (\zeta\cdot \p_{\zeta}) - \zeta^\beta (\eta\cdot\p_{\zeta}) \p_{\eta^\beta} \right] \widetilde{\Phi}^{\rho,1}_{x_0,\zeta,\eta}\\
    & = (\eta\cdot\p_{\zeta}) \widetilde{\Phi}^{\rho,1}_{x_0,\zeta,\eta} = \Phi^{\rho,1}_{x_0,\zeta,\eta}\, ,
\end{align*}
where we have used the fact that $\eta\cdot \zeta = \zeta^2 = (\eta\cdot\p_{\zeta})^2 = 0$ and that $\widetilde{\Phi}^{(\rho'),1}_{x_0,\zeta,\eta}$ is homogeneous of degree $S$ in $\zeta$.

\section{Graph-Building Operator for Section \ref{sec:mixing}}
\label{app:mixing}

In this appendix, we give the explicit expressions for the matrix elements $\widehat{\mathcal{B}}_{ij}$ of the matrix graph-building operator introduced in section \ref{sec:mixing}, see (\ref{hres FZJ}). In order to emphasise that the operators do not all act on the same space, we will use the notations $\Psi_\emptyset, \Psi_A,$ and $\Psi_X$ for functions belonging to different spaces. The first one is a function of $J$ variables $(x_1,\dots,x_J)$, but the other two also depend on $x_0$. Moreover, $\Psi_A$ is linear in $\theta$, i.e. carries one index.

There is a small freedom in the way we define the graph-building operator. For instance, one could replace $\widehat{\mathcal{B}}$ with $\mathcal{D} \widehat{\mathcal{B}} \mathcal{D}^{-1}$ for an arbitrary constant diagonal matrix $\mathcal{D}$. In general, we also expect the operator $\widehat{G}_0$ to be promoted to a matrix, and that different matrix elements of $\frac{1}{1-\widehat{\mathcal{B}}} \widehat{G}_0$ correspond to correlators of different non-local operators. In this paper, we only study the first few orders in $g$ of the graph-building operator, and it is possible to take $\widehat{G}_0$ to be diagonal. With our convention, we can then write

\begin{multline}
    \langle \tr(Z(x_1)\dots Z(x_J)) \,\tr(Z^\dagger(z_J)\dots Z^\dagger(z_1)) \rangle\\
    = \sum_{i=1}^J \int\frac{\prod_{i=1}^J \dd^4 y_i}{\pi^{2J}} \frac{ \bra{x_i,x_{i+1},\dots,x_{i-1}} \left(\frac{1}{1 - \widehat{\mathcal{B}}}\right)_{\emptyset\emptyset} \ket{y_1,\dots,y_J}}{(4\pi^2)^J \prod_{j=1}^J (y_j - z_j)^2} \, ,
\end{multline}
where $x_{J+1}=x_1$. Similarly, 
\begin{multline}
    \langle \tr(A^{\mu}(x_0) Z(x_1)\dots Z(x_J))\, \tr(Z^\dagger(z_J)\dots Z^\dagger(z_1))\rangle\\
    = -\frac{\ii}{2} \int\frac{\prod_{i=1}^J \dd^4 y_i}{\pi^{2J}} \frac{ \bra{x_0,x_1,\dots,x_J} \left(\frac{1}{1 - \widehat{\mathcal{B}}}\right)^\mu_{A\emptyset} \ket{y_1,\dots,y_J}}{(4\pi^2)^J \prod_{i=1}^J (y_i - z_i)^2} \, ,
\end{multline}
and
\begin{multline}
    \langle \tr(Z(x_1)\dots Z(x_J))\,\tr(X^\dagger(z_J) Z^\dagger(z_J)\dots Z^\dagger(z_1) X(z_1))\rangle\\
    = g\sum_{i=1}^J \int\frac{\prod_{i=0}^J \dd^4 y_i}{\pi^{2(J+1)}} \frac{ \bra{x_i,x_{i+1},\dots,x_{i-1}} \left(\frac{1}{1 - \widehat{\mathcal{B}}}\right)_{\emptyset X} \ket{y_0,y_1,\dots,y_J}}{(4\pi^2)^{J+2} (y_0 - z_1)^2 (y_0 - z_J)^2 \prod_{j=1}^J (y_j - z_{j})^2} \, ,
\end{multline}
for instance. Ultimately, the only important point is to define $\widehat{\mathcal{B}}$ and $\widehat{G}_0$ such that a given matrix element of $\frac{1}{1 - \widehat{\mathcal{B}}}  \widehat{G}_0$ contains all the Feynman diagrams appearing in the expansion of a specific correlator, and only those diagrams.

The fishnet graph-building operator is
\begin{equation}
    \left[\widehat{\mathcal{B}}_{\emptyset\emptyset} \Psi_\emptyset\right](x_1,\dots,x_J) = 8\pi^2 \int \frac{\prod_{i=1}^J \dd^4 y_i}{\pi^{2J}}\frac{\Psi_\emptyset(y_1,\dots,y_J)}{\prod_{i=1}^J (x_i - y_i)^2 y_{i,i+1}^2} 
\end{equation}
with inverse $\widehat{\mathcal{B}}_{\emptyset\emptyset}^{-1} = (8\pi^2)^{-1} \prod_{i=1}^J x_{i,i+1}^2 \prod_{i=1}^{J} \frac{\Box_{x_i}}{-4}$. The other operators are
\begin{align}
    \left[\widehat{\mathcal{B}}_{\emptyset A} \Psi_A\right](x_1,\dots,x_J) &= 4\pi^2 \int\frac{\prod_{i=0}^J \dd^4 y_i}{\pi^{2(J+1)}}\frac{\Psi_A^\rho(y_0,y_1,\dots,y_J)}{\prod_{i=1}^J (x_i - y_i)^2 y_{0J}^2 \prod_{i=0}^{J-1} y_{i,i+1}^2} \left(\frac{y_{J0,\rho}}{y_{J0}^2} - \frac{y_{10,\rho}}{y_{10}^2}\right)\,,\\
    \left[\widehat{\mathcal{B}}_{\emptyset X} \Psi_X\right](x_1,\dots,x_J) &= 2\pi^2 \int\frac{\prod_{i=0}^J \dd^4 y_i}{\pi^{2(J+1)}} \frac{\Psi_X(y_0,y_1,\dots,y_J)}{\prod_{i=1}^J (x_i - y_i)^2 y_{0J}^2 \prod_{i=0}^{J-1} y_{i,i+1}^2}\,,\\
    \left[\widehat{\mathcal{B}}_{A \emptyset} \Psi_\emptyset\right]^\mu(x_0,\dots,x_J) &= 4\pi^2\int\frac{\dd^4 z\prod_{i=1}^J \dd^4 y_i}{\pi^{2(J+1)}} \frac{\Delta^\mu_\sigma(x_0 - z) \Psi_\emptyset(y_1,\dots,y_J)}{\prod_{i=1}^J  y_{i,i+1}^2} \sum_{i=1}^J \frac{y_{i,i+1}^2}{\prod_{j=1}^J (x_j - y_{i+j})^2} \nonumber\\
    &\quad\times \frac{1}{(y_{i}-z)^2 (y_{i+1} - z)^2}\left( \frac{(y_{i}-z)^\sigma}{(y_{i}-z)^2} -  \frac{(y_{i+1}-z)^\sigma}{(y_{i+1}-z)^2}\right)\,,\\
    \left[\widehat{\mathcal{B}}_{X \emptyset} \Psi_\emptyset\right](x_0,\dots,x_J) &= 2 \int\frac{\prod_{i=1}^J \dd^4 y_i}{\pi^{2J}} \frac{\Psi_\emptyset(y_1,\dots,y_J)}{\prod_{i=1}^J y_{i,i+1}^2} \sum_{i=1}^J \frac{y_{i,i+1}^2}{\prod_{j=1}^J (x_j - y_{i+j})^2 (x_0 - y_i)^2 (x_0 - y_{i+1})^2} \label{HresXO}\,,
\end{align}
and
\begin{multline}\label{HresAA}
    \left[\widehat{\mathcal{B}}_{AA} \Psi_A\right]^\mu(x_0,\dots,x_J) = 4\pi^2\int\frac{\dd^4 z \prod_{i=1}^J \dd^4 y_i}{\pi^{2(J+1)}}\[\frac{\Delta^\mu_\sigma(x_0 - z) \Psi_A^\sigma(z,y_1,\dots,y_J)}{\prod_{i=1}^J (x_i-y_i)^2 \prod_{i=1}^{J-1} y_{i,i+1}^2 (z - y_1)^2 (z - y_J)^2}\right.\\
    +\frac{1}{2} \sum_{i=1}^{J-1}\int {\dd^4y_0\over\pi^2}\frac{\Delta^\mu_\sigma(x_0 - z) \Psi_A^\rho(y_0,\dots,y_J)}{\prod_{j=1}^J (x_j - y_{i+j})^2 y_{0J}^2 \prod_{i=0}^{J-1}  y_{j,j+1}^2} \left(\frac{y_{J0,\rho}}{y_{J0}^2} - \frac{y_{10,\rho}}{y_{10}^2}\right)\\
    \times \frac{y_{i,i+1}^2}{(y_{i}-z)^2 (y_{i+1}-z)^2} \left(\frac{(y_{i}-z)^\sigma}{(y_{i}-z)^2} - \frac{(y_{i+1} - z)^\sigma}{(y_{i+1} - z)^2}\right)\\ 
    + \frac{1}{4} \int{\dd^4y_0\over\pi^2}\frac{\Delta^\mu_\sigma(z-x_0) \Psi_A^\rho(y_0,\dots,y_J)}{\prod_{j=1}^J (x_j - y_{j})^2 \prod_{j=1}^{J-1}  y_{j,j+1}^2}\Bigg[\frac{1}{(y_J-z)^2} \frac{\p}{\p z_\sigma} \left(\frac{y_{10,\rho}}{y_{10}^4 (z-y_0)^2} - \frac{(z-y_0)_\rho}{(z-y_0)^4 y_{10}^2}\right) \\
   - \left(\frac{y_{10,\rho}}{y_{10}^4 (z-y_0)^2} - \frac{(z-y_0)_\rho}{(z-y_0)^4 y_{10}^2}\right) \frac{\p}{\p z_\sigma} \frac{1}{(y_J-z)^2} + \frac{1}{(y_1-z)^2} \frac{\p}{\p z_\sigma} \left(\frac{y_{J0,\rho}}{y_{J0}^4 (z-y_0)^2} - \frac{(z-y_0)_\rho}{(z-y_0)^4 y_{J0}^2}\right) \\
   \left.- \left(\frac{y_{J0,\rho}}{y_{J0}^4 (z-y_0)^2} - \frac{(z-y_0)_\rho}{(z-y_0)^4 y_{J0}^2}\right) \frac{\p}{\p z_\sigma} \frac{1}{(y_1-z)^2}\Bigg]\]\,.
\end{multline}
Similarly,
\begin{multline}
\left[\widehat{\mathcal{B}}_{AX} \Psi_X\right](x_0,\dots,x_J) = \pi^2 \int\frac{\dd^4 z \prod_{i=0}^J \dd^4 y_i}{\pi^{2(J+2)}}\frac{\Delta^\mu_\sigma(z-x_0) \Psi_X (y_0,\dots,y_J)}{y_{0J}^2 \prod_{i=0}^{J-1} y_{i,i+1}^2}\\
    \times \Bigg[\sum_{i=0}^{J-1} \frac{y_{i,i+1}^2}{\prod_{j=1}^J (x_j - y_{i+j})^2 (y_{i}-z)^2 (y_{i+1}-z)^2} \left(\frac{(y_{i}-z)^\sigma}{(y_{i}-z)^2} - \frac{(y_{i+1} - z)^\sigma}{(y_{i+1} - z)^2}\right)\\
    + \frac{y_{J0}^2}{\prod_{i=1}^J (x_i - y_{i})^2 (y_{0}-z)^2 (y_{J}-z)^2} \left(\frac{(y_{J}-z)^\sigma}{(y_{J}-z)^2} - \frac{(y_{0} - z)^\sigma}{(y_{0} - z)^2} \right) \Bigg] \, ,
\end{multline}
\begin{multline}
\left[\widehat{\mathcal{B}}_{XA} \Psi_A\right]^\mu(x_0,\dots,x_J) = \int\frac{\prod_{i=0}^J \dd^4 y_i}{\pi^{2(J+1)}}\frac{\Psi_A^\rho(y_0,\dots,y_J)}{\prod_{i=1}^{J-1} y_{i,i+1}^2} \Bigg[ \frac{1}{\prod_{i=1}^J (x_i-y_i)^2 (x_0 - y_J)^2 (x_0 - y_1)^2}\\
    \times \left( \frac{(x_0 - y_1)^2}{y_{10}^2 (x_0-y_0)^2}\left(\frac{(x_0-y_0)_\rho}{(x_0-y_0)^2} - \frac{y_{10,\rho}}{y_{10}^2} \right) + \frac{(x_0 - y_J)^2}{y_{J0}^2 (x_0-y_0)^2} \left( \frac{y_{J0,\rho}}{y_{J0}^2} - \frac{(x_0-y_0)_\rho}{(x_0-y_0)^2} \right) \right)\\
    + \frac{1}{y_{10}^2 y_{J0}^2}\left(\frac{y_{J0,\rho}}{y_{J0}^2} - \frac{y_{10,\rho}}{y_{10}^2} \right) \sum_{i=1}^{J-1} \frac{y_{i,i+1}^2}{\prod_{j=1}^J (x_j - y_{i+j})^2 (x_0 - y_i)^2 (x_0 - y_{i+1})^2}\Bigg]\, ,
\end{multline}
and
\begin{multline}
\left[\widehat{\mathcal{B}}_{XX} \Psi_X\right](x_0,\dots,x_J) = \frac{1}{2} \int\frac{\prod_{i=0}^J \dd^4 y_i}{\pi^{2(J+1)}}\frac{\Psi_X(y_0,\dots,y_J)}{y_{0J}^2 \prod_{i=0}^{J-1} y_{i,i+1}^2}\\
\times\Bigg[ \frac{1}{\prod_{i=0}^J (x_i-y_i)^2} \left(\frac{y_{01}^2}{(x_0 - y_1)^2} + \frac{y_{0J}^2}{(x_0 - y_J)^2}\right)
    + \sum_{i=1}^{J-1} \frac{y_{i,i+1}^2}{\prod_{j=1}^J (x_j - y_{i+j})^2 (x_0 - y_i)^2 (x_0 - y_{i+1})^2}\Bigg] \, ,
\end{multline}
Note that the dependence on $g^2$ and $\xi^2_{1+1/J}$ has been factored out as in (\ref{hres FZJ}) and (\ref{EtoDelta}). Note also that in the matrix elements involving a gluon, there are also diagrams where the gluon is absorbed or emitted by one of the $Z$ propagators. These are, however, corrections of order $g^2$ to $\mathcal{B}_{A\emptyset}$ and  $\mathcal{B}_{\emptyset A}$. Hence, they do not contribute at leading order in $g$. As explained in the main text, even if we had included them here, the subtraction performed in \eqref{HtildeHres} would have removed them. 

As explained around (\ref{HPhi mixing}), the spectrum is deducted from the eigenvalues of the matrix $\mathfrak{B}_{ij} = \widehat{\mathcal{B}}_{ij} - \widehat{\mathcal{B}}_{i\emptyset} \widehat{\mathcal{B}}_{\emptyset\emptyset}^{-1} \widehat{\mathcal{B}}_{\emptyset j}$ for $(i,j)\in\{A,X\}^2$. We shall now compute these matrix elements and show that they are intertwined with $\mathfrak{B}_{ij}$ for $(i,j)\in\{F,X\}^2$ introduced in section \ref{sec:mixing} in the following way:
\begin{equation}
    \begin{pmatrix}
    \theta\cdot\p_{x_0} & 0\\
    0 & \mathrm{Id}
    \end{pmatrix}
    \begin{pmatrix}
    \mathfrak{B}_{AA} & \mathfrak{B}_{AX}\\
    \mathfrak{B}_{XA} & \mathfrak{B}_{XX}
    \end{pmatrix} =
    \begin{pmatrix}
    \mathfrak{B}_{FF} & \mathfrak{B}_{FX}\\
    \mathfrak{B}_{XF} & \mathfrak{B}_{XX}
    \end{pmatrix}
    \begin{pmatrix}
    \theta\cdot\p_{x_0} & 0\\
    0 & \mathrm{Id}
    \end{pmatrix}\, .
\end{equation}

Using the previous expressions, it is clear that $\mathfrak{B}_{XX}$ is given by equation \eqref{HXX}. It is also straightforward to check that
\begin{equation}
    \theta\cdot\p_{x_0}\,  \mathfrak{B}_{AX} \Psi_X = \mathfrak{B}_{FX} \Psi_X  \, .
\end{equation}

One then computes
\begin{multline}
    \left[\mathfrak{B}_{XA} \Psi_A\right](x_0,\dots,x_J) = \int\frac{\prod_{i=0}^J \dd^4 y_i}{\pi^{2(J+1)}}\frac{\Psi_A^\rho(y_0,\dots,y_J)}{\prod_{i=1}^J (x_i-y_i)^2 \prod_{i=1}^{J-1} y_{i,i+1}^2 (x_0 - y_J)^2 (x_0 - y_1)^2}\\ \times \Bigg[\frac{y_{1J}^2}{y_{10}^2 y_{J0}^2}\left(\frac{y_{10,\rho}}{y_{10}^2} - \frac{y_{J0,\rho}}{y_{J0}^2}\right)
    + \frac{(x_0 - y_1)^2}{y_{10}^2 (x_0-y_0)^2}\left(\frac{(x_0-y_0)_\rho}{(x_0-y_0)^2} - \frac{y_{10,\rho}}{y_{10}^2} \right) + \frac{(x_0 - y_J)^2}{y_{J0}^2 (x_0-y_0)^2} \left( \frac{y_{J0,\rho}}{y_{J0}^2} - \frac{(x_0-y_0)_\rho}{(x_0-y_0)^2} \right) \Bigg] \, .
\end{multline}
If we define $\Psi_F^{\mu\nu} = \frac{1}{2}\left( \p_{x_0}^\mu \Psi_A^\nu - \p_{x_0}^\nu \Psi_A^\mu\right)$, i.e. $\Psi_F = \theta\cdot\p_{x_0} \Psi_A$, and use the integral formula \eqref{secondterm}, the previous equation can be rewritten
\begin{equation}
    \mathfrak{B}_{XA}\Psi_A =  \mathfrak{B}_{XF} \Psi_F\, .
\end{equation}

Finally,
\begin{multline}
    \left[\mathfrak{B}_{AA} \Psi_A\right]^\mu (x_0,\dots,x_J) = 4\pi^2 \int\frac{\dd^4 z \prod_{i=1}^J \dd^4 y_i}{\pi^{2(J+1)}}\frac{\Delta^\mu_\sigma(x_0 - z) \Psi_A^\sigma(z,y_1,\dots,y_J)}{\prod_{i=1}^J (x_i-y_i)^2 \prod_{i=1}^{J-1} y_{i,i+1}^2 (z-y_1)^2 (z-y_J)^2}  \\
    + \pi^2 \int\frac{\dd^4 z\prod_{i=0}^J \dd^4 y_i}{\pi^{2(J+2)}}\frac{\Delta^\mu_\sigma(x_0 - z) \Psi_A^\rho(y_0,\dots,y_J)}{\prod_{j=1}^J (x_j - y_{j})^2 \prod_{j=1}^{J-1}  y_{j,j+1}^2}\Bigg[\frac{1}{(y_J-z)^2} \frac{\p}{\p z_\sigma} \left(\frac{y_{10,\rho}}{y_{10}^4 (z-y_0)^2} - \frac{(z-y_0)_\rho}{(z-y_0)^4 y_{10}^2}\right) \\
   - \left(\frac{y_{10,\rho}}{y_{10}^4 (z-y_0)^2} - \frac{(z-y_0)_\rho}{(z-y_0)^4 y_{10}^2}\right) \frac{\p}{\p z_\sigma} \frac{1}{(y_J-z)^2} + \frac{1}{(y_1-z)^2} \frac{\p}{\p z_\sigma} \left(\frac{y_{J0,\rho}}{y_{J0}^4 (z-y_0)^2} - \frac{(z-y_0)_\rho}{(z-y_0)^4 y_{J0}^2}\right) \\
   - \left(\frac{y_{J0,\rho}}{y_{J0}^4 (z-y_0)^2} - \frac{(z-y_0)_\rho}{(z-y_0)^4 y_{J0}^2}\right) \frac{\p}{\p z_\sigma} \frac{1}{(y_1-z)^2}\\
   + \frac{2\, y_{1J}^2}{(y_1 - z)^2 (y_J - z)^2 y_{10}^2 y_{J0}^2} \left( \frac{(y_{1}-z)^\sigma}{(y_{1}-z)^2} -  \frac{(y_{J}-z)^\sigma}{(y_{J}-z)^2}\right) \left(\frac{y_{J0,\rho}}{y_{J0}^2} - \frac{y_{10,\rho}}{y_{10}^2}\right) \Bigg]\, .
\end{multline}
We first rewrite this expression as
\begin{multline}
    \left[\mathfrak{B}_{AA} \Psi_A\right]^\mu (x_0,\dots,x_J) = 4\pi^2 \int\frac{\dd^4 z \prod_{i=1}^J \dd^4 y_i}{\pi^{2(J+1)}}\frac{\Delta^\mu_\sigma(x_0 - z) \Psi_A^\sigma(z,y_1,\dots,y_J)}{\prod_{i=1}^J (x_i-y_i)^2 \prod_{i=1}^{J-1} y_{i,i+1}^2 (z-y_1)^2 (z-y_J)^2}  \\
    + \pi^2 \int \frac{\Delta^\mu_\sigma(x_0 - z) \Psi_A^\rho(y_0,\dots,y_J)}{\prod_{j=1}^J (x_j - y_{j})^2 \prod_{j=1}^{J-1}  y_{j,j+1}^2 (z-y_1)^2 (z-y_J)^2} \Bigg[ \frac{\p}{\p z_\sigma} \frac{(y_1-z)^2}{y_{10}^2 (z-y_0)^2} \left(\frac{y_{10,\rho}}{y_{10}^2} - \frac{(z-y_0)_\rho}{(z-y_0)^2}\right) \\
    + \frac{\p}{\p z_\sigma} \frac{(y_J-z)^2}{y_{J0}^2 (z-y_0)^2} \left(\frac{y_{J0,\rho}}{y_{J0}^2} - \frac{(z-y_0)_\rho}{(z-y_0)^2}\right) + 2 \left( \frac{(y_{1}-z)^\sigma}{(y_{1}-z)^2} -  \frac{(y_{J}-z)^\sigma}{(y_{J}-z)^2}\right)\bigg[\frac{y_{1J}^2}{y_{10}^2 y_{J0}^2}  \left(\frac{y_{J0,\rho}}{y_{J0}^2} - \frac{y_{10,\rho}}{y_{10}^2}\right)\\
    + \frac{(z-y_1)^2}{y_{10}^2 (z-y_0)^2}  \left(\frac{y_{10,\rho}}{y_{10}^2} - \frac{(z-y_0)_\rho}{(z-y_0)^2}\right) + \frac{(z-y_J)^2}{y_{J0}^2 (z-y_0)^2}  \left(\frac{(z-y_0)_\rho}{(z-y_0)^2} - \frac{y_{J0,\rho}}{y_{J0}^2}\right)\bigg] \Bigg] \frac{\dd^4 z\prod_{i=0}^J \dd^4 y_i}{\pi^{2(J+2)}}\, .
\end{multline}
Applying equations \eqref{ibp 1} and \eqref{secondterm} allows to write the right-hand side as an integral over $\Psi_F^{\mu\nu} = \frac{1}{2}\left(\p_{x_0}^\mu \Psi_A^\nu - \p_{x_0}^\nu \Psi_A^\mu\right)$, the result is
\begin{multline}
    \left[\mathfrak{B}_{AA} \Psi_A\right]^\mu(x_0,\dots,x_J) = 4\pi^2 \int\frac{\dd^4 z \prod_{i=0}^J \dd^4 y_i}{\pi^{2(J+2)}}\frac{\Delta^\mu_\sigma(x_0 - z) \Psi_F^{\rho\tau}(y_0,\dots,y_J)}{\prod_{i=1}^J (x_i - y_i)^2 \prod_{i=1}^{J-1} y_{i,i+1}^2 (z-y_1)^2 (z-y_J)^2}\\
    \times \Bigg[\frac{(z - y_0)_\rho}{(z-y_0)^4} \delta_\tau^\sigma + \left(\frac{(y_J - z)^\sigma}{(y_J - z)^2} - \frac{(y_1 - z)^\sigma}{(y_1 - z)^2}\right) \bigg( \frac{(y_0 - z)_\rho y_{0J,\tau}}{(y_0 - z)^2 y_{0J}^2} + \frac{y_{0J,\rho} y_{01,\tau}}{y_{0J}^2 y_{01}^2}\\
    + \frac{y_{01,\rho} (y_0 - z)_\tau}{y_{01}^2 (y_0 - z)^2}\bigg) + \frac{1}{2} \frac{\p}{\p z_\sigma} \frac{(y_0 - z)_\rho}{(y_0 - z)^2}\left(\frac{y_{01,\tau}}{y_{01}^2} + \frac{y_{0J,\tau}}{y_{0J}^2}\right)\Bigg] \, .
\end{multline}
This implies that, if $\Psi_F = \theta\cdot\p_{x_0} \Psi_A$, then
\begin{equation}
    \theta\cdot\p_{x_0}\,  \mathfrak{B}_{AA} \Psi_A = \mathfrak{B}_{FF} \Psi_F\, .
\end{equation}

Finally, we state without proof two useful properties:
\begin{equation}
    \p_{x_0} \cdot \p_{\theta} \mathfrak{B}_{AX}\Big|_{\alpha = 1} = \p_{x_0} \cdot \p_{\theta} \mathfrak{B}_{AA}\Big|_{\alpha = 1} = 0\,,
\end{equation}
where $\alpha$ parameterises the gluon propagator in the $R_\alpha$ gauge, (\ref{gluonprop}). These relations are useful when checking that $\mathfrak{B}^{-1}\mathfrak{B} = \mathrm{Id}$.

\end{appendix}

\bibliography{refs.bib}
\bibliographystyle{JHEP.bst}

\end{document}